\documentclass[twocolumn,notitlepage,english,aps,prb,nofootinbib]{revtex4-1}
\usepackage{CJK}

\usepackage{amsfonts}
\usepackage{amsmath}
\usepackage{ dsfont }
\usepackage{hyperref}
\usepackage{amssymb}
\usepackage{array,multirow}
\usepackage{longtable}
\usepackage{ragged2e}
\usepackage{xspace}
\usepackage{xcolor}
\usepackage{relsize}
\usepackage[bottom]{footmisc}
\usepackage{color}
\usepackage{graphicx}
\usepackage[space]{grffile}
\usepackage{verbatim}
\usepackage{amssymb}
\usepackage{mathrsfs}
\usepackage{wasysym}
\usepackage[caption=false]{subfig}
\usepackage{url}
\usepackage{bbold}
\usepackage{slashed}
\usepackage{epstopdf}
\usepackage{braket}
\usepackage{float}
\usepackage[percent]{overpic}

\usepackage{tikz}
\usetikzlibrary{calc}

\newcommand{\be}{\begin{equation}}
\newcommand{\ee}{\end{equation}}
\newcommand{\bea}{\begin{equation} \begin{aligned}}
\newcommand{\eea}{\end{aligned} \end{equation} }
\newcommand{\bi}{\begin{itemize}}
\newcommand{\ei}{\end{itemize}}

\newcolumntype{C}[1]{>{\centering\arraybackslash}p{#1}}

\usepackage{xspace}

\newcommand{\la}{\lambda}
\renewcommand{\be}{\beta}
\newcommand{\al}{\alpha}
\newcommand{\eps}{\epsilon}

\renewcommand{\th}{\theta}
\newcommand{\lp}{\left(}
\newcommand{\rp}{\right)}

\newcommand{\del}{\partial}

\newcommand{\Tr}{\text{Tr} \ }

\newcommand{\mbf}[1]{\mathbf{#1}}

\newcommand{\bpm}{\begin{pmatrix}}
\newcommand{\epm}{\end{pmatrix}}
\renewcommand\arraystretch{1.3}
\renewcommand{\u}{\uparrow}
\renewcommand{\d}{\downarrow}

\newcommand{\crea}[1]{#1^{\dag}}
\newcommand{\ani}[1]{#1^{\vphantom{\dag}}}
\newcommand{\brakett}[2]{\langle#1|#2\rangle}
\newcommand{\ave}[1]{\langle#1\rangle}
\renewcommand{\vec}[1]{\boldsymbol{#1}}

\DeclareRobustCommand{\App}[1]{App.~\ref{#1}}

\DeclareRobustCommand{\Tab}[1]{Table~\ref{#1}}

\DeclareRobustCommand{\Fig}[1]{Fig.~\ref{#1}}

\DeclareRobustCommand{\Eq}[1]{Eq.~(\ref{#1})}
\DeclareRobustCommand{\Eqs}[2]{Eqs.~(\ref{#1}) and (\ref{#2})}
\DeclareRobustCommand{\Ref}[1]{Ref.~\cite{#1}}
\DeclareRobustCommand{\Refs}[2]{Refs.~\cite{#1, #2}}
\DeclareMathAlphabet\mathbfcal{OMS}{cmsy}{b}{n}
\RequirePackage[normalem]{ulem} 
\RequirePackage{color}\definecolor{RED}{rgb}{1,0,0}\definecolor{BLUE}{rgb}{0,0,1} 

\newcommand{\aaa}{{\bf {a}}}

\newcommand{\kk}{{\bf{k}}}

\newcommand{\RR}{{\bf{R}}}
\newcommand{\qq}{{\bf{q}}}

\newcommand{\pp}{{\bf{p}}}

\begin{document}

\title{Many-Body Superconductivity in Topological Flat Bands}

\author{Jonah Herzog-Arbeitman$^{1}$}
\author{Aaron Chew$^{1}$}
\author{Kukka-Emilia Huhtinen$^2$}
\author{P\"aivi T\"orm\"a$^2$}
\author{B. Andrei Bernevig$^{1,3,4}$}

\affiliation{$^1$Department of Physics, Princeton University, Princeton, NJ 08544}
\affiliation{$^2$Department of Applied Physics, Aalto University School of Science, FI-00076 Aalto, Finland}
\affiliation{$^3$Donostia International Physics Center, P. Manuel de Lardizabal 4, 20018 Donostia-San Sebastian, Spain}
\affiliation{$^4$IKERBASQUE, Basque Foundation for Science, Bilbao, Spain}

\date{\today}

\begin{abstract}

In a flat band superconductor, bosonic excitations can disperse while unpaired electrons are immobile. To study this strongly interacting system, we construct a family of multi-band Hubbard models with exact eta-pairing ground states in all space groups. We analytically compute their many-body excitations and find that the Cooper pair bound states and density excitations obey an effective single-particle Hamiltonian written in terms of the non-interacting wavefunctions. These bound states possess a unique zero-energy excitation whose quadratic dispersion is determined by the minimal quantum metric. The rest of the bound state spectrum is classified by topological quantum chemistry, which we use to identify Cooper pairs with Weyl nodes, higher angular momentum pairing, and  fragile topology. We also add electron kinetic energy as a perturbation to show that the strongest pairing occurs at half filling and not at the highest density of states. This is similar in spirit to the superconductivity observed in twisted bilayer graphene.

\end{abstract}

\maketitle

Attractive interactions among electrons can lead to a pairing instability at the Fermi surface and create a superconducting ground state at low temperatures. Both the weakly and strongly interacting limits of these systems can be described by attractive Hubbard models. Such models, however, are rarely exactly solvable. Recent work\cite{2021arXiv211100807T,PhysRevB.83.220503,2016PhRvL.117d5303J,2018PhRvB..98v0511T,2018PhRvB..98m4513T,2020PhRvB.101f0505J,2021arXiv211213401T,2021PhRvL.126b7002P,PhysRevLett.123.237002,2021PhRvB.103v0502H,2021PhRvL.127x6403W,PhysRevA.103.053311,https://doi.org/10.48550/arxiv.2204.12610,2022PhRvR...4b3232K} has generated great interest in flat band superconductivity where electrons are completely immobile without interactions. A mean-field calculation in the multiband Hubbard model shows~\cite{2015NatCo...6.8944P} that the superfluid weight can be non-zero due to quantum geometry\cite{2011EPJB...79..121R}, indicating mobile Cooper pairs despite the infinite effective mass of the unpaired electrons. Lower bounds on the superfluid weight have been obtained from topology \cite{2015NatCo...6.8944P,2020PhRvL.124p7002X,PhysRevB.94.245149} and symmetry\cite{2022PhRvL.128h7002H}. While numerical beyond-mean-field work exists \cite{2016PhRvL.117d5303J,2021PhRvL.126b7002P,2022PhRvL.128h7002H,Chan2022,2022arXiv220610651C,2022arXiv220402994H,2020PhRvB.102t1112H,Orso2022} and confirms the mean-field picture, the few exact approaches on the attractive Hubbard model governing flat band superconductivity have been limited to ground state properties~\cite{PhysRevB.94.245149} or one dimension \cite{2018PhRvB..98m4513T}. In this Letter, we show that the \emph{many-body} problem is analytically tractable in any dimension subject to a simple requirement called the uniform pairing condition~\cite{2015NatCo...6.8944P,PhysRevB.94.245149}. Our exact results characterize not only the superconducting ground state, but also its excitations which control the low-temperature behavior.

\begin{figure*}
\includegraphics[width=\textwidth]{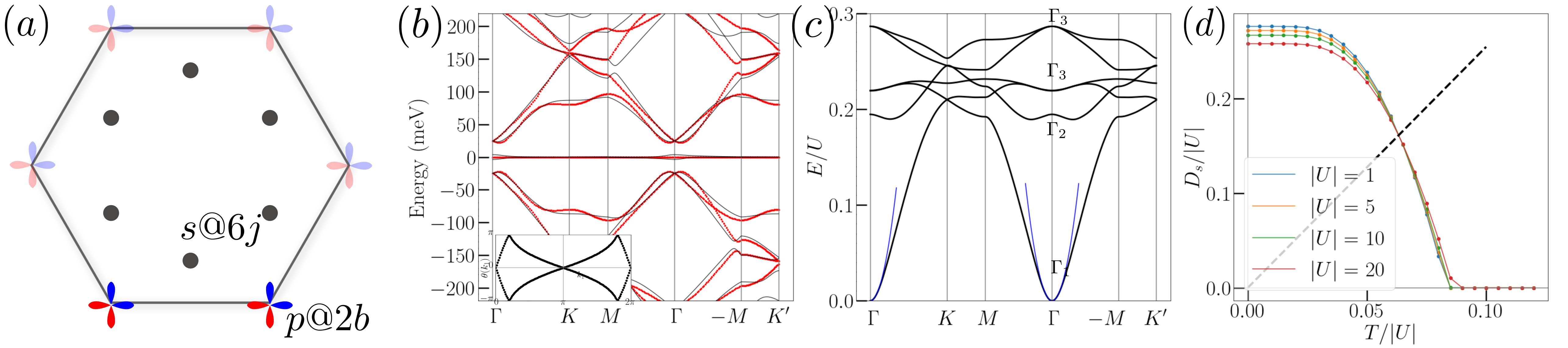}
\caption{Ten band model inspired by TBG. ($a$) A bipartite lattice of $s$ ($p$) orbitals at the 6j (2b) Wyckoff positions reproduces the momentum space irreps \cite{2018arXiv180710676S,2019PhRvX...9b1013A,2018arXiv180802482P} of the Bistritzer-MacDonald (BM) model\cite{2011PNAS..10812233B} with uniform pairing. ($b$) Band structure of the 10-band model (red) and BM model (black) with Wilson loop inset. ($c$) Cooper pair spectrum with the minimal quantum metric approximation in blue. $p$-wave pairing Leggett modes appear above the zero-energy $s$-wave mode (\App{app:Aaron}). ($d$) Superfluid weight $D_s$ from mean-field theory in units where $e,\hbar,k_B = 1$. The BKT temperature $T_c = \frac{\pi}{8} D_s(T_c)$ is estimated at $T_c \approx 0.06 |U|$.}
\label{figure1}
\end{figure*}

We consider tight-binding models with electron operators $c^\dag_{\mbf{k},\al,\sigma}$ with spin $\sigma= \u,\d$, orbital $\al = 1,\dots, N_{orb}$, and momentum $\mbf{k}$ in the Brillouin zone (BZ). Throughout, we assume $S_z$ spin conservation, e.g. no spin-orbit coupling. We consider kinetic terms with flat bands at zero energy labeled by $m=1,\dots,N_f$ separated by a large gap to the other bands (see Fig.~\ref{figure1}b for example). We assume the gap is much larger than the interaction strength $|U|$. Under this approximation, the kinetic term is completely characterized by the single-particle eigenvectors of the flat bands which form the $N_{orb} \times N_f$ matrix $[U_\sigma(\mbf{k})]_{\al m}$. By orthonormality, $U^\dag_\sigma(\mbf{k}) U_\sigma(\mbf{k}) = \mathbb{1}_{N_f}$, and $P^\sigma_{\al \be}(\mbf{k}) = [U_\sigma(\mbf{k}) U^\dag_\sigma(\mbf{k})]_{\al \be}$ is a rank $N_f$ projector, or $\mbf{k}$-space correlation function \cite{andreibook}. The gap to the other bands can be sent adiabatically to infinity so all operators can be projected into the flat bands spanned by
\bea
\label{eq:anticomP}
\bar{c}^\dag_{\mbf{k},\al,\sigma} &= \sum_\be c^\dag_{\mbf{k},\be,\sigma} P^\sigma_{\be \al}(\mbf{k}), \  \{\bar{c}_{\mbf{k}',\al,\sigma}, \bar{c}^\dag_{\mbf{k},\be,\sigma}\} = \delta_{\mbf{k}\mbf{k}'} P^\sigma_{\al \be}(\mbf{k})
\eea
which form an overcomplete basis due to the projection (\App{app:singlepart}). The canonical flat band operators $\gamma^\dag_{\mbf{k},m,\sigma} = \sum_{\al} c^\dag_{\mbf{k},\al,\sigma} U_{\al,m}^\sigma(\mbf{k})$ alternatively form a complete basis of the projected Hilbert space. For brevity, we define $P(\mbf{k}) = P^\u(\mbf{k})$. We assume  $S_z$ conservation and spinful time-reversal symmetry $\mathcal{T}$ which yields $P^\d(\mbf{k}) = P(-\mbf{k})^*$.

\textit{Hamiltonian.} The $N_f$ flat bands have zero kinetic energy and the remaining bands are projected out, so the kinetic term is zero and only interactions remain. We study an attractive Hubbard model proposed in \Ref{PhysRevB.94.245149},
\bea
\label{eq:Ham}
H &= \frac{|U|}{2} \sum_{\mbf{R}\al} (\bar{n}_{\mbf{R},\al,\u}-\bar{n}_{\mbf{R},\al,\d})^2, \ \ \bar{n}_{\mbf{R},\al,\sigma} = \bar{c}^\dag_{\mbf{R},\al,\sigma}\bar{c}_{\mbf{R},\al,\sigma} \\
\eea
where $\bar{c}^\dag_{\mbf{R},\al,\sigma}$ is the Fourier transform of $\bar{c}^\dag_{\mbf{k},\al,\sigma}$. If we impose the uniform pairing condition (UPC)~\cite{2015NatCo...6.8944P,PhysRevB.94.245149}
\bea
\label{eq:UPC}
\frac{1}{\mathcal{N}} \sum_\mbf{k} P_{\al \al}(\mbf{k}) = \frac{N_f}{N_{orb}} \equiv \eps \leq 1\\
\eea
where the $\al$-independent constant $\eps$ is fixed by $\Tr P(\mbf{k}) = N_f$, and $\mathcal{N}$ is the number of unit cells, then we find
\bea
H = \frac{\eps |U|}{2} \bar{N} - \sum_{\mbf{R}\al} \bar{n}_{\mbf{R},\al,\u}\bar{n}_{\mbf{R},\al,\d}
\eea
which is the usual Hubbard model up to a chemical potential (note that $\bar{N} = \sum_{\mbf{R}\al\sigma} \bar{n}_{\mbf{R},\al,\sigma}$ is the projected number operator). The UPC thus allows to relate the flat band Hubbard Hamiltonian to the positive-definite Hamiltonian \Eq{eq:Ham}, which is
crucial to analytically solving the ground states and excitations.

Intuitively, if the orbitals are related by symmetry, the UPC should hold \cite{PhysRevB.94.245149}. Indeed, we prove that \Eq{eq:UPC} is enforced by symmetry. Our proof (\App{app:UPC}) uses Schur's lemma and some basic facts from topological quantum chemistry \cite{2017Natur.547..298B,2020arXiv200604890C,bacry1988symmetry,Aroyo:firstpaper,Aroyo:xo5013}. We show that if the orbitals of the model can all be related by a symmetry operator (technically, if they form an irrep of a Wyckoff position), then the UPC holds. Thus the class of flat bands with the UPC is very large, and encompasses existing examples \cite{2021PhRvL.126b7002P,2022PhRvL.128h7002H,PhysRevB.78.125104,PhysRevB.99.045107,rsis,2021PhRvB.103v0502H,2018PhRvB..98v0511T,2018PhRvB..98m4513T}.

There is a generalization\cite{PhysRevB.94.245149} of \Eq{eq:UPC}: if the flat band wavefunction $[U_\sigma(\mbf{k})]_{\al m}$ vanishes on some orbitals $\tilde{\al}$, then $\bar{n}_{\mbf{R},\tilde{\alpha},\sigma}= 0$ and the UPC is only required on $\al \neq \tilde{\al}$. Calugaru et al.\cite{2021NatPh..18..185C} construct models with $N_f = N_L-N_{\tilde{L}}$ perfectly flat bands on bipartite lattices consisting of two sublattices $L, \tilde L$ with $[U_\sigma(\mbf{k})]_{\tilde{\al} m} = 0$ where $N_{L} \, (N_{\tilde{L}})$ is the number of orbitals in $L \, (\tilde{L})$ per unit cell. Choosing the $L$ lattice to have the UPC, their method constructs exactly flat bands with the UPC in all space groups. In this case, \Eq{eq:UPC} holds with $\eps = N_f/N_L$. In this manner, we construct a 10 band model (see Fig.~\ref{figure1}) with the momentum space irreps of twisted bilayer graphene (TBG)\cite{2011PNAS..10812233B,2018arXiv180710676S,2018arXiv180802482P} and the UPC (see Fig.~\ref{figure1}), but with Wilson loop winding \cite{2019PhRvX...9b1013A,2018arXiv180409719B,2022arXiv220202353Y} 2 (instead of 1 as in TBG).

\textit{Superconducting Ground States.} Originally, $H$ in \Eq{eq:Ham} was proposed to exhibit an exact BCS-type groundstate despite the ill-defined Fermi surface of the flat band\cite{PhysRevB.94.245149}. In this work, we instead focus on the more physical ground states at \emph{fixed} particle number. We define the local spin operator $\bar{S}^z_{\mbf{R},\al} = \bar{n}_{\mbf{R},\al,\u}-\bar{n}_{\mbf{R},\al,\d}$ appearing in \Eq{eq:Ham}. Because of the projection, $\bar{c}^\dag_{\mbf{R},\al,\sigma}$ is extended on the scale of the correlation length, which cannot be adiabatically sent to zero (the atomic limit) if the bands are topological or obstructed \cite{2022PhRvL.128h7002H}. This is crucial for superconductivity in transport since trivial flat bands in the atomic limit would make $H = \frac{|U|}{2} \sum_{\mbf{R},\al} (\bar{S}^z_{\mbf{R},\al})^2$ a sum of commuting local terms with trivial dynamics and macroscopic degeneracies. Away from the trivial limit, $H$ retains a single nontrivial symmetry operator $\eta$
\bea
\label{eq:etaoperator}
\null [\eta^\dag, \bar{S}^z_{\mbf{R},\al}] = 0, \ \eta^\dag=  \sum_{\mbf{k}\al} \bar{c}^\dag_{\mbf{k},\al, \u} \bar{c}^\dag_{-\mbf{k},\al, \d} = \sum_{\mbf{R}\al} \bar{c}^\dag_{\mbf{R},\al, \u} \bar{c}^\dag_{\mbf{R},\al, \d} \ .
\eea
This result relies on the UPC, global $S_z$ conservation, and $\mathcal{T}$ (\App{app:Hubermodel}). Here $\eta^\dag$ is  simply the creation operator of a zero-momentum $s$-wave Cooper pair, and it forms an $su(2)$ algebra, $[\bar{N}, \eta^\dag] = 2 \eta^\dag$, which extends $U(1)$ charge conservation \cite{PhysRevB.94.245149}.  We define the states $\ket{n} \propto \eta^{\dag n} \ket{0}$ which have fixed particle number $\bar{N} \ket{n} = 2 n \ket{n}$ for $n = 0, \dots, \mathcal{N} N_f$. Yang first introduced similar eta-pairing states\cite{PhysRevLett.63.2144} on the square lattice, which is also bipartite. However, his operators carried momentum $\pi$ and were not ground states. In contrast, $\ket{n}$ must be ground states because $H$ is positive semi-definite, and $H \ket{n} \propto \eta^{\dag n} H \ket{0} = 0$.

Because $\ket{n}$ differs from the BCS-BEC\cite{2005PhR...412....1C,2013arXiv1306.5785R} states $e^{z \eta^\dag}\ket{0}$ (which are also exact ground states\cite{PhysRevB.94.245149}), it may not be apparent that they describe a bosonic condensate. To establish this, we show that $\ket{n}$ has off-diagonal long-range order in the two-particle density matrix\cite{RevModPhys.34.694}; this is intuitive since the $\eta$ pairs are superpositions of pairs on \textit{all} sites. In the states $\ket{n}$, a generalized Wick's theorem \cite{2022arXiv220412163P,2017arXiv171009248M,1971qtmp.book.....F} holds for the correlators of the Wannier state operators $w^\dag_{\mbf{R}m\sigma} = \frac{1}{\sqrt{\mathcal{N}}} \sum_{\mbf{k}\al} e^{i \mbf{R}\cdot \mbf{k}} \gamma^\dag_{\mbf{k},m,\sigma}$ obeying $\{w_{\mbf{R},m,\sigma},w^\dag_{\mbf{R}',m',\sigma'} \} = \delta_{\mbf{R},\mbf{R}'} \delta_{m,m'} \delta_{\sigma,\sigma'}$. We find (\App{app:generatingfunctions})
\bea
\label{eq:ODLRO}
\lim_{|\mbf{R}-\mbf{R}'| \to \infty} \braket{n| w^\dag_{\mbf{R}' m' \d}w^\dag_{\mbf{R}' m' \u} w_{\mbf{R}m \u} w_{\mbf{R} m \d} |n} = \nu(1-\nu)
\eea
in the thermodynamic limit where $\nu = n/{N_f\mathcal{N}}$ is the filling. Clearly the strongest condensate appears at half-filling of the flat bands. Similar dependence on the filling appears in mean-field results on the multi-band Hubbard model \cite{2015NatCo...6.8944P} which also indicate half-filling is favorable for the superconductor. We remark that the Wannier operators have a gauge freedom corresponding to the redefinition of eigenvectors $U_\sigma(\mbf{k}) \to U_\sigma(\mbf{k}) \mathcal{W}_\sigma(\mbf{k})$ where $\mathcal{W}_\sigma(\mbf{k}) \in U(N_f)$ \cite{2012RvMP...84.1419M,PhysRevLett.98.046402,2006Sci...314.1757B,2018CMaPh.tmp....8M,2016arXiv161209557M,2016PhRvB..93c5453W,PhysRevB.83.035108,2020PhRvX..10c1001S,2020Sci...367..794S}. Nevertheless, \Eq{eq:ODLRO} shows that the resulting correlator is gauge-invariant. This condensate ground state is a universal result in all UPC lattices.

\textit{Electron Excitations.} The presence of an enlarged symmetry algebra and exact ground states suggests that more features of $H$ may be accessible. We now show that the electron (and hole) excitations above the $\ket{n}$ ground states are exactly solvable. Our procedure\cite{2021PhRvB.103t5415B,2021PhRvL.127z6402K,2021PhRvB.103t5415B,2021arXiv211111434H,PhysRevLett.122.246401} is to add a single electron $\gamma^\dag_{\mbf{k},m,\sigma}$ to the ground state and calculate
\bea
\label{eq:Rham}
\null H \gamma^\dag_{\mbf{k},m,\sigma}\ket{n} = [H, \gamma^\dag_{\mbf{k},m,\sigma}]\ket{n} &= \sum_{m'} \gamma^\dag_{\mbf{k},m',\sigma} \ket{n} [R^\sigma(\mbf{k})]_{m'm}
\eea
since $H\ket{n}=0$. Diagonalizing $R^\sigma(\mbf{k})$, one obtains the charge $+1$ (electron) excitation spectrum and the eigenstates. Using $H = \frac{|U|}{2} \sum_{\mbf{q}\al} \bar{S}^z_{-\mbf{q},\al} \bar{S}^z_{\mbf{q},\al}$ where $\bar{S}^z_{\mbf{q},\al}$ is the Fourier transform of $\bar{S}^z_{\mbf{R},\al}$ and $\bar{S}^z_{\mbf{q},\al}\ket{n}=0$, we find
\begin{align}
\label{eq:Rham2}
\null [H, \gamma^\dag_{\mbf{k},m, \sigma}] \ket{n} &= \frac{|U|}{2} \sum_{\mbf{q}\al}   [\bar{S}^z_{-\mbf{q},\al} , [ \bar{S}^z_{\mbf{q},\al}, \gamma^\dag_{\mbf{k}, m, \sigma}]] \ket{n}  \\
&\!\!\!\!\!\!\!\!\!\!\!\!\!\!\!\!\!\!\!\!\!\!\!\!\!= \frac{|U|}{2\mathcal{N}}\sum_{\mbf{q} \al,m'l} M^{m'l}_{\sigma,\al}(\mbf{k}+\mbf{q},-\mbf{q}) M^{lm}_{\sigma,\al}(\mbf{k},\mbf{q})  \gamma^\dag_{\mbf{k},m',\sigma} \ket{n} \nonumber
\end{align}
defining $M^{mn}_{\al,\sigma}(\mbf{k},\mbf{q}) = U^{\sigma*}_{\al m}(\mbf{k}+\mbf{q}) U^\sigma_{\al n}(\mbf{k})$ as the orbital-resolved form factors (\App{app:charge1}). Matching \Eq{eq:Rham2} to \Eq{eq:Rham} and simplifying the form factors yields
\begin{align}
\label{eq:chargegap}
\null [R^\sigma(\mbf{k})]_{m'm} &= \frac{|U|}{2\mathcal{N}}\sum_{\mbf{q} \al} U^{\sigma *}_{\al m'}(\mbf{k}) P^\sigma_{\al\al}(\mbf{k}+\mbf{q}) U^\sigma_{\al m}(\mbf{k}) \\
&= \frac{\eps |U|}{2} \sum_\al U^{\sigma *}_{\al m'}(\mbf{k})U^\sigma_{\al m}(\mbf{k}) = \frac{\eps |U|}{2} \delta_{m'm} \nonumber
\end{align}
where the UPC, \Eq{eq:UPC}, was crucial to compute the $\mbf{q}$ sum. Hence, the electron excitations are independent of $\nu$, $N_f$-degenerate per spin, and exactly flat meaning that the single-electron excitations are immobile. Indeed, a mean-field treatment assuming a $\mbf{k}$-independent pairing gap reproduces the $\eps |U|/2$ $s$-wave gap \cite{2022arXiv220311133H}. The temperature scale $\eps |U|$ describes the onset of pairing, but is expected to be somewhat above the superconducting critical temperature with a pseudogap intervening\cite{2020PhRvB.102r4504W,2021arXiv211110018Z,2014NatPh..10..483H,2018PhRvB..98m4513T}.

The electronic excitations characterize the stability of the superconducting phase, as in BCS theory. When dispersion is added to the flat bands, as in realistic systems, we determine the filling where superconductivity is strongest through the Richardson criterion\cite{1964NucPh..52..221R,1977JMP....18.1802R,2021PhRvB.103t5415B} $E_\Delta(N+2) = E(N+2)-2E(N+1) +E(N)$ where $E(N)$ is the ground state energy at particle number $N$. Here $E_\Delta$ is an estimate of the Cooper pair binding energy: $E_\Delta < 0$ indicates the favorability of pair formation (at $T=0$). We can compute $E_\Delta(N)$ perturbatively when dispersion $\tilde{E}_{n}(\mbf{k})$ is added to the flat bands. For convenience, we assume $\sum_{\mbf{k},n} \tilde{E}_{n}(\mbf{k}) = 0$ in which case (\App{app:Richcrit})
\bea
\label{eq:Edelta}
- E_\Delta(\nu) = \eps |U| - |1-2\nu| \tilde{E}_\nu
\eea
where $\tilde{E}_\nu > 0$ is $\max_{\mbf{k},n} \tilde{E}_{n}(\mbf{k})$ ($|\min_{\mbf{k},n} \tilde{E}_{n}(\mbf{k})|)$ for $\nu \geq 1/2$ ($\nu < 1/2$). Hence the strongest Cooper pair binding occurs at $\nu = \frac{1}{2}$, where the condensate is also strongest (see \Eq{eq:ODLRO}) in agreement with multi-band mean-field results \cite{2015NatCo...6.8944P}, but in contrast to conventional BCS theory where pairing is maximized at the highest density of states. This echoes the phenomenology of twisted bilayer graphene where critical temperature is highest \emph{not} at the van Hove singularities of the flat bands \cite{2018Natur.556...43C,2021Natur.600..240O,2020arXiv200812296P, 2020Natur.583..375S,2019Natur.572..101X}. \Eq{eq:Edelta} is not valid when spin-orbit coupling is added, but we provide a general formula in \App{app:Richcrit}.

\textit{Cooper Pair Excitations and Bound States.} We now show that the two-particle excitations, containing the Cooper pair and density excitation spectrum, are analytically solvable. We will also find a set of excitations (Leggett modes \cite{1966PThPh..36..901L}) in other pairing channels. We find that these modes can carry higher angular momentum, e.g. $d$-wave symmetry, and fragile topology. Our calculation shows in microscopic detail the formation of bosons from the underlying fermion bands.

The two-body excitations are characterized by their global quantum numbers, $S_z$ spin and charge. We start in the charge $+2$ sector and compute the scattering matrix
\begin{align}
\label{eq:Rcharge2}
\null &[H, \gamma^\dag_{\mbf{p}+\mbf{k},m,\sigma}\gamma^\dag_{-\mbf{k},n,\sigma'}] \ket{n} \\
&=  \sum_{\mbf{k}'m'n'} \gamma^\dag_{\mbf{p}+\mbf{k}',m',\sigma}\gamma^\dag_{-\mbf{k}',n',\sigma'}  \ket{n} [R^{\sigma \sigma'}(\mbf{p})]_{\mbf{k}'m'n',\mbf{k}mn} \nonumber
\end{align}
which is a function of the total momentum $\mbf{p}$. The commutator can be computed since $\bar{S}^z_{\mbf{R},\al}\ket{n}=0$. The full scattering matrix can be written (\App{app:charge2})
\bea
R^{\sigma \sigma'}(\mbf{p}) &=  \eps |U|  \mathbb{1} + s^z_\sigma  s^z_{\sigma'}\frac{ |U| }{\mathcal{N}} \mathcal{U}_{\sigma \sigma'}(\mbf{p})\mathcal{U}_{\sigma \sigma'}^\dag(\mbf{p}) \\
\eea
where $s^z_{\u/\d} = \pm1$ and $[\mathcal{U}_{\sigma \sigma'}(\mbf{p})]_{\mbf{k}mn,\al} = U^{\sigma *}_{m \al}(\mbf{p}+\mbf{k}) U^{\sigma' *}_{n \al}(-\mbf{k})$ is a $\mathcal{N}N_f^2 \times N_{L}$ matrix ($\mathcal{N}$ is the number of $\mbf{k}$ points). If $\sigma = \sigma'$, then $\mathcal{U}_{\sigma \sigma}(\mbf{p})$ is symmetric under $\mbf{p}+\mbf{k},m \leftrightarrow -\mbf{k},n$ whereas the fermion operators in \Eq{eq:Rcharge2} are anti-symmetric. Hence the anti-symmetrization of the fermions annihilates the $\mathcal{U}_{\sigma \sigma}(\mbf{p})\mathcal{U}^\dag_{\sigma \sigma}(\mbf{p})$ term, so all spin-aligned states have energy $\eps |U|$ for all $\mbf{p}$ and are completely unpaired.

We now consider $\sigma = -\sigma'$. Denote $R^{\u \d}(\mbf{p}) = \eps|U| \mathbb{1} - \mathcal{U}(\mbf{p})\mathcal{U}^\dag(\mbf{p})/\mathcal{N}$, dropping spin labels for brevity. First, note that $\text{rank } \mathcal{U} \mathcal{U}^\dag \leq \text{rank }\mathcal{U} \leq N_{L}$ and hence $\mathcal{U}\mathcal{U}^\dag$ has a very large kernel of dimension at least $\mathcal{N} N_f^2 - N_{L}$, corresponding to the particle-particle continuum. Its nonzero eigenvalues are identical to the eigenvalues of the $N_{L} \times N_{L}$ matrix $h(\mbf{p}) = \mathcal{U}^\dag(\mbf{p})\mathcal{U}(\mbf{p})/\mathcal{N}$. This pairing Hamiltonian takes the universal form (\App{app:charge2})
\bea
\label{eq:hpairing}
h_{\al\be}(\mbf{p}) &= \frac{1}{\mathcal{N}} \sum_\mbf{k} P_{\al \be}(\mbf{k}+\mbf{p})P_{\be \al}(\mbf{k}) \\
\eea
which can be interpreted as the effective single-particle Hamiltonian describing the hopping of a tightly bound Cooper pair (\App{app:spectrumbounds}). However, we emphasize that the \Eq{eq:hpairing} derives from an exact many-body calculation, and its eigenvalues are precisely the binding energies of the various pairing channels.

We denote the eigen-decompositon of \Eq{eq:hpairing} as $h(\mbf{p}) u_\mu(\mbf{p}) = \eps_\mu(\mbf{p}) u_\mu(\mbf{p})$ for $\mu = 0,\dots,N_{L}-1$. The many-body energies of $R^{\u \d}(\mbf{p})$ are $\eps |U|$ for the $\mathcal{N}N_f^2-N_{L}$ unpaired states and $E_\mu(\mbf{p}) = |U|(\eps - \eps_\mu(\mbf{p}))$ for the $N_{L}$ bound states ( \Fig{figure1}c). Returning to \Eq{eq:Rcharge2}, the pair excitations can be written as $\eta^\dag_{\mbf{p},\mu} \ket{n}$ where (\App{app:charge2})
\bea
\label{eq:etaeigenstates}
\eta^\dag_{\mbf{p},\mu} &= \frac{1}{\sqrt{\mathcal{N}}} \sum_{\mbf{k}\al} \frac{u^\al_\mu(\mbf{p})}{\sqrt{\eps_\mu(\mbf{p})}} \bar{c}^\dag_{\mbf{p}+\mbf{k},\al,\u} \bar{c}^\dag_{-\mbf{k},\al,\d}
\eea
which is the many-body analogue of the BCS pairing, with $\mu$ enumerating the pairing channels. We will show that $\eta^\dag_{\mbf{p},\mu}$ can carry nontrivial topology protected by $G$.

We remark that the charge-0 density excitations are also exactly solvable. We determine (\App{app:goldstones}) that the spin $\pm1$ density excitations are gapped with energies above the $\eps |U|$ pairing threshold. The two spin-0 density excitations, of the form $\gamma^\dag_{\mbf{p}+\mbf{k},\sigma}\gamma_{\mbf{k},\sigma}$, in fact have a spectrum \emph{identical} to the spin-0 Cooper pairs. We will show momentarily that there is guaranteed to be a gapless Goldstone mode in this spectrum, similar to the Anderson-Bogoliubov phonon \cite{pitaevskii2003bose,PhysRev.94.262,zwerger2011bcs}. 
However our calculation shows this Goldstone mode is quadratic (\App{app:quantummetric}), whereas
the Anderson-Bogoliubov phonon typically has linear dispersion. The quadratic dispersion can be attributed to the breaking of the $\eta, \eta^\dag$ generators by fixing the particle number\cite{2019arXiv190400569W,2021arXiv211110018Z}.

\textit{Boson Bands and Topology.} We have reduced the two-body scattering matrix to an effective band theory problem given by the Hamiltonian in \Eq{eq:hpairing}. Notably, this result is generic: it is valid for all lattices with the UPC. We now prove a number of universal features of $h(\mbf{p})$.

First, $\eps_\mu(\mbf{p}) \geq 0 $. This can be proved from the Schur product theorem\cite{prasolov1994problems}, which states that the Hadamard product $[A \circ B]_{\al \be} = A_{\al \be} B_{\al \be}$ of positive semi-definite matrices is positive semi-definite. Therefore $h(\mbf{p})$ is the sum of positive semi-definite terms, and is also positive semi-definite. Secondly, $\eps_\mu(\mbf{p}) \leq \eps$ follows since \Eq{eq:Ham} is the sum of squares and hence $E_\mu(\mbf{p}) = |U|(\eps - \eps_\mu(\mbf{p}))>0$. Note that the largest eigenvalue of $h(\mbf{p})$, i.e. the strongest pairing, corresponds to the lowest energy many-body state. We number the bands such that $\eps_0(\mbf{p}) \geq \eps_\mu(\mbf{p})$.

Because $\eta^\dag$ is a zero-energy, $s$-wave, two-body excitation with $E_0(\mbf{0}) = 0$, it follows that $\eps_0(\mbf{0}) = \eps$ (\App{app:spectrumbounds}). To study the small $\mbf{p}$ behavior around $\eps_0(\mbf{0}) = \eps$, we apply perturbation theory. The result is $E_0(\mbf{p}) = \frac{|U|}{N_{L}} g_{ij} p_i p_j + O(p^4)$ where $g_{ij}$ is the integrated minimal quantum metric \cite{2022arXiv220311133H} (\App{app:quantummetric}). This amounts to a many-body proof of the result first reported in mean-field\cite{2015NatCo...6.8944P} and by a two-body calculation\cite{2018PhRvB..98v0511T}. Notably, the bipartite construction\cite{2021NatPh..18..185C} produces many examples of fragile or obstructed single-particle bands, for which $g_{ij}$ is lower bounded\cite{2022PhRvL.128h7002H,2022arXiv220202353Y}. \textit{Hence, the low-energy spectrum is universally governed by the quantum geometry of the flat bands.} The presence of tightly bound pairs with quadratic dispersion suggests these models are in the Bose-Einstein condensate regime of superconductivity \cite{2013arXiv1306.5785R,2005PhR...412....1C,PhysRevLett.127.170404,zwerger2011bcs}. Transport in the superconductor can be characterized by the $T=0$ mean-field superfluid weight $D_s$, shown in Fig.~\ref{figure1}(d). For degenerate flat
bands, a mean-field calculation shows (\App{app:meanfield})
\begin{equation}
  [D_s]_{ij} = 8 \frac{\eps |U|}{V_c}\nu(1-\nu)g_{ij}, \quad g_{ij} = \frac{1}{\mathcal{N}} \sum_\mbf{k} g_{ij}(\mbf{k})
 \label{eq.sfw}
\end{equation}
where $V_c$ is the volume of a unit cell and $g_{ij}(\mbf{k})$ is the minimal quantum metric of the degenerate bands \cite{2020PhRvL.124p7002X,2022arXiv220311133H,Mera:2022ocf,2022PhRvL.128h7002H}. Eq.~\eqref{eq.sfw} predicts $D_s/|U|\approx 0.277$ for our model, in good agreement with \Fig{figure1}d.

The eigenvalues of $h_{\al \be}(\mbf{p})$ correspond to the pairing channels, with $s$-wave pairing at low energies, but higher energy modes with nontrivial pairing symmetries appear (see \Fig{figure1}c) and contribute to the finite temperature behavior. Although the zero-energy $s$-wave mode is universal across all space groups $G$, the higher energy spectrum varies. Nevertheless, we can use topological quantum chemistry to determine all possible Cooper pair band connectivities\cite{PhysRevB.97.035138,2021arXiv210610267C,2020PhRvB.102c5110E,2018SciA....4.8685W,2017PhRvX...7d1069K} and topology for each $G$ (\App{app:Cooperpairspectrum}).

The pairing Hamiltonian $h(\mbf{p})$ in an effective tight-binding model which inherits \emph{bosonic} single-particle symmetries from the underlying electron space group $G$. \Eq{eq:hpairing} shows $h_{\al \be}(\mbf{p}) = h^*_{\al \be}(-\mbf{p})$ so $h(\mbf{p})$ has a spinless time-reversal symmetry. We can also show (\App{eq:CPsymmetry}) that $h(\mbf{p})$ inherits the space group $G$ but in the real representation $|D_{\al \be}[g]|, g\in G$, where $D_{\al\be}[g]$ is the representation of the spin-$\u$ electrons \cite{PhysRevB.102.115117}. The possible band structures with these orbitals are highly constrained by topological quantum chemistry since, by enforcing the UPC within our construction, the orbitals of $h(\mbf{p})$ form a single Wyckoff position\cite{2020arXiv200604890C} and are symmetry-related. If the orbitals are at low-symmetry (non-maximal Wyckoff) positions, then the bands of $h(\mbf{p})$ are composite and can be deformed into disconnected atomic representations\cite{2018PhRvB..97c5139C}. If the orbitals are located at high-symmetry (maximal Wyckoff) positions, the bands are either all connected through high-symmetry points of the BZ or decomposable into \emph{topological} bands\cite{2018arXiv180709729B,2018PhRvL.120z6401C}. However, because $h(\mbf{p})$ has a spinless time-reversal symmetry, no stable topology is possible \cite{PhysRevLett.98.046402,2012PhRvB..86k5112F}. Instead, a non-zero symmetry indicator\cite{2017Natur.547..298B,2020JPCM...32z3001P,2017NatCo...8...50P,2017PhRvX...7d1069K} identifies the presence of topologically protected Weyl nodes \cite{2011PhRvB..83x5132H,2020arXiv201000598E}. In 2D, we check exhaustively that the only groups with fragile topological Cooper pairs are $G =p6, p6mm$. In $p6mm$, placing $s$ or $p$ orbitals at the kagome (3c) position induces a unique decomposition: the low-energy branch is a topologically trivial $s$-wave Cooper pair, and there is a gap to two connected bands with fragile topology where the Cooper pair wavefunction interpolates between $d$-wave and $p$-wave character across the BZ (\App{app:examples}). This decomposition is also possible in $p6$, but there is also a second decomposition with a nonzero symmetry indicator $\theta_2 \in\mathds{Z}_2$. In this case, the Cooper pair bands are connected by Weyl nodes. An exhaustive search of the 3D Wyckoff positions with the UPC yields 22 space groups with decomposable bands (\App{app:bosontopology}). We find instances where the low-energy branch is fragile topological. All 22 are kagome/pyrochlore groups where many promising flat band candidates \cite{2022Natur.603..824R} have been identified. Our classification identifies flat band superconductors which host topological Cooper pairs and higher angular momentum spin-singlet pairing due to the multi-orbital lattice. 

\textit{Conclusions.} Although the BCS paradigm is successful in describing weakly-coupled superconductors, the numerous unresolved questions in high-$T_c$ materials\cite{RevModPhys.74.235,RevModPhys.63.239,RevModPhys.78.17} and moir\'e heterostructures\cite{2020NatPh..16..725B,2021arXiv211100807T,2021arXiv211209270Z,2021arXiv211210760P,https://doi.org/10.48550/arxiv.2204.12579,2021PhRvB.103b4506W,2019PhRvL.122y7002L,2021arXiv211110018Z}  demonstrate a pressing need for a strong-coupling theory of superconductivity\cite{2021arXiv211110018Z,2020PhRvL.124y7002L,2020PhRvR...2c3309L,richardson1963restricted,2009PhRvB..79r0501I,PhysRevB.93.220503}. To this end, we introduced a broad family of flat band Hubbard models constructed within the framework of topological quantum chemistry. Their groundstates are bosonic condensates featuring maximal off-diagonal long-range order at half-filling and a linear-in-$|U|$ pairing gap. Computing the mass of the condensing Cooper pair, we established that symmetry-protected wavefunction topology coheres the superconductor with a nonzero superfluid weight. We also found exact Cooper pair bound states with $p,d,\dots$-wave symmetries below the pairing gap and completely classified Cooper pair bands with fragile topology. Finally, we solved the Goldstone and gapped spin-wave excitations which hint at BEC-like behavior. Further study of these exactly solvable models may provide a new route to high-temperature superconductivity. 

B.A.B. and A.C. were supported by the European Research Council (ERC) under the European Union's Horizon 2020 research and innovation programme (grant agreement No. 101020833),
the ONR Grant No. N00014-20-1-2303, 
the Schmidt Fund for Innovative Research, 
Simons Investigator Grant No. 404513, 
the Packard Foundation, 
the Gordon and Betty Moore Foundation through the EPiQS Initiative, Grant GBMF11070 and Grant No. GBMF8685 towards the Princeton theory program.
Further support was provided by the NSF-MRSEC Grant No. DMR-2011750, 
BSF Israel US foundation Grant No. 2018226, 
and the Princeton Global Network Funds.
 JHA is supported by a Hertz Fellowship.
P.T. and K-E.H. acknowledge support by the Academy of Finland under project numbers 303351 and 327293. K-E.H. acknowledges support from the Magnus Ehrnrooth Foundation.

\let\oldaddcontentsline\addcontentsline
\renewcommand{\addcontentsline}[3]{}
\bibliography{finalbib}
\bibliographystyle{aipnum4-1}
\bibliographystyle{unsrtnat}
\let\addcontentsline\oldaddcontentsline

\setcitestyle{numbers,square}

\cleardoublepage
\onecolumngrid
\appendix

\tableofcontents

\section{Single-Particle Symmetry and Topology}
\label{app:singlepart}

In this section, we define the notation used in the rest of the work for the single-particle projectors, projected many-body operators, and space group symmetries (\App{app:projbands}). We prove
that the uniform pairing condition can be symmetry enforced by choosing orbitals to transform as irreps of a single Wyckoff positions (\App{app:UPC}), and extend this result to bipartite lattices where perfectly flat bands can be systematically constructed.

\subsection{Projected Bands}
\label{app:projbands}

Our problem projects the main electronic degrees of freedom into the quasi-flat bands of the non-interacting Hamiltonian. In a strong coupling approximation, the quasi-flat bands are set to zero energy and the remaining bands are sent to infinite energy and projected out. As such, the physics depends entirely on the flat band wavefunctions.

We now introduce our notation which differs slightly from \Ref{PhysRevB.94.245149} where only nondegenerate flat bands (doubly degenerate when including spin) were considered. Let $\mbf{R}$ index the unit cells and $\al = 1, \dots, N_{orb}$ index the orbitals at positions $\mbf{r}_\al$. For now we neglect spin, but it will be introduced later with $\sigma = \u,\d$. The electron operators (obeying canonical anti-commutation relations) are
\bea
\label{eq:cRck}
c^\dag_{\mbf{k},\al} = \frac{1}{\sqrt{\mathcal{N}}} \sum_\mbf{R} e^{- i \mbf{k} \cdot (\mbf{R} + \mbf{r}_\al)} c^\dag_{\mbf{R},\al}, \qquad c^\dag_{\mbf{R},\al} = \frac{1}{\sqrt{\mathcal{N}}}\sum_{\mbf{k}} e^{i \mbf{k} \cdot (\mbf{R} + \mbf{r}_\al)} c^\dag_{\mbf{k},\al} \\
\eea
where $\mathcal{N}$ is the total number of unit cells, which is the number of terms in the $\mbf{R},\mbf{k}$ sums. The general single-particle Hamiltonian reads
\bea
\label{eq:gammasing}
\tilde{H} &=  \sum_{\mbf{k},\al \be}c^\dag_{\mbf{k},\al} \tilde{h}_{\al \be}(\mbf{k}) c_{\mbf{k},\be} = \sum_{\mbf{k},n} \tilde{E}_n(\mbf{k}) \gamma^\dag_{\mbf{k},n}  \gamma_{\mbf{k},n}, \qquad \gamma_{\mbf{k},n} &= \sum_\al [\tilde{U}^\dag(\mbf{k})]_{n,\al} c_{\mbf{k},\al}
\eea
and $\tilde{h}(\mbf{k}) = \tilde{U} \tilde{E} \tilde{U}^\dag$ where $\tilde{E}$ is the diagonal matrix of energies, $\tilde{U}$ is the $N_{orb} \times N_{orb}$ eigenvector matrix, and the diagonalized electron operators obey
$\{\gamma_{\mbf{k},m},\gamma^\dag_{\mbf{k}',n} \} = \delta_{\mbf{k},\mbf{k}'} [\tilde{U}^\dag \tilde{U}]_{mn} =  \delta_{\mbf{k},\mbf{k}'} \delta_{mn}$. For brevity, we often suppress the $\mbf{k}$ dependence.

We assume there is a gap around the $N_f$ quasi-flat bands, which we will refer to as the flat bands; in the strong coupling approximation, we do not require them to be exactly flat. We project all operators into the flat bands by setting $\gamma_{\mbf{k},n} = 0$ if $n$ is not one of the flat bands. Now denote $U(\mbf{k})$ as the $N_{orb} \times N_f$ occupied eigenstate matrix which obeys $U^\dag(\mbf{k}) U(\mbf{k}) = \mathbb{1}_{N_f}$ and $U(\mbf{k})U^\dag(\mbf{k}) = P(\mbf{k})$ where $P$ is the $N_{orb} \times N_{orb}$ projector matrix obeying $P^2 = P, P^\dag = P, \Tr P = N_f$. We recall that in the exactly flat case, the eigenstates are degenerate and hence are only defined up to the gauge freedom $U(\mbf{k}) \to U(\mbf{k}) \mathcal{W}(\mbf{k})$ where $\mathcal{W}(\mbf{k}) \in U(N_f)$ is a unitary matrix with arbitrary $\mbf{k}$ dependence \cite{2022PhRvL.128h7002H}. Both $U^\dag(\mbf{k}) U(\mbf{k})$ and $U(\mbf{k})U^\dag(\mbf{k}) $ are invariant under this gauge freedom. The projector $P(\mbf{k})$ encodes the topology and quantum geometry of the bulk bands. It will play an essential role in the many-body problem, highlighting the interplay of topology and interactions.

We now define the projected operators. Because we have set all $\gamma_{\mbf{k},n}$ to zero other than the flat band operators, an (over)complete basis of the projected states is given by
\bea
\label{eq:bargamma}
\bar{c}_{\mbf{k}, \al} \equiv  \sum_{n=1}^{N_{orb}} [\tilde{U}(\mbf{k})]_{\al,n} \gamma_{\mbf{k},n} &= \sum_{n=1}^{N_f} [U(\mbf{k})]_{\al,n} \gamma_{\mbf{k},n} = \sum_{\be} [U(\mbf{k}) U^\dag(\mbf{k})]_{\al \be} c_{\mbf{k},\be} = \sum_{\be} [P(\mbf{k})]_{\al \be} c_{\mbf{k},\be}
\eea
using \Eq{eq:gammasing}. Because the projector $P(\mbf{k})$ is Hermitian, we have
\bea
\label{eq:cdatPc}
 \bar{c}^\dag_{\mbf{k},\al}  &= \sum_\be c^\dag_{\mbf{k},\be}  P^*_{\al \be}(\mbf{k}) = \sum_\be c^\dag_{\mbf{k},\be}  P_{\be \al}(\mbf{k}) \ .
\eea
The $\bar{c}^\dag_{\mbf{k},\al}$ basis is invariant under the eigenvector gauge freedom (unlike the $\gamma^\dag_{\mbf{k},n}$ operators). However, the barred operators do not obey canonical anti-commutation relations because they are an overcomplete basis, whereas the $\gamma^\dag_{\mbf{k},n}$ operators form a complete basis. We compute
\bea
\label{eq:anticom}
\{\bar{c}_{\mbf{k},\al}, \bar{c}^\dag_{\mbf{k}',\be} \} = \sum_{\al'\be'} P_{\al \al'}(\mbf{k}) \{c_{\mbf{k},\al'},c^\dag_{\mbf{k}',\be'} \} P_{\be' \be}(\mbf{k}') = \delta_{\mbf{k}\mbf{k}'} [P(\mbf{k})^2]_{\al \be} = \delta_{\mbf{k}\mbf{k}'} P_{\al \be}(\mbf{k}) \ . \\
\eea
We now need the projected position operators in real space. They are defined in analogy to \Eq{eq:cRck} as
\bea
\label{eq:barcexpansoin}
\bar{c}_{\mbf{R},\al}  &=  \frac{1}{\sqrt{\mathcal{N}}}\sum_{\mbf{k}} e^{-i \mbf{k} \cdot (\mbf{R} + \mbf{r}_\al)} \bar{c}_{\mbf{k},\al} \\
&= \frac{1}{\sqrt{\mathcal{N}}}\sum_{\mbf{k}\be} e^{-i \mbf{k} \cdot (\mbf{R} + \mbf{r}_\al)} P_{\al \be}(\mbf{k}) c_{\mbf{k},\be} \\
&= \frac{1}{\sqrt{\mathcal{N}}}\sum_{\mbf{R}'} \lp \frac{1}{\sqrt{\mathcal{N}}} \sum_{\mbf{k}\be} e^{-i \mbf{k} \cdot (\mbf{R} + \mbf{r}_\al - \mbf{R}'-\mbf{r}_\be)} P_{\al \be}(\mbf{k}) \rp c_{\mbf{R}',\be} \\
&= \sum_{\mbf{R}'\be}p_{\al \be}(\mbf{R} - \mbf{R}') c_{\mbf{R}',\be} \\
\eea
where we defined the Fourier transform of the projector, i.e. the real-space correlation function of the non-interacting flat bands, as
\bea
\label{eq:pRandembeddingV}
p_{\al \be}(\mbf{R} -\mbf{R}') = \frac{1}{\mathcal{N}} \sum_{\mbf{k}} e^{-i \mbf{k} \cdot (\mbf{R} + \mbf{r}_\al - \mbf{R}'-\mbf{r}_\be)} P_{\al \be}(\mbf{k})= \frac{1}{\mathcal{N}} \sum_{\mbf{k}} e^{- i \mbf{k} \cdot (\mbf{R} - \mbf{R}')} [V(\mbf{k}) P(\mbf{k}) V^\dag(\mbf{k})]_{\al \be}, \quad  [V(\mbf{k})]_{\al \be} = e^{-i \mbf{k} \cdot \mbf{r}_\al} \delta_{\al \be} \ . \\
\eea
The embedding matrix $V(\mbf{k})$ defined in \Eq{eq:pRandembeddingV} is a unitary diagonal matrix. Although it is sometimes useful\cite{2022PhRvL.128h7002H} to take the thermodynamic limit $\mathcal{N}\to \infty$, we will keep $\mathcal{N}$ finite throughout this work.

We prove some properties about $p(\mbf{R})$. It is obvious that $p(\mbf{R}) = p^\dag(-\mbf{R})$ by Hermiticity of $P(\mbf{k})$ and $\Tr p(\mbf{R}) = N_f \delta_{\mbf{R},0}$ follows from $\Tr P(\mbf{k}) = N_f$. We also have the Fourier inversion
\bea
\label{eq:projdef}
P(\mbf{k}) = \sum_{\mbf{R}} e^{i \mbf{k} \cdot \mbf{R}} V^\dag(\mbf{k})p(\mbf{R})V(\mbf{k}) \ .
\eea
From the projector property $P^2 = P$, we find the convolution identity\cite{2022PhRvL.128h7002H}
\bea
\label{eq:convolution}
p(\mbf{R}) &= \sum_{\mbf{R}'}  p(\mbf{R}-\mbf{R}')p(\mbf{R}') 
\eea
which we use to derive the real space anti-commutation relations (and can also be recovered from Eq.~\ref{eq:anticom}). We compute
\bea
\label{eq:realspaceanticom}
\{\bar{c}_{\mbf{R},\al}, \bar{c}^\dag_{\mbf{R}',\be} \} &= \sum_{\mbf{S}\mbf{S}',\al'\be'} p_{\al \al'}(\mbf{R}-\mbf{S})  [p^\dag(\mbf{R}'-\mbf{S}')]_{\be' \be} \{c_{\mbf{S},\al'}, c^\dag_{\mbf{S}',\be'} \} \\
&= \sum_\mbf{S} [p(\mbf{R}-\mbf{S})p^\dag(\mbf{R}'-\mbf{S}) ]_{\al \be} \\
&= \sum_\mbf{S} [p(\mbf{R}-\mbf{R}'+\mbf{S})p^\dag(\mbf{S}) ]_{\al \be} \\
&=p_{\al \be}(\mbf{R}-\mbf{R}') \ . \\
\eea
In the thermodynamic limit where $\mathcal{N} \to \infty$, the flat bands are gapped so $P(\mbf{k})$ is continuous on the BZ and the real space correlation function $p(\mbf{R})$ decays at least exponentially fast.

We now look at the projected density operators $\bar{n}_{\mbf{k},\al} = \bar{c}^\dag_{\mbf{k},\al} \bar{c}_{\mbf{k},\al}$. We use the anti-commutation relations \Eq{eq:anticom}, finding
\bea
\bar{n}_{\mbf{k},\al}^2 &= \bar{c}^\dag_{\mbf{k},\al} \bar{c}_{\mbf{k},\al} \bar{c}^\dag_{\mbf{k},\al} \bar{c}_{\mbf{k},\al} = \bar{c}^\dag_{\mbf{k},\al} P_{\al \al}(\mbf{k}) \bar{c}_{\mbf{k},\al} - \bar{c}^\dag_{\mbf{k},\al} \bar{c}^\dag_{\mbf{k},\al} \bar{c}_{\mbf{k},\al}  \bar{c}_{\mbf{k},\al} =  P_{\al \al}(\mbf{k})  \bar{n}_{\mbf{k},\al} \\
\eea
where the second term vanishes because $\bar{c}^\dag_{\mbf{k},\al}$ anti-commutes with itself. In real space, the density $\bar{n}_{\mbf{R},\al} = \bar{c}^\dag_{\mbf{R},\al} \bar{c}_{\mbf{R},\al}$ obeys
\bea
\label{eq:nR2}
\bar{n}_{\mbf{R},\al}^2 &= \bar{c}^\dag_{\mbf{R},\al} \bar{c}_{\mbf{R},\al} \bar{c}^\dag_{\mbf{R},\al} \bar{c}_{\mbf{R},\al} = \bar{c}^\dag_{\mbf{R},\al}p_{\al \al}(\mbf{R}-\mbf{R}) \bar{c}_{\mbf{R},\al} -\bar{c}^\dag_{\mbf{R},\al} \bar{c}^\dag_{\mbf{R},\al}  \bar{c}_{\mbf{R},\al} \bar{c}_{\mbf{R},\al} = p_{\al \al}(0) \bar{n}_{\mbf{R},\al}  \\
\eea
using \Eq{eq:realspaceanticom}. The correlation function elements $p_{\al \al}(0)$ are onsite and measure local density of each orbital. They will play an important role in the many-body problem.

One may also derive these expressions from the electron operators in the kinetic energy band basis, $\gamma_{\kk,m}$.  In this basis, projection simply involves restricting the bands to $m \in M$, where $M$ labels the low-energy flat bands.  The projected density reads
\begin{align}
{\bar n}_{\kk, \alpha} &= \sum_{m,n} U(\kk)^*_{\alpha, m}
 U (\kk)_{\alpha, n} \gamma^\dagger_{\kk,m} \gamma_{\kk,n} \\
 {\bar n}_{\kk, \alpha}^2 &= \sum_{m,n} U(\kk)^*_{\alpha, m}
  U (\kk)_{\alpha, n} \gamma^\dagger_{\kk,m} \gamma_{\kk,n} \sum_{m',n'} U(\kk)^*_{\alpha, m'}
   U (\kk)_{\alpha, n'} \gamma^\dagger_{\kk,m'} \gamma_{\kk,n'} \nonumber \\
&= \sum_{m,n} U(\kk)^*_{\alpha, m}
    U (\kk)_{\alpha, n} \sum_{m',n'} U(\kk)^*_{\alpha, m'}
    U (\kk)_{\alpha, n'}  \gamma^\dagger_{\kk,m} [\{\gamma_{\kk,n}, \gamma^\dagger_{\kk,m'}\} - \gamma^\dagger_{\kk,m'} \gamma_{\kk,n}] \gamma_{\kk,n'} \nonumber \\
&= \sum_{m,n} U(\kk)^*_{\alpha, m}
        U (\kk)_{\alpha, n} \sum_{m',n'} U(\kk)^*_{\alpha, m'}
        U (\kk)_{\alpha, n'}  \gamma^\dagger_{\kk,m} [\delta_{n,m'} - 0] \gamma_{\kk,n'} \nonumber \\
&= P_{\alpha\alpha}(\kk) {\bar n}_{\kk,\alpha}.
\label{}
\end{align}
The zero in the second to last line arises from the antisymmetry upon interchange of the $m,m'$ indices.

\subsection{Uniform Pairing Condition from the Space group}
\label{app:UPC}

As Tovmasyan, Peotta, T\"orm\"a, and Huber identified in \Ref{PhysRevB.94.245149}, the solvability of the Hubbard model groundstates and excitations hinges on enforcing a simple criterion called the \textit{uniform pairing condition} (UPC):  $\sum_{\mbf{R}\al} \bar{n}^2_{\mbf{R},\al} \propto \sum_{\mbf{R}\al} \bar{n}_{\mbf{R},\al} + \bar{N}$ where $\bar{N}$ is the total (projected) number operator. The UPC was originally proposed to ensure equal mean-field pairing strength on all the orbital sites, but the UPC is simply a property of the single-particle wavefunctions as we will show. Using \Eq{eq:nR2}, we expand $\bar{n}^2_{\mbf{R},\al}$ to find
\bea
\label{eq:UPCMB}
\sum_{\mbf{R}\al} \bar{n}^2_{\mbf{R},\al} = \sum_{\mbf{R}\al} p_{\al\al}(0) \bar{n}_{\mbf{R},\al} 
\eea
To enforce \Eq{eq:UPCMB}, we must have
\bea
\label{eq:UPCapps}
p_{\al \al}(0) = \frac{1}{\mathcal{N}} \sum_\mbf{k} P_{\al\al}(\mbf{k})  = \eps \text{ if } \bar{n}_{\mbf{R},\al} \neq 0,~~\forall \alpha
\eea
where the constant $\eps$ is independent of $\al$ and is fixed by $\Tr p(0) = \sum_\al p_{\al \al}(0) = N_f$. We will derive the UPC (\Eq{eq:UPCMB}) from the appearance of the eta-pairing symmetry in the following section \App{app:Hubermodel}. 

The case of $\bar{n}_{\mbf{R},\al} = 0$ can occur for a given $\al$ at all $\mbf{R}$ if the wavefunction of the flat bands is exactly zero on the orbital $\al$ (see \Eq{eq:cdatPc}). We will see that this is the case in the bipartite crystalline lattices we consider shortly\cite{2021NatPh..18..185C}.

We now show that space group symmetries can enforce the uniform pairing condition \emph{generically}: arbitrary hoppings (preserving $G$) can be chosen as long as the underlying orbitals obey simple symmetry conditions which we make precise shortly. We consider unitary space group symmetries $g \in G, \ g = T_{\pmb{\delta}} g_s$ consisting of a symmorphic part $g_s$ (a rotation or reflection) and non-symmorphic part $T_{\pmb{\delta}}$ (a translation by a fraction of a lattice vector). The vector representation of $g$ on real-space vectors $\mbf{r}$ is defined to be $g\mbf{r} = g_s\mbf{r} + \pmb{\delta}$ \cite{2018Sci...361..246W}. The representations in momentum space take the form $D[g] = e^{i g_s \mbf{k} \cdot \pmb{\delta}} D[g_s]$ where $D[g_s]$ is the $N_{orb} \times N_{orb}$ representation on the orbitals. On the single-particle Hamiltonian, $g$ enforces
\bea
\label{eq:simhtilde}
e^{i g_s \mbf{k} \cdot \pmb{\delta}} D[g_s] \tilde{h}(\mbf{k}) \lp e^{i g_s \mbf{k} \cdot \pmb{\delta}} D[g_s] \rp^\dag = D[g_s] \tilde{h}(\mbf{k}) D^\dag[g_s] = \tilde{h}(g_s \mbf{k}) , \\
\eea
canceling the non-symmorphic phases. From \Eq{eq:simhtilde}, $D[g_s] P(\mbf{k}) D^\dag[g_s] =  P(g_s\mbf{k})$\cite{2022PhRvL.128h7002H,2018Sci...361..246W}. It is understood that the representation matrix $D[g]$ only depends on the symmorphic part of $g$, and henceforth we drop the $s$ subscript.

\begin{figure*}
\includegraphics[height=.3\textwidth]{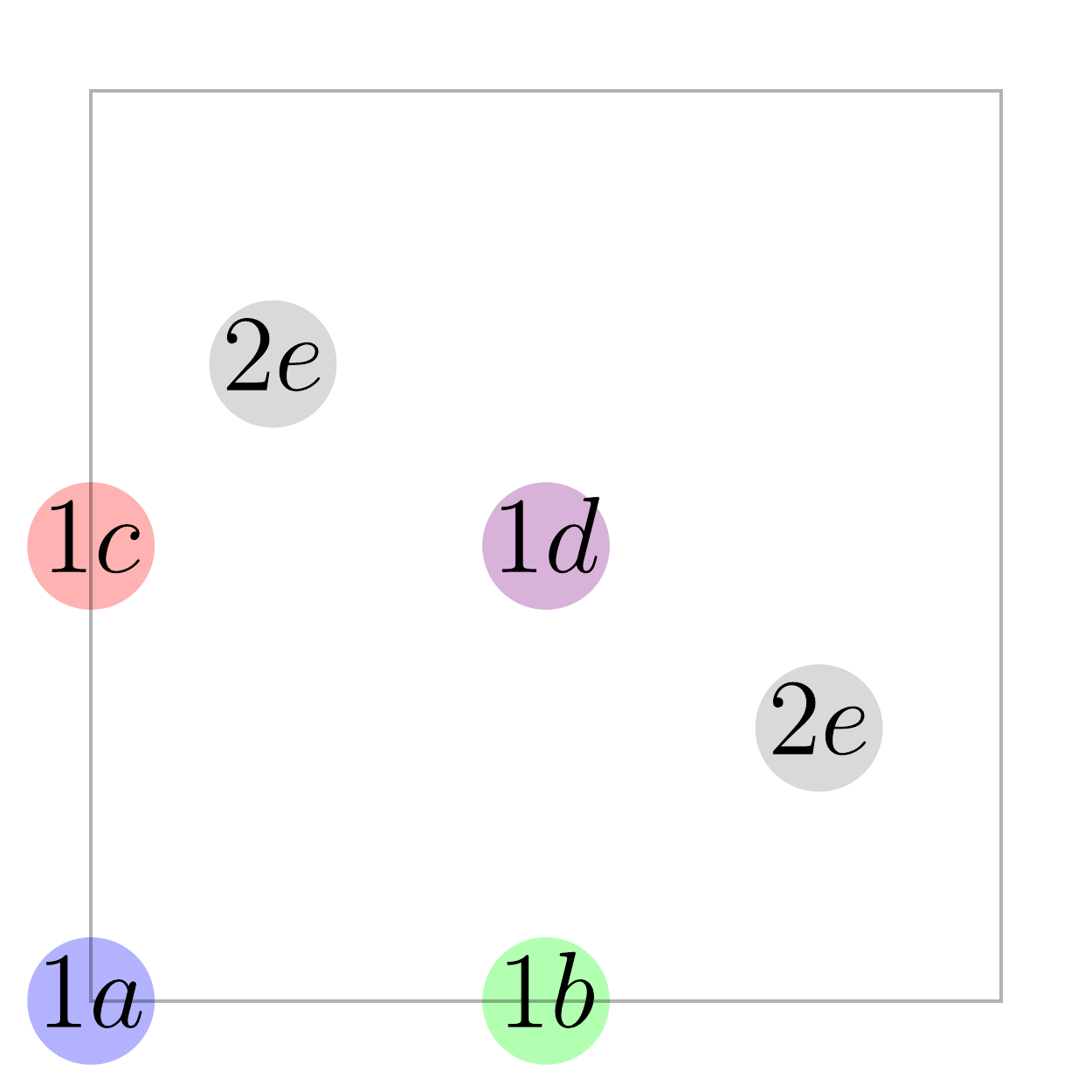}
\caption{Unit cell of the wallpaper group $p2$ generated by $C_2$ and translations (we take $\mbf{a}_1 = (1,0),\mbf{a}_2 = (0,1)$ for ease). There are four maximal Wyckoff positons 1a = $(0,0)$,1b= $(0,1/2)$,1c= $(0,1/2)$,1d= $(1/2,1/2)$ which all have multiplicity 1 and site symmetry group $G_x = 2$. At 1a for instance, the site symmetry group is $\{1, C_2\}$. The non-maximal Wyckoff position 2e$=\{(x,y),(1-x,1-y)\}$ has multiplicity 2 and consists of two general inversion-related points in the unit cell. Since $C_2$ exchanges the two points in 2e, it is not part of the site symmetry group and thus $G_{2e} = 1$ containing only the identity.}
\label{fig:wyckoff}
\end{figure*}

We also need some simple notions from topological quantum chemistry. In a lattice with space group $G$, a Wyckoff position \cite{2020arXiv200604890C,bacry1988symmetry} $x$ of multiplicity $n$ consists of the points $\mbf{x}_1, \dots, \mbf{x}_n$ in the unit cell which form an orbit of $G$: there exists $g\in G$ such that $\mbf{x}_i = g \mbf{x}_j$ for each $\mbf{x}_i,\mbf{x}_j \in x$. The site symmetry group of a point $\mbf{x}$ is defined by $G_\mbf{x} = \{g\in G| g\mbf{x} = \mbf{x}\} \in G$, and all $G_{\mbf{x}}$ for $\mbf{x}\in x$ are isomorphic. Hence in any lattice with $G$, we can organize the orbitals into irreps of $G_x$ at some Wyckoff position $x$. All such irreps are tabulated on the \href{https://www.cryst.ehu.es/cgi-bin/cryst/programs/bandrep.pl}{Bilbao Crystallographic Server} \cite{Aroyo:firstpaper,Aroyo:xo5013}. For instance in $G = p2$ generated by $C_2$ rotation and translations, the $1a$ position (of multiplicity 1) is the origin of the unit cell and $G_{1a} = \{1, C_2\}$ with two 1D irreps (even and odd under $C_2$). The $2e$ position (of multiplicity 2) is the generic nonmaximal position consisting of a pair of $C_2$-related points, so $G_{2e}$ is the trivial group $\{1\}$ (see \Fig{fig:wyckoff}).

We can now state our claim.

\textbf{Lemma}. Let $x$ be a Wyckoff position (not necessarily maximal) with site-symmetry group $G_x$ and irreps $\chi$. The uniform pairing condition holds between all orbitals forming an irrep $\chi$ induced from the Wyckoff position $x$.

\textbf{Proof.}  We define $\tilde{G}$ as the group of symmetries of the unit cell, i.e. the space group mod lattice translations. We define the site-symmetry group $H = G_{\mbf{x}_1} \cong G_x$ where for concreteness $G_{\mbf{x}_1}$ is the point group of the site $\mbf{x}_1 \in x$. For a Wyckoff position of multiplicity $n$ and an irrep $\chi$ of $G_x$ with dimension $|\chi|$ and (irreducible) representations $D^\chi[h]$ for $h\in H$, the Wyckoff position contains $n |\chi|$ orbitals per unit cell. We index the orbitals in a tensor product basis $\al = i,a$ where $i = 1, \dots, n$ and $a =1,\dots, |\chi|$. It is useful to note that in this basis, the embedding matrices are block diagonal, $V_{ia,jb}[\mbf{G}] = e^{- i \mbf{r}_i \cdot \mbf{G}} \delta_{ij} \delta_{ab}$. It is natural to employ a coset decomposition separating the onsite symmetries of $H$ and the symmetries that connect different points in the Wyckoff position. The formal expression is $\tilde{G} = \tilde{g}_1 H + \dots + \tilde{g}_n H$,  $n = \frac{|\tilde{G}|}{|H|}$ such that irreps of $H$ form an induced representation in $\tilde{G}$ \cite{2020Sci...367..794S}. Hence the representation matrices of $G$ can be chosen as
\bea
\label{eq:symmetryforms}
D[h] &= \mathbb{1}_n \otimes D^\chi[h], \quad h \in H \\
D[\tilde{g}_i] &= P_{\tilde{g}_i} \otimes \mathbb{1}_{|\chi|}, \qquad \tilde{g}_i \in \{\tilde{g}_1, \dots, \tilde{g}_n\}
\eea
where $[P_{\tilde{g}}]_{ij} = \delta_{\tilde{g}\mbf{x}_i,\mbf{x}_j}$ is a permutation matrix\cite{2020Sci...367..794S} on the $n$ sites of the Wyckoff position, and the Kronecker delta is understood mod lattice vectors. First we prove uniform pairing holds within the $|\chi|$ orbitals at a given $\mbf{x}_i$.
Choose an arbitrary diagonal block of $P(\mbf{k})$ denoted $[P_i(\mbf{k})]_{ab} = P_{ia,ib}(\mbf{k})$. We have denoted $\alpha = ia$ and $\beta = ib$.  Note that $P_i(\mbf{k}) = P_i(\mbf{k}+\mbf{G})$ because the embedding matrix acts as $V_{ia,ib}[\mbf{G}] = e^{-i \mbf{x}_i \cdot \mbf{G}} \delta_{ab}$, proportional to the identity. Then because the BZ is symmetric under $H$, we have
\bea
\frac{1}{\mathcal{N}} \sum_\mbf{k} P_i(\mbf{k}) = \frac{1}{\mathcal{N}} \sum_\mbf{k} P_i(h\mbf{k}) =  D^\chi[h] \lp \frac{1}{\mathcal{N}} \sum_\mbf{k} P_i(\mbf{k}) \rp D^\chi[h]^\dag
\eea
but because $D^\chi[h]$ is irreducible, by Schur's lemma we find $\frac{1}{\mathcal{N}} \sum_\mbf{k} P_i(\mbf{k}) = \eps_i \mathbb{1}_{|\chi|}$ where $\eps_i$ is an arbitrary constant.  That is, the only matrix invariant under an irreducible representation is proportional to the identity.  If there were another invariant matrix, call it $M$, that matrix $M$ would commute with all symmetries and thus will block-diagonalize the representation matrices, implying the representation was reducible.

To prove that $\eps_i = \eps_j$, relating different blocks, we use the fact that there exists $\tilde{g} \notin H$ such that $\tilde{g} \mbf{x}_i = \mbf{x}_j$. Then because $D[\tilde{g}] P(\mbf{k}) D^\dag[\tilde{g}] = P(\tilde{g}\mbf{k})$, we find that
\bea
\frac{1}{\mathcal{N}} \sum_\mbf{k} P_i(\mbf{k}) = \frac{1}{\mathcal{N}} \sum_\mbf{k} P_i(g\mbf{k}) = \frac{1}{\mathcal{N}} \sum_\mbf{k} P_j(\mbf{k})
\eea
such that $\eps_i = \eps_j$. Thus $\frac{1}{\mathcal{N}} \sum_\mbf{k} P_{\al\al}(\mbf{k}) = \eps$ for all $\al$ in a the irrep $\chi$ of the Wyckoff position $x$. Because $\Tr P(\mbf{k}) = N_f$, the number of projected bands, we have
\bea
\label{eq:UPCint}
\frac{1}{\mathcal{N}} \sum_\mbf{k} P_{\al\al}(\mbf{k}) = \frac{1}{\mathcal{N}} \sum_\mbf{k} \frac{1}{n |\chi|}\Tr P(\mbf{k}) = \frac{N_f}{n |\chi|} \equiv \eps \leq 1  .
\eea
The value of $\eps$ has physical meaning. From \Eq{eq:projdef}, we see that $\eps = p_{\al\al}(0)$, which is the diagonal of the onsite correlation function. Although we have discussed only unitary symmetry groups in this proof, anti-unitary spatial symmetries, which do not act on spin, also enforce uniform pairing. This is easily seen because $P(\mbf{k})$ is Hermitian and its diagonal elements must be real, so complex conjugation acts trivially. In \App{app:Aaron}, we use the spinless symmetry $C_{6z} T$ to enforce uniform pairing between 6 orbitals. \\

We now extend this result to the case of a bipartite crystalline lattice consisting of the sublattices $L$ and $\tilde{L}$. The integers $N_{L}$ and $N_{\tilde{L}}$ denote the number of orbitals on each sublattice, so $N_{orb} =  N_{L} + N_{\tilde{L}}$. Without loss of generality we take $N_{L} > N_{\tilde{L}}$. We consider single-particle Hamiltonians of the form
\bea
\label{eq:starterham}
\tilde{h}(\mbf{k}) &= \bpm 0 & S_\mbf{k} \\ S_\mbf{k}^\dag & B_\mbf{k}  \epm
\eea
where $S_\mbf{k}$ is a $N_{L} \times N_{\tilde{L}}$ rectangular matrix and $B_\mbf{k}$ is an $N_{\tilde{L}} \times N_{\tilde{L}}$ Hermitian matrix. The number of flat bands is $N_f = N_{L} - N_{\tilde{L}} > 0$ and their wavefunctions obey \cite{2021NatPh..18..185C}
\bea
 \bpm 0 & S_\mbf{k} \\ S_\mbf{k}^\dag & B_\mbf{k}  \epm \bpm \psi_\mbf{k} \\ 0\epm &= \bpm 0 \\ S^\dag_\mbf{k} \psi_\mbf{k} \epm = 0
\eea
where the existence of $N_f = N_{L} - N_{\tilde{L}}$ wavefunctions $\psi_{\mbf{k},n}, n = 1,\dots, N_f$ in the nullspace of $S^\dag_\mbf{k}$ is guaranteed by rank deficiency. Thus the flat band projector
\bea
P(\mbf{k}) &= \sum_{n=1}^{N_f} \bpm \psi_{\mbf{k},n} \\ 0\epm \bpm \psi^\dag_{\mbf{k},n} & 0\epm = \sum_{n=1}^{N_f} \bpm \psi_{\mbf{k},n} \psi^{\dag}_{\mbf{k},n} & 0 \\ 0 & 0 \epm
\eea
has nonzero entries only in the $L$ sublattice. As such, the projected operators $\bar{c}_{\mbf{k},\al} = P_{\al \be}(\mbf{k})c_{\mbf{k},\be}$ are only nonzero when $\al$ is an orbital in the $L$ sublattice, and so $\bar{n}_{\mbf{R},\al \in \tilde{L}} = 0$. Hence for \Eq{eq:UPCMB} to hold, we only require $p_{\al\al}(0)$ to be independent of $\al$ within the $L$ sublattice. As proven in the Lemma above, this is accomplished by requiring the $L$ sublattice to consist of orbitals within a single irrep. More than one irrep would no longer guarantee the uniform pairing condition.  Again because $\Tr p(0) = N_f = N_{L} - N_{\tilde{L}}$, we have
\bea
p_{\al \al}(0) = \frac{N_f}{N_{L}} \equiv \eps < 1 \ .
\eea
For simplicity of notation, we will not distinguish between sums over $\al \in L \oplus \tilde{L}$ and $\al \in L$. This is because the flat band wavefunctions have no weight in the $\tilde{L}$ sublattice, and so the sums are interchangable (any additional terms belonging to $\tilde L$ are $0$).

The notable feature of our construction is the total freedom to choose the hopping elements once the space group and orbitals are fixed. Using the bipartite crystalline lattice construction, we obtain perfectly flat bands obeying the uniform pairing condition in all space groups. The topology of the flat bands can be determined directly as the formal subtraction of the band representations of the two sublattices, and in many cases is fragile topological or obstructed atomic\cite{2021NatPh..18..185C}. If the formal subtraction results in a ``negative" irrep at high symmetry momentum $\bf K$, then the flat bands are gapless with the dispersive bands at $\bf K$. In this case, projection into the flat bands is not justified, but this case can be analyzed very generally in mean field theory \cite{2022arXiv220311133H}.

\Ref{2021NatPh..18..185C} also constructs auxiliary models from the bipartite Hamiltonian in \Eq{eq:starterham}. By integrating out the $\tilde{L}$ sublattice, one obtains the positive semi-definite $N_L\times N_L$ Hamiltonian $\tilde{h}_L(\mbf{k}) = S_\mbf{k}S^\dag_\mbf{k}$ which retains the $N_L - N_{\tilde{L}}$ zero-energy flat bands with wavefunctions $\psi_{\mbf{k},n}$, as well as $N_{\tilde{L}}$ dispersive bands. These models $\tilde{h}_L(\mbf{k})$ also possess the uniform pairing condition.

\subsubsection{Spin}
\label{app:spin}

To add spin, we impose $S_z$ conservation and spinful time-reversal symmetry $\mathcal{T}$. Together, $S_z$ and $\mathcal{T}$ ensure the existence of an eta-pairing operator as we show in \App{app:Hubermodel}. Enforcing these symmetries, the single-particle Hamiltonian takes the form
\bea
\tilde{h}_s(\mbf{k}) &= \bpm \tilde{h}_\u(\mbf{k}) & \\ &  \tilde{h}_\d(\mbf{k}) \\ \epm = \bpm \tilde{h}(\mbf{k}) & \\ &  \tilde{h}^*(-\mbf{k}) \\ \epm
\eea
where we used $D[\mathcal{T}] = i \sigma_y K$, so in particular $\mathcal{T}$ acts diagonally on the orbital index $\al,\be$.  Here $\sigma_y$ acts on the spin indices and $K$ is complex conjugation. Because of the global symmetry $S_z$, we can fix the gauge of $\mathcal{T}$ on the eigenstates so that $U^\d(\mbf{k}) = U^\u(-\mbf{k})^*$  and $P^\d(\mbf{k}) = P^\u(-\mbf{k})^*$. As such we have
\bea
\label{eq:TRSpR}
p_{\al \be}^\d(\mbf{R} -\mbf{R}')
 = \frac{1}{\mathcal{N}} \sum_\mbf{k} e^{-i \mbf{k} \cdot (\mbf{R}+\mbf{r}_\al-\mbf{R}'-\mbf{r}_\be)} P_{\al \be}^\u(-\mbf{k})^* = \lp \frac{1}{\mathcal{N}} \sum_\mbf{k} e^{-i \mbf{k} \cdot (\mbf{R}+\mbf{r}_\al-\mbf{R}'-\mbf{r}_\be)} P_{\al \be}^\u(\mbf{k}) \rp ^*= p^\u_{\al \be}(\mbf{R} -\mbf{R}')^*
\eea
which is just the statement that $\mathcal{T}$ acts locally in real space. If the uniform pairing condition is obeyed by $p^\d(0)$, it is also by $p^\u(0)$ using $\mathcal{T}$ and $S_z$.

\section{Hubbard Model and Eta-pairing Symmetry}
\label{app:Hubermodel}

In this section, we construct a class of positive semi-definite attractive Hubbard models using the uniform pairing condition generalizing the model of \Ref{PhysRevB.94.245149} (\App{eq:hubbardmodels}). The existence of an eta-pairing symmetry in the models we consider is protected by $S_z$ and $\mathcal{T}$. We derive the form of the eta-pairing symmetry (\App{app:etaops}) and show that it expands the $U(1) \times U(1)$ charge-spin symmetry group to a non-abelian $SU(2)$ symmetry group (\App{app:symalg}).

\subsection{Attractive Hubbard Models}
\label{eq:hubbardmodels}

We now add a Hubbard term $H$ to the non-interacting Hamiltonian $\tilde{H}$ in \Eq{eq:gammasing}. The form of the Hubbard term was discovered in \Ref{PhysRevB.94.245149} and yields an eta-pairing symmetry, $[H, \eta] = 0$. The first step is to define a projected Hermitian spin operator
\bea
\bar{S}^z_{\mbf{R},\al} &= \bar{c}^\dag_{\mbf{R},\al,\u}\bar{c}_{\mbf{R},\al,\u}-\bar{c}^\dag_{\mbf{R},\al,\d}\bar{c}_{\mbf{R},\al,\d} = \bar{n}_{\mbf{R},\al,\u}-  \bar{n}_{\mbf{R},\al,\d} \ .\\
\eea
With the uniform pairing condition \Eq{eq:UPCapps}, the form of this operator naturally gives an attractive Hubbard model. We define
\bea
\label{eq:HuberN}
H = \frac{1}{2}|U| \sum_{\mbf{R},\al} (\bar{S}^z_{\mbf{R},\al})^2  \ .
\eea
The crucial feature of $H$ is that it is the sum of squares and thus is positive semi-definite. Hence a state $\ket{\psi}$ obeying $H\ket{\psi} = 0$ must be a ground state. Now we show that $H$ is an attractive Hubbard model. Using \Eq{eq:nR2}, we compute
\bea
\label{eq:Hhubbardform}
H &= \frac{1}{2} |U| \sum_{\mbf{R},\al} (\bar{n}_{\mbf{R},\al,\u}-  \bar{n}_{\mbf{R},\al,\d} )^2  \\
&= \frac{1}{2}|U| \sum_{\mbf{R},\al} -2 \bar{n}_{\mbf{R},\al,\u} \bar{n}_{\mbf{R},\al,\d} + p^\u_{\al \al}(0) \bar{n}_{\mbf{R},\al,\u}+ p^\d_{\al \al}(0) \bar{n}_{\mbf{R},\al,\d}  \\
&= -|U| \sum_{\mbf{R},\al} \bar{n}_{\mbf{R},\al,\u} \bar{n}_{\mbf{R},\al,\d} + \frac{\eps |U|}{2} \sum_{\mbf{R} ,\al} \bar{n}_{\mbf{R},\al,\u}+ \bar{n}_{\mbf{R},\al,\d}  \\
&= -|U| \sum_{\mbf{R},\al} \bar{n}_{\mbf{R},\al,\u} \bar{n}_{\mbf{R},\al,\d} + \frac{\eps |U|}{2} \bar{N}\\
\eea
which is a Hubbard interaction with strength $|U|$ and a chemical potential term $\frac{\eps |U|}{2} \bar{N}$. To get from line 2 to line 3, we have employed the uniform pairing condition $p^\u_{\al \al}(0) = p^\d_{\al \al}(0) = \frac{\eps |U|}{2}$.  We have also used the fact that
\bea
\sum_{\mbf{R},\al,\sigma} \bar{n}_{\mbf{R},\al,\sigma} &= \frac{1}{\mathcal{N}} \sum_{\mbf{k},\mbf{k}'} \sum_{\mbf{R},\al,\sigma}  e^{- i (\mbf{k}-\mbf{k}') \cdot (\mbf{R}+ \mbf{r}_\al)} \bar{c}^\dag_{\mbf{k}' \al,\sigma} \bar{c}_{\mbf{k} \al,\sigma}  = \sum_{\mbf{k}\al,\sigma}  \bar{c}^\dag_{\mbf{k} \al,\sigma} \bar{c}_{\mbf{k} \al,\sigma}  = \sum_{\mbf{k},n,\sigma} \gamma^\dag_{\mbf{k},n,\sigma} \gamma_{\mbf{k},n,\sigma}  = \bar{N} \\
\eea
where the final sum is over the $N_f$ flat bands. Let us make a brief comment about the form of the attractive Hubbard term. The form of $\sum_{\al} \bar{n}_{\mbf{R},\al,\u} \bar{n}_{\mbf{R},\al,\d}$ is somewhat special because each orbital $\al$ only couples to itself (but opposite in spin). If there is only one orbital per Wyckoff position (i.e. a 1D irrep of $G_x$), this is the same type of term that one would get from a purely local Coulomb interaction. However if the irrep $\chi$ is multi-dimensional, like a pair of $p_x, p_y$ orbitals per site, then a Coulomb interaction coupling the \emph{total} densities would be $\sum_i (\sum_{a} \bar{n}_{\mbf{R},ia,\u})(\sum_{a} \bar{n}_{\mbf{R},ia,\d})$ in the tensor product notation of \App{app:UPC} where $a$ indexes the orbitals at a given point $\mbf{x}_i \in x$.  Given that the mechanism of generating attractive Hubbard interactions is typically phonons with a complicated coupling structure, we view \Eq{eq:Hhubbardform} as a simple model capturing the essential physics. Although we focus here on an attractive interaction, a particle-hole transformation\cite{PhysRevLett.62.1201} on the $\d$ spins maps all our results to the repulsive case with ferromagnetism.

%
%

\subsection{Eta Operators}
\label{app:etaops}

Generically, the $\bar{S}^z_{\mbf{R},\al}$ operators do not commute among themselves because of the projection (see \Eq{eq:realspaceanticom}). However, in the trivial atomic limit case where the orbitals are totally decoupled, $\bar{S}^z_{\mbf{R},\al} \to S_{\mbf{R},\al}$ is commuting for all $\mbf{R},\al$, yielding an infinite number of conserved charges. This is the classical limit where the model contains only strictly local density operators. This limit does not have superconductivity because there is no coherence -- every site is decoupled. This is an indication that some obstruction to atomic localization, like topology or symmetry protected obstructed states, is necessary for the projected strong-coupling model to have a superconducting phase.

There is still a nontrivial symmetry operator when the projected bands are not atomic, the eta-pairing symmetry \cite{PhysRevB.94.245149}. We use the most general two-body, charge $+2$, spin 0, momentum 0 operator for our ansatz of the $\eta^\dag$ operator. It takes the form
\bea
\label{eq:etadef}
\eta^\dag &= \sum_{\mbf{q}} \gamma^\dag_{\mbf{q},m,\u} \Omega_{mn}(\mbf{q}) \gamma^\dag_{-\mbf{q},n,\d} \\
\eea
involving the yet undetermined matrix $\Omega_{mn}(\mbf{q})$. We now must discuss the invariance of $\eta^\dag$ under the eigenvector gauge freedom $U^\sigma(\mbf{k}) \to U^\sigma(\mbf{k}) \mathcal{W}^\sigma(\mbf{k})$. Clearly for $\eta^\dag$ to be a physical symmetry, it must be invariant under the eigenvector gauge freedom which severely restricts the form of $\eta$ to a simple ansatz. We will find that with $S_z$ and $\mathcal{T}$, this ansatz is successful. Since $S_z$ and $\mathcal{T}$ are the symmetries relevant for real materials, we only consider this case explicitly in our work. However, we mention that the combination of particle-hole symmetry $\mathcal{P}$ and $SU(2)$ spin symmetry also protects the existence of an $\eta$ operator.

The operators $\gamma^\dag_{\mbf{q},n,\sigma}$ are not gauge invariant. Under the eigenvector gauge freedom, we have $\gamma^\dag_{\mbf{q},n,\sigma} \to \sum_{n'}\gamma^\dag_{\mbf{q},n',\sigma} \mathcal{W}^\sigma_{n'n}(\mbf{q})$ where $\mathcal{W}^\sigma(\mbf{q})$ is a unitary matrix. Hence
\bea
\eta^\dag &\to \sum_{\mbf{q},mnm'n'} \gamma^\dag_{\mbf{q},m',\u} \mathcal{W}_{m' m}^\u(\mbf{q}) \Omega_{mn}(\mbf{q})  \gamma^\dag_{-\mbf{q},n',\d} \mathcal{W}_{n' n}^\d(-\mbf{q}) \\
&= \sum_{\mbf{q},m'n'} \gamma^\dag_{\mbf{q},m',\u} [\mathcal{W}^\u(\mbf{q}) \Omega(\mbf{q}) \mathcal{W}^{\d \dag}(-\mbf{q})^*]_{m' n'} \gamma^\dag_{-\mbf{q},n',\d} \ .  \\
\eea
In order for $\eta^\dag$ to be invariant, we must have
\bea
\label{eq:etainvariance}
\mathcal{W}^\u(\mbf{q}) \Omega(\mbf{q}) \mathcal{W}^{\d \dag}(-\mbf{q})^* &=   \Omega(\mbf{q}) \ .
\eea
The natural object $\Omega$ transforming in this way is a sewing matrix\cite{2017PhRvB..96x5115B,Benalcazar_2017} relating states at $\mbf{q}$ and $-\mbf{q}$. Because of the conjugation in \Eq{eq:etainvariance}, the sewing matrix must be that of an anti-unitary symmetry, like time-reversal.

Fixing the gauge of the $S_z$ and $\mathcal{T}$ sewing matrices so that $U^\u(-\mbf{q})^* = U^\d(\mbf{q})$, we can choose eigenstates such that $\mathcal{W}^\u(\mbf{q}) = \mathcal{W}^\d(-\mbf{q})^*$ and thus $\mathcal{W}^\u(\mbf{q}) \Omega(\mbf{q}) \mathcal{W}^{\d \dag}(-\mbf{q})^* = \mathcal{W}^\u(\mbf{q}) \Omega(\mbf{q}) \mathcal{W}^{\u}(\mbf{q})^\dag$. Thus $\eta^\dag$ is invariant under the gauge freedom if $\mathcal{W}^\u(\mbf{q}) \Omega(\mbf{q}) \mathcal{W}^{\u}(\mbf{q})^\dag = \Omega(\mbf{q}) $ which is satisfied for all arbitrary $\mathcal{W}^\u(\mbf{q})$ iff $\Omega_{mn}(\mbf{q}) \propto \delta_{mn}$.  Thus we arrive at the ansatz
\bea
\label{eq:etasymant}
\eta^\dag &= \sum_{\mbf{q},m} \gamma^\dag_{\mbf{q},m,\u} \gamma^\dag_{-\mbf{q},m,\d} \eea
in the gauge-fixed wavefunctions where $U^\u(\mbf{k}) = U^\d(-\mbf{k})^*$. The $\eta^\dag$ operator takes a simpler form in terms of the projected operators (which are gauge invariant). Recalling $\gamma^\dag_{\mbf{q},m,\sigma} = \sum_\al \bar{c}^\dag_{\mbf{q},\al,\sigma} U^\sigma_{\al, m}(\mbf{q})$, we have
\bea
\label{eq:etadagniceform}
\eta^\dag &= \sum_{\mbf{q},m\al\be} \bar{c}^\dag_{\mbf{q},\al,\u} U^\u _{\al m}(\mbf{q}) U^\d_{\be m}(-\mbf{q}) \bar{c}^\dag_{-\mbf{q},\be,\d} = \sum_{\mbf{q},\al\be} \bar{c}^\dag_{\mbf{q},\al,\u} P^\u_{\al \be}(\mbf{q}) \bar{c}^\dag_{-\mbf{q},\be,\d} =  \sum_{\mbf{q}\al} \bar{c}^\dag_{\mbf{q},\al,\u} \bar{c}^\dag_{-\mbf{q},\al,\d} =  \sum_{\mbf{R}\al} \bar{c}^\dag_{\mbf{R},\al,\u} \bar{c}^\dag_{\mbf{R},\al,\d} .
\eea
In the last equality of \Eq{eq:etadagniceform}, we Fourier transformed to real space. Physically, $\eta^\dag$ can be understood as creating a Cooper pair at zero momentum. Note that the wavefunction of $\bar{c}^\dag_{\mbf{R},\al,\u} \bar{c}^\dag_{\mbf{R},\al,\d}$ decays exponentially outside the correlation length of $p(\mbf{R})$, which describes the ``size" of the Cooper pair bound state.

We now proceed to the commutator of $\eta^\dag$ and $\bar{S}_{\mbf{R},\al}$:
\bea
\null [\bar{S}^z_{\mbf{R},\al},\eta^\dag] &= \sum_{\mbf{R}'\be} [\bar{c}^\dag_{\mbf{R},\al,\u}\bar{c}_{\mbf{R},\al,\u}-\bar{c}^\dag_{\mbf{R},\al,\d}\bar{c}_{\mbf{R},\al,\d},\bar{c}^\dag_{\mbf{R}',\be,\u} \bar{c}^\dag_{\mbf{R}',\be,\d}] \\
&=   \sum_{\mbf{R}'\be} \bar{c}^\dag_{\mbf{R},\al,\u} [\bar{c}_{\mbf{R},\al,\u},\bar{c}^\dag_{\mbf{R}',\be,\u} \bar{c}^\dag_{\mbf{R}',\be,\d}] -\bar{c}^\dag_{\mbf{R},\al,\d}[\bar{c}_{\mbf{R},\al,\d},\bar{c}^\dag_{\mbf{R}',\be,\u} \bar{c}^\dag_{\mbf{R}',\be,\d}] \\
&= \sum_{\mbf{R}'\be} \bar{c}^\dag_{\mbf{R},\al,\u} p_{\al \be}^\u(\mbf{R}-\mbf{R}') \bar{c}^\dag_{\mbf{R}',\be,\d} + \bar{c}^\dag_{\mbf{R},\al,\d} \bar{c}^\dag_{\mbf{R}',\be,\u} p_{\al \be}^\d(\mbf{R}-\mbf{R}') \\
\eea
where we used the anti-commutator $\{\bar{c}_{\mbf{k},\al,\sigma}, \bar{c}^\dag_{\mbf{k},\be,\sigma'} \} = \delta_{\mbf{k}\mbf{k}'} P_{\al \be}(\mbf{k}) \delta_{\sigma \sigma'}$ (see \Eq{eq:realspaceanticom}). We expand the $\bar{c}^\dag_{\mbf{R}',\be,\sigma}$ operators using \Eq{eq:barcexpansoin} to find
\bea
\null [\bar{S}^z_{\mbf{R},\al},\eta^\dag] &=  \sum_{\mbf{R}'\mbf{R}'',\be \be'} \bar{c}^\dag_{\mbf{R},\al,\u} c^\dag_{\mbf{R}'',\be',\d} p_{\al \be}^\u(\mbf{R}-\mbf{R}')   p^{\d*}_{\be \be'}(\mbf{R}'-\mbf{R}'')+ \bar{c}^\dag_{\mbf{R},\al,\d} c^\dag_{\mbf{R}'',\be',\u} p_{\be \be'}^{\u *}(\mbf{R}'-\mbf{R}'') p_{\al \be}^\d(\mbf{R}-\mbf{R}') \\
&=  \sum_{\mbf{R}'\mbf{R}'',\be \be'} \bar{c}^\dag_{\mbf{R},\al,\u} c^\dag_{\mbf{R}'',\be',\d} p_{\al \be}^\u(\mbf{R}-\mbf{R}')   p^{\u}_{\be \be'}(\mbf{R}'-\mbf{R}'')+ \bar{c}^\dag_{\mbf{R},\al,\d} c^\dag_{\mbf{R}'',\be',\u} p_{\be \be'}^{\d}(\mbf{R}'-\mbf{R}'') p_{\al \be}^\d(\mbf{R}-\mbf{R}') \\
&= \sum_{\mbf{R}'\mbf{R}'',\be'} \bar{c}^\dag_{\mbf{R},\al,\u} c^\dag_{\mbf{R}'',\be',\d} [p^\u(\mbf{R}-\mbf{R}') p^{\u}(\mbf{R}'-\mbf{R}'')]_{\al \be'}+ \bar{c}^\dag_{\mbf{R},\al,\d} c^\dag_{\mbf{R}'',\be',\u} [p^\d(\mbf{R}-\mbf{R}') p^{\d}(\mbf{R}'-\mbf{R}'')]_{\al \be'}  \\
\eea
where we used $p^\u_{\al\be}(\mbf{R})^* = p^\d_{\al\be}(\mbf{R})$ thanks to $\mathcal{T}$ in \App{app:spin}. Finally, the convolution identity of $p(\mbf{R})$ in \Eq{eq:convolution} gives
\bea
\null [\bar{S}^z_{\mbf{R},\al},\eta^\dag] &=  \sum_{\mbf{R}'',\be'} \bar{c}^\dag_{\mbf{R},\al,\u} c^\dag_{\mbf{R}'',\be',\d} p_{\al \be'}^{\u}(\mbf{R}-\mbf{R}'')+ \bar{c}^\dag_{\mbf{R},\al,\d} c^\dag_{\mbf{R}'',\be',\u} p_{\al \be'}^{\d}(\mbf{R}-\mbf{R}'')\\
&=  \sum_{\mbf{R}'',\be'} \bar{c}^\dag_{\mbf{R},\al,\u} c^\dag_{\mbf{R}'',\be',\d} p_{\al \be'}^{\d *}(\mbf{R}-\mbf{R}'')+ \bar{c}^\dag_{\mbf{R},\al,\d} c^\dag_{\mbf{R}'',\be',\u} p_{\al \be'}^{\u *}(\mbf{R}-\mbf{R}'')\\
&= ( \bar{c}^\dag_{\mbf{R},\al,\u} \bar{c}^\dag_{\mbf{R},\al,\d} + \bar{c}^\dag_{\mbf{R},\al,\d} \bar{c}^\dag_{\mbf{R},\al,\u} )\\
&= 0
\eea
again using \Eq{eq:barcexpansoin} to recover the projected operators in the penultimate line. Thus we have proven $\eta^\dag$ commutes with $\bar{S}^z_{\mbf{R},\al}$, so it is trivially established that $[H, \eta^\dag] = 0$. It follows that the eta-pairing states are all eigenstates:
\bea
H \eta^{\dag n} \ket{0} &= \eta^{\dag n} H \ket{0}= 0
\eea
and in particular must be groundstates because $H$ is positive semi-definite. Note that because $[\eta^\dag, \bar{S}^z_{\mbf{R},\al}] = 0$, one could consider much more general (though not necessarily physical) Hamiltonians with an eta-pairing symmetry than the onsite $ H = \frac{|U|}{2} \sum_{\mbf{R},\al}\bar{S}^z_{\mbf{R},\al}\bar{S}^z_{\mbf{R},\al}$ Hamiltonian studied in this work.

One may also calculate the commutators using the $\gamma$ basis. We define the momentum space spin operator
\begin{align}
\bar{S}^z_{\mbf{q},\al} &= \sum_\RR e^{-i\mbf{q}\cdot\RR} \bar{S}^z_{\RR,\alpha} \\
&= \sum_{\RR, \alpha,\sigma} (-1)^\sigma e^{-i\mbf{q}\cdot\RR} {\bar n}_{\RR,\alpha,\sigma} \\
&= \sum_{\kk, \alpha,\sigma} (-1)^\sigma {\bar c}^\dagger_{\kk-\qq,\alpha,\sigma} {\bar c}_{\kk,\alpha,\sigma} \\
&= \sum_{\kk, \alpha,\sigma} (-1)^\sigma [U^*(\kk-\qq)]_{\alpha,m,\sigma} [U(\kk)]_{\alpha,n,\sigma} \gamma^\dagger_{\kk-\qq,m,\sigma} \gamma_{\kk,n,\sigma}.
\label{}
\end{align}
The commutator becomes
\begin{align}
  \null [\bar{S}^z_{\pp,\al},\eta^\dag] &= \sum_{\kk,m,n,\sigma} \sum_{\mbf{q},l} [(-1)^\sigma [U^*(\kk-\pp)]_{\alpha,m,\sigma} [U(\kk)]_{\alpha,n,\sigma} \gamma^\dagger_{\kk-\pp,m,\sigma} \gamma_{\kk,n,\sigma} ,\gamma^\dag_{\mbf{q},l,\u} \gamma^\dag_{-\mbf{q},l,\d} ] \\
   &=  \sum_{\kk,m,n,\sigma} \sum_{\mbf{q},l} [U^*(\kk-\pp)]_{\alpha,m,\sigma} [U(\kk)]_{\alpha,n,\sigma} \left( \gamma^\dagger_{\kk-\pp,m,\u} \delta_{\kk,\mbf{q}} \delta_{ln} \delta_{\sigma,\u} \gamma^\dag_{-\mbf{q},l,\d} - \gamma^\dagger_{\kk-\pp,m,\d} \delta_{\kk,-\mbf{q}} \delta_{ln} \delta_{\sigma,\d}  \gamma^\dagger_{\mbf{q},l,\u} \right) \\
&= \sum_{\kk,m,n,\sigma} [U^*(\kk-\pp)]_{\alpha,m,\sigma} [U(\kk)]_{\alpha,n,\sigma} \left(  \delta_{\sigma,\u} \gamma^\dagger_{\kk-\pp,m,\u} \gamma^\dag_{-\kk,n,\d} -  \delta_{\sigma,\d} \gamma^\dagger_{\kk-\pp,m,\d} \gamma^\dagger_{-\kk,n,\u} \right) \\
&= \sum_{\kk,m,n,\sigma} \left( [U^*(\kk-\pp)]_{\u,m,\sigma} [U(\kk)]_{\u,n,\sigma} - [U^*(-\kk)]_{\d,n,\sigma} [U(\pp-\kk)]_{\d,m,\sigma} \right) \gamma^\dagger_{\kk-\pp,m,\u} \gamma^\dag_{-\kk,n,\d} = 0.
\label{}
\end{align}

\subsection{Symmetry Algebra}
\label{app:symalg}

The eta-pairing states $\eta^{\dag n}\ket{0}$ are naturally interpreted as zero-momentum states of charge $+2$ Cooper pairs, with $\eta^\dag$ being their creation operator. Because the states $\eta^{\dag n}\ket{0}$ are zero energy for each $n$, we can obtain eigenstates of $H$ by taking superpositions with different numbers of particles. This is the simply the Cooper pair condensate that appears in the mean-field description. However, we work at fixed particle number in this work.

We now show that the eta-pairing operators form an $su(2)$ charge algebra which extends the usual $u(1)$ charge algebra. We compute
\bea
\label{eq:etadagcom}
\null [\eta, \eta^\dag] &= \sum_{\mbf{k}\mbf{q}\al\be} [\bar{c}_{-\mbf{k},\be,\d} \bar{c}_{\mbf{k},\be,\u} , \bar{c}^\dag_{\mbf{q},\al,\u} \bar{c}^\dag_{-\mbf{q},\al,\d}] \\
 &= \sum_{\mbf{k}\mbf{q}\al\be} \delta_{\mbf{k},\mbf{q}} \bar{c}_{-\mbf{k},\be,\d} P^\u_{\be \al}(\mbf{k})\bar{c}^\dag_{-\mbf{q},\al,\d} - \delta_{\mbf{k},\mbf{q}} \bar{c}^\dag_{\mbf{q},\al,\u}  P^\d_{\be \al}(-\mbf{k}) \bar{c}_{\mbf{k},\be,\u}\\
  &=  \sum_{\mbf{k}\al\be} \bar{c}_{\mbf{k},\be,\d} P^\u_{\be \al}(-\mbf{k})\bar{c}^\dag_{\mbf{k},\al,\d} - \bar{c}^\dag_{\mbf{k},\al,\u}  P^\d_{\be \al}(-\mbf{k}) \bar{c}_{\mbf{k},\be,\u}\\
    &= \sum_{\mbf{k}\al\be} P^\u_{\be \al}(-\mbf{k}) P^\d_{\be \al}(\mbf{k}) -\bar{c}^\dag_{\mbf{k},\al,\d} P^\d_{\al \be}(\mbf{k}) \bar{c}_{\mbf{k},\be,\d} - \bar{c}^\dag_{\mbf{k},\al,\u}  P^\u_{\al \be}(\mbf{k}) \bar{c}_{\mbf{k},\be,\u}\\
    &= \sum_{\mbf{k}\al\be} P^\u_{\be \al}(-\mbf{k}) P^\u_{\al \be}(-\mbf{k}) -\bar{c}^\dag_{\mbf{k},\al,\d} P^\d_{\al \be}(\mbf{k}) \bar{c}_{\mbf{k},\be,\d} - \bar{c}^\dag_{\mbf{k},\al,\u}  P^\u_{\al \be}(\mbf{k}) \bar{c}_{\mbf{k},\be,\u}\\
  &= \sum_\mbf{k} \Tr P^\u(-\mbf{k}) \ - \bar{N} \\
 &= \mathcal{N}N_f - \bar{N}
\eea
where we used that $P^\d_{\be \al}(\mbf{k})  = P^\u_{\al \be}(-\mbf{k})$ and $\sum_\be P_{\al \be}(\mbf{k}) \bar{c}_{\mbf{k},\be} = \bar{c}_{\mbf{k},\al}$. $\bar{N}$ is the total number operator so
\bea
\label{eq:etadagcomN}
[\bar{N}, \eta^\dag] = 2 \eta^\dag, \qquad [\bar{N}, \eta] = -2 \eta \ .
\eea
Alternatively, in the $\gamma$ band basis,
\begin{align}
[\eta, \eta^\dag] &= \sum_{\mbf{q},m} \sum_{\mbf{q}',m'} [\gamma_{-\mbf{q}',m',\d} \gamma_{\mbf{q}',m',\u}, \gamma^\dag_{\mbf{q},m,\u} \gamma^\dag_{-\mbf{q},m,\d}] \\
&= \sum_{\mbf{q},m} \sum_{\mbf{q}',m'} \gamma_{-\mbf{q}',m',\d} \{\gamma_{\mbf{q}',m',\u}, \gamma^\dag_{\mbf{q},m,\u}\} \gamma^\dag_{-\mbf{q},m,\d} - \gamma^\dag_{\mbf{q},m,\u} \gamma_{\mbf{q}',m',\u} \{\gamma^\dag_{-\mbf{q},m,\d},\gamma_{-\mbf{q}',m',\d}\} \\
&= \sum_{\mbf{q},m} \sum_{\mbf{q}',m'} \gamma_{-\mbf{q}',m',\d} \delta_{mm'} \delta_{\mbf{q}\mbf{q}'} \gamma^\dag_{-\mbf{q},m,\d} - \gamma^\dag_{\mbf{q},m,\u} \gamma_{\mbf{q}',m',\u} \delta_{mm'} \delta_{\mbf{q}\mbf{q}'} \\
&= \sum_{\mbf{q},m} \gamma_{-\mbf{q},m,\d} \gamma^\dag_{-\mbf{q},m,\d} - \gamma^\dag_{\mbf{q},m,\u} \gamma_{\mbf{q},m,\u} = {\cal N} N_f - {\bar N}.
\label{}
\end{align}

To make the $su(2)$ Lie algebra explicit, the correct normalizations are
\bea
\null [\eta^a,\eta^b] = i \eps_{abc} \eta^c, \qquad \eta^x = (\eta + \eta^\dag)/2, \quad  \eta^y = i (\eta-\eta^\dag)/2, \quad \eta^z = (\bar{N} - \mathcal{N} N_f)/2 \ . \\
\eea
Thus we have shown that the $\eta$ operators form a representation of $SU(2)$. It is simple to find the dimension of the representation. For a dimension $d$ irrep of $SU(2)$, $\eta^{\dag d} = 0$. Indeed, we know that $\eta^{\dag (\mathcal{N} N_f +1)} = 0$ because the projected Hilbert space only contains $2 \mathcal{N} N_f$ electrons, and $\eta^\dag$ adds two electrons to the state. It is possible to use the explicit representations of $SU(2)$ to obtain normalized states $\ket{n} \propto \eta^{\dag n}\ket{0}$, but we will use a more direct method involving a generating function in the next section.



We now briefly address the spin symmetries. In the case of $S_z$ and $\mathcal{T}$, the many-body $S_z$ operator
\bea
\bar{S}_z = \sum_{\mbf{k},n} \gamma^\dag_{\mbf{k},n,\u}\gamma_{\mbf{k},n,\u} - \gamma^\dag_{\mbf{k},n,\d}\gamma_{\mbf{k},n,\d}
\eea
commutes with $H$. To see this, we use $[\bar{S}_z, \gamma^\dag_{\mbf{k},n,\sigma}] = s^z_\sigma \gamma^\dag_{\mbf{k},n,\sigma}$ (where $s^z_\sigma = \pm 1$ for $\sigma = \u/\d$) to check that  $[\bar{S}_z, \bar{c}^\dag_{\mbf{R},\al, \sigma}] = s^z_\sigma  \bar{c}^\dag_{\mbf{R},\al, \sigma}$. Thus $[\bar{S}_z, \bar{S}_{\mbf{R},\al}^z] = 0$ (so $[\bar{S}_z, H] = 0$) and also $[\bar{S}_z, \eta^\dag] = 0$. Thus we have a $U(1) \times SU(2)$ symmetry group. We emphasize that $H$ breaks spin $SU(2)$, having only $U(1)$ from conservation of $S_z$. One sign of this, which we will see later, is that the spin $\pm1$ excitations are gapped. \\

\section{Off-diagonal Long Range Order in the Ground State and Generating Functions}
\label{app:generatingfunctions}

In this section, we show that the eta-pairing ground states have ODLRO and thus display superconductivity at zero temperature. This calculation is tractable using a generating function of the BCS form (\App{app:odlro}). In an effort to be self-contained, we then show that Wick's theorem holds in the ground state (\App{app:wick}) using a similar generating function with Grassman variables.  Our calculations resemble previous calculations involving eta-pairing, for example \Refs{PhysRevLett.63.2144}{2017ScPP....3...43V}. 

\subsection{ODLRO}
\label{app:odlro}

To assess the superconducting properties of the groundstate, we will compute the off-diagonals of the two-body density matrix:
\bea
\label{eq:odlroGS}
\lim_{|\mbf{R}- \mbf{R}'| \to \infty} \braket{GS| w^\dag_{\mbf{R}' m \d}w^\dag_{\mbf{R}' m \u} w_{\mbf{R}n \u} w_{\mbf{R} n \d} |GS}
\eea
in an eta-pairing groundstate $\ket{GS}$. Here $w^\dag_{\mbf{R},n,\sigma}$ is the Wannier function obtained from the canonical electron operators
\bea
\label{eq:Wannierspin}
w^\dag_{\mbf{R},n,\sigma} = \frac{1}{\sqrt{\mathcal{N}}} \sum_{\mbf{k}} e^{i \mbf{R} \cdot \mbf{k}} \gamma^\dag_{\mbf{k},n,\sigma} .
\eea
Such operators can be chosen to be exponentially decaying in obstructed atomic limit bands, and also in fragile bands if we allow their symmetries to not be represented locally \cite{2016PhRvB..93c5453W}. This is because the only obstruction to exponential localization is Chern number \cite{PhysRevLett.98.046402}. In the Wannier basis (choosing a time-reversal symmetric gauge for the $\gamma$ operators where $U^\u(\mbf{k})=U^\d(-\mbf{k})^*$), the $\eta$ operator reads
\bea
\eta^\dag &= \sum_{\mbf{q}m} \gamma^\dag_{\mbf{q},m,\u} \gamma^\dag_{-\mbf{q},m,\d} = \sum_{\mbf{R},\mbf{R}',m}  w^\dag_{\mbf{R},m,\u} w^\dag_{\mbf{R}',m,\d} \frac{1}{\mathcal{N}}\sum_{\mbf{q}} e^{-i (\mbf{R}-\mbf{R}') \cdot \mbf{q}} = \sum_{\mbf{R}}  w^\dag_{\mbf{R},m,\u} w^\dag_{\mbf{R},m,\d} \ . \\
\eea
To calculate the correlator in \Eq{eq:odlroGS}, we will employ a BCS-type ground state as a generating function:
\bea
\label{eq:wangen}
\ket{\{z\}} &= \exp \lp \sum_{\mbf{R},m} z_{\mbf{R},m} w^\dag_{\mbf{R},m,\u} w^\dag_{\mbf{R},m,\d} \rp \ket{0}  \\
\eea
with free parameters $z_{\mbf{R},m}$. First we observe that setting $z_{\mbf{R},m} = z$ gives
\bea
\label{eq:genorm}
\braket{\{z\}=z|\{z\}=z} &=  \braket{0|e^{\bar{z} \eta} e^{z \eta^\dag}|0} = \sum_n \frac{|z|^{2n}}{n!^2} \braket{0|\eta^n \eta^{\dag n}|0} \\
\eea
which we will use to determine the norms of the eta-pairing states. Second, we note the ODLRO correlators appear when taking derivatives:
\bea
\label{eq:ODLROcorr}
\left. \frac{\del}{\del \bar{z}_{\mbf{R},n}}\frac{\del}{\del z_{\mbf{R}',m}} \braket{\{z\}|\{z\}} \right|_{z_{\mbf{R},m} = z} &=  \braket{0|e^{\bar{z} \eta} w_{\mbf{R},n,\d}  w_{\mbf{R},n,\u} w^\dag_{\mbf{R}',m,\u} w^\dag_{\mbf{R}',m,\d} e^{z \eta^\dag}|0} \\
&= \sum_n \frac{|z|^{2n}}{n!^2} \braket{0|\eta^n w_{\mbf{R},n,\d}  w_{\mbf{R},n,\u} w^\dag_{\mbf{R}',m,\u} w^\dag_{\mbf{R}',m,\d} \eta^{\dag n}|0} \ . \\
\eea
We now must compute $\braket{\{z\}|\{z\}}$. We use the orthogonality of the Wannier basis, $\{w_{\mbf{R},m,\sigma},w^\dag_{\mbf{R}',n,\sigma'} \}= \delta_{\mbf{R}\mbf{R}'} \delta_{mn} \delta_{\sigma,\sigma'}$ to factor \Eq{eq:wangen} into commuting terms
\bea
\braket{\{z\}|\{z\}} &= \braket{0| \prod_{\mbf{R}m} e^{\bar{z}_{\mbf{R},m} w_{\mbf{R}, m,\d} w_{\mbf{R},m,\u} } \prod_{\mbf{R}'n} e^{z_{\mbf{R}',n} w^\dag_{\mbf{R}',n,\u} w^\dag_{\mbf{R}', n,\d}} |0} \\
&= \braket{0| \prod_{\mbf{R}m} (1+\bar{z}_{\mbf{R},m} w_{\mbf{R}, m,\d} w_{\mbf{R},m,\u}) \prod_{\mbf{R}'n} (1+ z_{\mbf{R}',n} w^\dag_{\mbf{R}',n,\u} w^\dag_{\mbf{R}', n,\d} ) |0} \\
\eea
where we used that fact that $(w^\dag_{\mbf{R}',n,\u} w^\dag_{\mbf{R}', n,\d})^2 = 0$. Commuting through the different factors and recalling $w^\dag_{\mbf{R},m}$ obey canonical commutation relations, we see that
\bea
\label{eq:zz}
\braket{\{z\}|\{z\}}  &= \braket{0| \prod_{\mbf{R}m} (1+\bar{z}_{\mbf{R},m} w_{\mbf{R},m,\d} w_{\mbf{R},m,\u} )  (1+ z_{\mbf{R},m}  w^\dag_{\mbf{R},m,\u} w^\dag_{\mbf{R},m,\d} ) |0} \\
&= \prod_{\mbf{R}m} (1+|z_{\mbf{R},m} |^2) \ .
\eea
Comparing to \Eq{eq:genorm} and using the binomial theorem, using \Eq{eq:genorm} we obtain
\bea
\braket{\{z\}|\{z\}}|_{\{z\}=z} &= (1+|z|^2)^{N_f \mathcal{N}} = \prod_{\mbf{R}m} (1+|z|^2) = (1+|z|^2)^{N_f\mathcal{N}} \\\sum_n \frac{|z|^{2n}}{n!^2} \braket{0|\eta^n \eta^{\dag n}|0} &= \sum_{n} \binom{N_f \mathcal{N}}{n} |z|^{2n}
\eea
so equating coefficients, the normed eta-pairing states are
\bea
\label{eq:normedetastates}
\ket{n} &= \frac{1}{n!} \binom{N_f\mathcal{N}}{n}^{-\frac{1}{2}} \eta^{\dag n}\ket{0} \ .
\eea
Then to obtain the ODLRO correlators, we observe that
\bea
\left. \frac{\del}{\del \bar{z}_{\mbf{R},n}}\frac{\del}{\del z_{\mbf{R}',m}} \braket{\{z\}|\{z\}} \right|_{\{z\} = z} &= |z|^2 (1+|z|^2)^{N_f\mathcal{N}-2} = \sum_{n} \binom{N_f \mathcal{N}-2}{n} |z|^{2n+2}, \qquad \mbf{R},n \neq \mbf{R}',m \ .\\
\eea
Comparing to \Eq{eq:ODLROcorr} and using, we obtain
\bea
\binom{N_f\mathcal{N}}{n} \braket{0|\eta^n w_{\mbf{R},n,\d}  w_{\mbf{R},n,\u} w^\dag_{\mbf{R}',m,\u} w^\dag_{\mbf{R}',m,\d} \eta^{\dag n}|0}  &=  \binom{N_f \mathcal{N}-2}{n-1} \ .
\eea
For $\mbf{R},n \neq \mbf{R}',m$, we commute $w_{\mbf{R},n,\d}  w_{\mbf{R},n,\u}$ through $ w^\dag_{\mbf{R}',m,\u} w^\dag_{\mbf{R}',m,\d} $ and obtain
\bea
\label{eq:ODLROexp}
 \braket{n| w^\dag_{\mbf{R} m \d}w^\dag_{\mbf{R} m \u} w_{\mbf{R}' n \u} w_{\mbf{R} n' \d} |n} &=  \frac{(N_f \mathcal{N}-2)!}{(N_f\mathcal{N}-n-1)!(n-1)!}  \frac{(N_f\mathcal{N}-n)!n!}{(N_f \mathcal{N})!} \\
&=  \frac{n(N_f\mathcal{N}-n)}{N_f \mathcal{N}(N_f \mathcal{N}-1)} \\
&=  \nu (1-\nu) + O(1/\mathcal{N}) \\
\eea
where we defined the filling $\nu = n/N_f\mathcal{N} \in (0,1)$. The other notable feature of this calculation is that for $\mbf{R} \neq \mbf{R}'$, there is no dependence on $|\mbf{R}-\mbf{R}'|$ in the correlator. This is reminiscent of the ODRLO in a $T=0$ Bose-Einstein condensate, which also has no $|\mbf{R}-\mbf{R}'|$ dependence and is equal to the boson density. Indeed, comparing to mean field calculations of the superfluid weight\cite{2022arXiv220311133H}, $\nu (1-\nu)$ is also obtained as the Cooper pair density.

For exponentially decaying Wannier states, \Eq{eq:ODLROexp} is a good local measure of off-diagonal long range order. The ODLRO has the interpretation of correlators between two particles at position $\RR$ and another two at position $\RR'$:  should the Wannier states fail to be localized \Eq{eq:odlroGS} loses this interpretation.  We thus must address the question: are the $w$ degrees of freedom employed in the ODLRO correlation exponentially localized?  In the uniform pairing, where bands are constructed from the $S$-matrix method, the bands are fragile or obstructed atomic and thus exponentially localized Wannier states can be found \cite{PhysRevLett.98.046402}, but the symmetries cannot be represented locally on them if there is fragile topology. We remark that it is possible to obtain perfectly flat bands with a nonzero Chern number $C_\sigma$ with uniform pairing from a single Wyckoff position sublattice, but this requires exponentially decaying (infinite range) hoppings. The total Chern number $C_\u + C_\d$ will vanish due to spin-ful time reversal symmetry, and thus it is possible to find exponentially localized Wannier states by mixing the bands of different spin. However, the calculation in this section relies on the spin-polarized Wannier states \Eq{eq:Wannierspin}, and thus the optimal decay of the Wannier states, for nonzero $C_\sigma$, is a power law \cite{2016arXiv161209557M}. An adaptation of \Eq{eq:ODLROexp} in terms of exponentially localized states when $C_\sigma \neq 0$ is left for future work.

\subsection{Wick's theorem}
\label{app:wick}

We now extend the generating function method of \App{app:odlro} to show that a version of Wick's theorem holds in the eta -pairing states. In this section we work in the momentum basis using the $\gamma^\dag_{\mbf{k},m,\sigma}$ operators. Wick's theorem shows that arbitrary correlators can be determined by enumerating the contractions, which allows us to compute expressions for the norms of the excitations in App.~\ref{app:excit} and calculate expectation values.

The generating function we introduce is the norm $N(z,\xi)$ of the following state:
\bea
\ket{\xi,z} = \exp \lp \sum_{\mbf{k}n\sigma} \xi_{\mbf{k},n,\sigma} \gamma^\dag_{\mbf{k}, n,\sigma}  \rp e^{z \eta^\dag} \ket{0}, \qquad N(z,\xi) = \braket{\xi,z|\xi,z} \\
\eea
where $\xi_{\mbf{k},n,\sigma}$ are anti-commuting (fermionic) Grassman variables, and bar denotes the complex conjugate ${\bar \xi} = \xi^\dagger$. Taking $\xi$ derivatives of $N(\xi,z)$ and then setting $\xi = 0$ pulls down $\gamma^\dag$ operators which build the correlation function. We observe that
\bea
\label{eq:Wicks}
\gamma^\dag_{\mbf{k}, n,\sigma} \ket{\xi,z} &= \del_{\xi_{\mbf{k},n,\sigma}} \exp \lp \sum_{\mbf{k}'n'\sigma'} \xi_{\mbf{k}',n',\sigma'} \gamma^\dag_{\mbf{k}', n',\sigma'}  \rp e^{z \eta^\dag} \ket{0} =  \del_{\xi_{\mbf{k},n,\sigma}} \ket{\xi,z} \\
\bra{\xi,z} \gamma_{\mbf{k}, n,\sigma} &= -\del_{\bar{\xi}_{\mbf{k},n,\sigma}} \bra{0} e^{\bar{z} \eta} \exp \lp \sum_{\mbf{k}'n'\sigma'} \gamma_{\mbf{k}', n',\sigma'}  \bar{\xi}_{\mbf{k}',n',\sigma'}   \rp  = -\del_{\bar{\xi}_{\mbf{k},n,\sigma}}  \bra{\xi,z} \ .
\eea
We see for instance that
\bea
\label{eq:deldelxixix}
\bra{\xi,z} \gamma_{\mbf{k}', n',\sigma'} \gamma^\dag_{\mbf{k}, n,\sigma} \ket{\xi,z} &=  -\del_{\bar{\xi}_{\mbf{k}',n',\sigma'}}  \bra{\xi,z}  \, \del_{\xi_{\mbf{k},n,\sigma}} \ket{\xi,z} =  \del_{\xi_{\mbf{k},n,\sigma}} \del_{\bar{\xi}_{\mbf{k}',n',\sigma'}}  N(z,\xi) \\
\bra{z} \gamma_{\mbf{k}', n',\sigma'} \gamma^\dag_{\mbf{k}, n,\sigma} \ket{z} &= \left. \del_{\xi_{\mbf{k},n,\sigma}} \del_{\bar{\xi}_{\mbf{k}',n',\sigma'}}  N(z,\xi) \right|_{\xi = 0}\\
\eea
and thus by matching powers of $|z|^2$ in $\del_{\xi_{\mbf{k}m \sigma}} \del_{\bar{\xi}_{\mbf{k}m \sigma}} N(z, \xi)$ and $ \bra{z}\gamma_{\mbf{k}, m,\sigma} \gamma^\dag_{\mbf{k} m \sigma} \ket{z}$, the correlators $ \bra{n}\gamma_{\mbf{k}, m,\sigma} \gamma^\dag_{\mbf{k} m \sigma} \ket{n}$ can be determined. Generally, any correlator with creation operators to the right and annihilation operators to the left can be reduced by repeated application of \Eq{eq:Wicks} to derivatives of $\braket{\xi,z|\xi,z} = N(\xi,z)$ (and then setting $\xi \to 0$). This defines a normal ordering.

We can compute $N(z, \xi)$ directly. Because $\xi$ and $\gamma^\dag$ are all anti-commuting, we have
\bea
\ket{\xi,z} &= \exp \lp \sum_{\mbf{k}n \sigma} \xi_{\mbf{k},n,\sigma} \gamma^\dag_{\mbf{k},n,\sigma} \rp e^{z \eta^\dag} \ket{0} \\
&= \exp \lp \sum_{\mbf{k}n} \xi_{\mbf{k},n,\u} \gamma^\dag_{\mbf{k}, n,\u} + \xi_{-\mbf{k},n,\d} \gamma^\dag_{-\mbf{k}, n,\d} + z \gamma^\dag_{\mbf{k},n,\u} \gamma^\dag_{-\mbf{k},n,\d} \rp \ket{0} \\
&= \prod_{\mbf{k},n} \exp \lp \xi_{\mbf{k},n,\u} \gamma^\dag_{\mbf{k},n,\u}+\xi_{-\mbf{k},n,\d} \gamma^\dag_{-\mbf{k},n,\d} + z \gamma^\dag_{\mbf{k},n,\u} \gamma^\dag_{-\mbf{k},n,\d} \rp \ket{0} \\
&= \prod_{\mbf{k},n}  \big( 1 + \xi_{\mbf{k},n,\u} \gamma^\dag_{\mbf{k}, n,\u}+ \xi_{-\mbf{k},n,\d} \gamma^\dag_{-\mbf{k}, n,\d} + (z+\xi_{-\mbf{k},n,\d} \xi_{\mbf{k},n,\u}) \gamma^\dag_{\mbf{k},n,\u} \gamma^\dag_{-\mbf{k},n,\d}  \big) \ket{0} \\
\eea
using $\xi^2 = \gamma^{\dag 2} = 0$ and $[\xi \gamma^\dagger, \eta^\dagger]=0$.  The $\xi$ operators must be Grassmanns, else $[\xi_{\kk,n,\uparrow} \gamma^\dagger_{\kk,n,\uparrow}, \xi_{\kk',n',\uparrow} \gamma^\dagger_{\kk',n',\uparrow}] \neq 0$.  The only term at second order is the cross term from $\frac{1}{2!}(\xi_{-\mbf{k},n,\d} \gamma^\dag_{-\mbf{k}, n,\d}+\gamma^\dag_{\mbf{k}, n,\u} \xi_{\mbf{k},n,\u})^2$. Hence we find
\bea
\label{eq:Nzxiprod}
N(z, \xi) = \braket{\xi,z|\xi,z} &= \prod_{\mbf{k},n} \lp 1+ \bar{\xi}_{\mbf{k},n,\u}  \xi_{\mbf{k},n,\u}  + \bar{\xi}_{-\mbf{k},n,\d}  \xi_{-\mbf{k},n,\d}  +  (\bar{z}+\bar{\xi}_{\mbf{k},n,\u} \bar{\xi}_{-\mbf{k},n,\d} )  (z+\xi_{-\mbf{k},n,\d} \xi_{\mbf{k},n,\u}) \rp  \\
\eea
which importantly is decoupled in $\mbf{k}$ and $n$. Let us expand out a single factor of \Eq{eq:Nzxiprod}:
\bea
1+ |z|^2  + \bar{\xi}_{\mbf{k},n,\u}  \xi_{\mbf{k},n,\u}  + \bar{\xi}_{-\mbf{k},n,\d}  \xi_{-\mbf{k},n,\d}  + z \bar{\xi}_{\mbf{k},n,\u} \bar{\xi}_{-\mbf{k},n,\d} +   \bar{z} \xi_{-\mbf{k},n,\d} \xi_{\mbf{k},n,\u} +
\bar{\xi}_{\mbf{k},n,\u} \bar{\xi}_{-\mbf{k},n,\d} \xi_{-\mbf{k},n,\d} \xi_{\mbf{k},n,\u}
\eea
which contains only quadratic and quartic terms in $\xi$. Because we take $\xi$ derivatives and then set $\xi \to 0$, we see that only two- and four- term derivatives (for each $\mbf{k},m$) are nonzero. From \Eq{eq:Wicks}, we compute
\bea
\label{eq:cor1}
\braket{z|\gamma_{\mbf{k}'m'\sigma'}\gamma^\dag_{\mbf{k}m\sigma} |z} = \del_{\xi_{\mbf{k},m,\sigma}} \del_{\bar{\xi}_{\mbf{k}',m',\sigma'}}  N  |_{\xi=0}&= \frac{\delta_{\mbf{k}\mbf{k}'}\delta_{mm'} \delta_{\sigma \sigma'}}{1+|z|^2} \braket{z|z} \\
\braket{z|\gamma_{\mbf{k}m\d}\gamma_{\mbf{k}'m'\u} |z} =  \del_{\bar{\xi}_{\mbf{k},m,\d}}  \del_{\bar{\xi}_{\mbf{k}',m',\u} }N  |_{\xi=0}&= \frac{z \delta_{\mbf{k},-\mbf{k}'} \delta_{mm'} }{1+|z|^2} \braket{z|z} \\
\braket{z|\gamma^\dag_{\mbf{k}m\u} \gamma^\dag_{\mbf{k}'m'\d} |z}  =\del_{\xi_{\mbf{k},m,\u} }\del_{\xi_{\mbf{k}',m',\d}}  N  |_{\xi=0}&= \frac{\bar{z} \delta_{\mbf{k},-\mbf{k}'} \delta_{mm'}}{1+|z|^2} \braket{z|z} \\
\eea
with all other two derivative terms being zero, and $\braket{z|z} = (1+|z|^2)^{N_f \mathcal{N}}$. We observe in the second and third line a nonzero anomalous BCS-type contraction between $\gamma \gamma$ or $\gamma^\dag \gamma^\dag$. The last nonzero term comes from the four-body derivative:
\bea
\label{eq:cor2}
\braket{z|\gamma_{-\mbf{k}m\d}\gamma_{\mbf{k}m\u} \gamma^\dag_{\mbf{k}m\u} \gamma^\dag_{-\mbf{k}m\d}  |z} =  \del_{\xi_{\mbf{k},m,\u}} \del_{\xi_{-\mbf{k},m,\d}}  \del_{\bar{\xi}_{-\mbf{k},m,\d}} \del_{\bar{\xi}_{\mbf{k},m,\u}}  N  |_{\xi=0}&= \frac{1}{1+|z|^2} \braket{z|z} \ . \\
\eea

So far we have given a recipe to compute all normal ordered correlators. Using the canonical anti-commutation relations of $\gamma$, all correlators can be manipulated into normal order and then evaluated. However, it is of practical interest to evaluate correlators directly in any order. We now calculate the one-body contractions, \Eqs{eq:cor1}{eq:cor2}, in which a $\gamma^\dag \gamma$ contraction is introduced to be consistent with the fermion anti-commutation relations. This is a standard result (see \Ref{1971qtmp.book.....F}) but we include a proof here to set the notation and be self-contained.  All many-body contractions will be built from these one-body contractions.

We compute the non-normal ordered correlator
\bea
\label{eq:cor3}
\braket{z|\gamma^\dag_{\mbf{k}m\sigma} \gamma_{\mbf{k}'m'\sigma'} |z} &= -\braket{z|\gamma_{\mbf{k}'m'\sigma'} \gamma^\dag_{\mbf{k}m\sigma}  |z} + \delta_{\mbf{k}\mbf{k}'} \delta_{mm'} \delta_{\sigma \sigma'}\braket{z|z} \\
&= \delta_{\mbf{k}\mbf{k}'} \delta_{mm'} \delta_{\sigma \sigma'} \braket{z|z} \lp - \frac{1}{1+|z|^2} + 1 \rp \\
&=  \delta_{\mbf{k}\mbf{k}'} \delta_{mm'} \delta_{\sigma \sigma'} \frac{|z|^2}{1+|z|^2} \braket{z|z} \ . \\
\eea
This correlator will define a two-particle contraction to be included in Wick's theorem:  correlators are equal to the signed sum of all possible two-particle contractions defined in \Eqs{eq:cor1}{eq:cor3}. Because only two- and four-particle contractions are nonzero and the two-particle contractions satisfy Wick's theorem automatically, we only need to check that \Eq{eq:cor2} is consistent with this Wick's theorem. Consider the only non-zero normal ordered four-particle correlator $\braket{z|\gamma_{-\mbf{k}m\d}\gamma_{\mbf{k}m\u} \gamma^\dag_{\mbf{k}m\u} \gamma^\dag_{-\mbf{k}m\d}  |z}$. There are three possible contractions from Wick's theorem (one vanishing):
\bea
\dfrac{\braket{z|\gamma_{-\mbf{k}m\d}\gamma_{\mbf{k}m\u} \gamma^\dag_{\mbf{k}m\u} \gamma^\dag_{-\mbf{k}m\d}  |z}}{\braket{z|z}} &= \dfrac{\braket{z|\gamma_{-\mbf{k}m\d}\gamma_{\mbf{k}m\u}|z}}{\braket{z|z}}  \dfrac{\braket{z| \gamma^\dag_{\mbf{k}m\u} \gamma^\dag_{-\mbf{k}m\d}|z}}{\braket{z|z}}  - \dfrac{\braket{z|\gamma_{-\mbf{k}m\d} \gamma^\dag_{\mbf{k}m\u}|z}}{\braket{z|z}}  \dfrac{\braket{z|\gamma_{\mbf{k}m\u} \gamma^\dag_{-\mbf{k}m\d}|z}}{\braket{z|z}}  \nonumber \\
&+ \dfrac{\braket{z|\gamma_{-\mbf{k}m\d} \gamma^\dag_{-\mbf{k}m\d}|z}}{\braket{z|z}}  \dfrac{\braket{z|\gamma_{\mbf{k}m\u} \gamma^\dag_{\mbf{k}m\u}|z}}{\braket{z|z}}  \nonumber \\
&= \lp \frac{z}{1+|z|^2} \rp \lp  \frac{\bar{z}}{1+|z|^2}\rp -  (0) (0) + \lp \frac{1}{1+|z|^2} \rp \lp \frac{1}{1+|z|^2} \rp\\
 &=  \frac{|z|^2+1}{(1+|z|^2)^2} =  \frac{1}{1+|z|^2}
\eea
matching \Eq{eq:cor2}.

For clarity, we list the nonzero elementary contractions of Wick's theorem:
\bea
\label{eq:wickcontractions}
\frac{1}{\braket{z|z}} \braket{z|\gamma_{\mbf{k}'m'\sigma'}\gamma^\dag_{\mbf{k}m\sigma} |z} &= \frac{1}{1+|z|^2} \delta_{\mbf{k}\mbf{k}'}\delta_{mm'} \delta_{\sigma \sigma'} \\
\frac{1}{\braket{z|z}}\braket{z|\gamma^\dag_{\mbf{k}m\sigma} \gamma_{\mbf{k}'m'\sigma'} |z}  &= \frac{|z|^2}{1+|z|^2} \delta_{\mbf{k}\mbf{k}'}\delta_{mm'} \delta_{\sigma \sigma'} \\
\frac{1}{\braket{z|z}} \braket{z|\gamma_{\mbf{k}m\sigma}\gamma_{\mbf{k}'m'\sigma'} |z} &= \frac{z}{1+|z|^2} \delta_{\mbf{k},-\mbf{k}'}\delta_{mm'} s_z^{\sigma'} \delta_{\sigma,-\sigma'} \\
\frac{1}{\braket{z|z}} \braket{z|\gamma^\dag_{\mbf{k}m\sigma} \gamma^\dag_{\mbf{k}'m'\sigma'} |z} &= \frac{\bar{z}}{1+|z|^2} \delta_{\mbf{k},-\mbf{k}'}\delta_{mm'} s_z^{\sigma} \delta_{\sigma,-\sigma'}  \\
\eea
and $\braket{z|z} = (1+|z|^2)^{N_f \mathcal{N}}$, $s_z^{\u/\d} = \pm 1$. We will use these results in \App{app:Richcrit} to evaluate Richardson's criterion.

\section{Charge excitations}
\label{app:excit}

In this section, we compute the spectrum and eigenstates of excitations above the eta-pairing ground states. We show that the fermionic excitations are gapped and perfectly flat (\App{app:charge1}), while the bosonic excitations are gapless and dispersive. We evaluate Richardson's criterion to show that kinetic energy generically favors superconductivity at half filling (\App{app:Richcrit}). We then study the bosonic (charge +2 and 0) excitations and find that the spin-0 Cooper pair spectrum splits into a set of low-lying bands described exactly by a single-particle pairing Hamiltonian and a flat continuum at the two-electron gap (\App{app:charge2}). The spin $\pm 1$ Cooper pair spectrum is entirely flat: all states are at the two-electron gap and are effectively unpaired (\App{app:charge2}). We provide explicit examples of the pairing Hamiltonian (\App{app:examples}) for obstructed atomic and fragile bands. Finally, we show that the spin-0 density mode (Goldstone mode) has the same spectrum as the spin-0 Cooper pairs. The spin $\pm1$ density excitations are all energetically above the two-electron gap, but have nontrivial dispersion also described by a single-particle Hamiltonian (\App{app:goldstones}).

\subsection{Charge +1}
\label{app:charge1}

Given an eta-pairing pair groundstate $\ket{GS}=\ket{n}$, we can compute the charge $+1$, spin $\pm \frac{1}{2}$ excitations from the effective Hamiltonian $R(\mbf{k})$ defined
\bea
\label{eq:charge1excit}
\null [H, \gamma^\dag_{\mbf{k},n,\sigma}] \ket{GS} &=  \sum_m \gamma^\dag_{\mbf{k},m,\sigma} R^\sigma_{mn}(\mbf{k})  \ket{GS} \ .
\eea
By diagonalizing $R^\sigma(\mbf{k})$, we obtain exact eigenstates and their energies. Note that $\sigma$ is not summed over. The excitation matrix is diagonal in spin because of $S_z$ conservation.

It is simplest to perform the excitation calculations in momentum space using the canonical $\gamma^\dag_{\mbf{k},n,\sigma}$ operators, which form a complete basis of the projected Hilbert space. First we define momentum space spin operators $\bar{S}^z_{\mbf{R},\al} = \sum_\sigma s^z_\sigma \bar{n}_{\mbf{R},\al,\sigma}$ (recall $s^z_\sigma = \pm1$ for $\sigma = \u,\d$) according to
\bea
\bar{S}^z_{\mbf{k},\al} &= \frac{1}{\sqrt{\mathcal{N}}} \sum_{\mbf{R}} e^{-i \mbf{k} \cdot (\mbf{R}+\mbf{r}_\al)} \bar{S}^z_{\mbf{R},\al}, \qquad \bar{S}^z_{\mbf{k},\al} = (\bar{S}^z_{-\mbf{k},\al})^\dag, \qquad \bar{S}^z_{\mbf{R},\al} = \frac{1}{\sqrt{\mathcal{N}}} \sum_\mbf{k} e^{i \mbf{k} \cdot (\mbf{R}+\mbf{r}_\al)}  \bar{S}^z_{\mbf{k},\al} \\
\eea
such that
\bea
H &= \frac{1}{2}|U| \sum_{\mbf{R},\al} (\bar{S}^z_{\mbf{R},\al})^2 = \frac{1}{2}|U| \sum_\mbf{k, \al} \bar{S}^z_{-\mbf{k},\al} \bar{S}^z_{\mbf{k},\al} \ . \\
\eea
$H$ can be written in terms of the $\gamma$ modes using
\bea
\bar{n}_{\mbf{R},\al,\sigma} &= \bar{c}^\dag_{\mbf{R},\al,\sigma}\bar{c}_{\mbf{R},\al,\sigma} = \frac{1}{\mathcal{N}} \sum_{\mbf{k},\mbf{k}'} e^{- i (\mbf{k} - \mbf{k}') \cdot (\mbf{R}+\mbf{r}_\al)} \bar{c}^\dag_{\mbf{k}',\al,\sigma}  \bar{c}_{\mbf{k},\al,\sigma} \\
&= \frac{1}{\mathcal{N}} \sum_{\mbf{k},\mbf{k}'mn} e^{- i (\mbf{k} - \mbf{k}') \cdot (\mbf{R}+\mbf{r}_\al)} \gamma^\dag_{\mbf{k}',m,\sigma} [U^\sigma_{\al m}(\mbf{k}')^*  U_{\al n}^\sigma(\mbf{k})]\gamma_{\mbf{k},n,\sigma} \\
&\equiv \frac{1}{\mathcal{N}} \sum_{\mbf{k},\mbf{k}',mn} e^{- i (\mbf{k} - \mbf{k}') \cdot (\mbf{R}+\mbf{r}_\al)} \gamma^\dag_{\mbf{k}',m,\sigma} M^{mn}_{\sigma,\al}(\mbf{k},\mbf{k}'-\mbf{k}) \gamma_{\mbf{k},n,\sigma} \\
\eea
where we defined the sublattice form factor
\bea
M^{mn}_{\sigma,\al}(\mbf{k},\mbf{q}) &= U^\sigma_{\al m}(\mbf{k}+\mbf{q})^*  U_{\al n}^\sigma(\mbf{k}) \ .
\eea
Thus the Fourier transform of the spin operator is
\bea
\bar{S}^z_{\mbf{q},\al} &= \frac{1}{\sqrt{\mathcal{N}}} \sum_{\mbf{R}} e^{-i \mbf{q} \cdot (\mbf{R}+\mbf{r}_\al)} \bar{S}^z_{\mbf{R},\al} \\
&= \frac{1}{\sqrt{\mathcal{N}}} \sum_{\mbf{R}mn} e^{-i \mbf{q} \cdot (\mbf{R}+\mbf{r}_\al)}  \frac{1}{\mathcal{N}} \sum_{\mbf{k},\mbf{k}',\sigma} e^{- i (\mbf{k} - \mbf{k}') \cdot (\mbf{R}+\mbf{r}_\al)} s^z_\sigma \gamma^\dag_{\mbf{k}',m,\sigma} M^{mn}_{\sigma,\al}(\mbf{k},\mbf{k}'-\mbf{k})\gamma_{\mbf{k},n,\sigma} \\
&= \frac{1}{\sqrt{\mathcal{N}}} \sum_{\mbf{k},\mbf{k}',mn,\sigma}  \frac{1}{\mathcal{N}} \sum_{\mbf{R}}  e^{- i (\mbf{k} +\mbf{q} - \mbf{k}') \cdot (\mbf{R}+\mbf{r}_\al)} s^z_\sigma \gamma^\dag_{\mbf{k}',m,\sigma} M^{mn}_{\sigma,\al}(\mbf{k},\mbf{k}'-\mbf{k})\gamma_{\mbf{k},n,\sigma} \\
&= \frac{1}{\sqrt{\mathcal{N}}} \sum_{\mbf{k},mn,\sigma} s^z_\sigma \gamma^\dag_{\mbf{q}+\mbf{k},m,\sigma} M^{mn}_{\sigma,\al}(\mbf{k},\mbf{q})\gamma_{\mbf{k},n,\sigma} \\
\eea
and hence it follows that
\bea
\label{eq:formfactorS}
\null [\bar{S}^z_{\mbf{q},\al}, \gamma^\dag_{\mbf{k}, n, \sigma}] &=  \frac{1}{\sqrt{\mathcal{N}}} \sum_m s^z_\sigma \gamma^\dag_{\mbf{k}+\mbf{q},m,\sigma} M^{mn}_{\sigma,\al}(\mbf{k},\mbf{q}) \ .
\eea
To prepare to compute the excitation spectrum, we observe
\bea
\label{eq:SSgamma}
\sum_{\mbf{q}\al} [\bar{S}^z_{-\mbf{q},\al} \bar{S}^z_{\mbf{q},\al}, \gamma^\dag_{\mbf{k}, n, \sigma}] &= \sum_{\mbf{q}\al}   \bar{S}^z_{-\mbf{q},\al}  [ \bar{S}^z_{\mbf{q},\al}, \gamma^\dag_{\mbf{k}, n, \sigma}] +[\bar{S}^z_{-\mbf{q},\al}, \gamma^\dag_{\mbf{k}, n, \sigma}]  \bar{S}^z_{\mbf{q},\al} \\
&= \sum_{\mbf{q}\al}   \bar{S}^z_{-\mbf{q},\al}  [ \bar{S}^z_{\mbf{q},\al}, \gamma^\dag_{\mbf{k}, n, \sigma}] +[\bar{S}^z_{\mbf{q},\al}, \gamma^\dag_{\mbf{k}, n, \sigma}]  \bar{S}^z_{-\mbf{q},\al} \\
&= \sum_{\mbf{q}\al}   [\bar{S}^z_{-\mbf{q},\al} , [ \bar{S}^z_{\mbf{q},\al}, \gamma^\dag_{\mbf{k}, n, \sigma}]] +2 [\bar{S}^z_{\mbf{q},\al}, \gamma^\dag_{\mbf{k}, n, \sigma}]  \bar{S}^z_{-\mbf{q},\al} \\
\eea
Acting on the groundstate where the second term is zero (recall that $S^z_{\mbf{q},\al}$ annihilates $\ket{n}$ because $S^z_{\mbf{q},\al} \ket{0} = 0$ and $[S^z_{\mbf{q},\al},\eta^\dag]=0$) and using \Eq{eq:formfactorS}, we derive
\bea
\sum_{\mbf{q}\al} [\bar{S}^z_{-\mbf{q},\al} \bar{S}^z_{\mbf{q},\al}, \gamma^\dag_{\mbf{k}, n, \sigma}] \ket{GS} &=  \frac{1}{\sqrt{\mathcal{N}}} s^z_\sigma \sum_{\mbf{q}\al m'} [\bar{S}^z_{-\mbf{q},\al} , \gamma^\dag_{\mbf{k}+\mbf{q},m',\sigma}] M^{m'n}_{\sigma,\al}(\mbf{k},\mbf{q}) \ket{GS} \\
&=  \frac{1}{\mathcal{N}}\sum_{\mbf{q}\al m m'}  \gamma^\dag_{\mbf{k},m,\sigma}  M^{mm'}_{\sigma,\al}(\mbf{k}+\mbf{q},-\mbf{q}) M^{m'n}_{\sigma,\al}(\mbf{k},\mbf{q}) \ket{GS} \\
\eea
where we used that $(s^z_\sigma)^2 = 1$. Thus we obtain
\bea
\label{eq:Rsteps}
R^\sigma_{mn}(\mbf{k}) &=  \frac{|U|}{2} \frac{1}{\mathcal{N}} \sum_{\mbf{q} \al m'} M^{mm'}_{\sigma,\al}(\mbf{k}+\mbf{q},-\mbf{q}) M^{m'n}_{\sigma,\al}(\mbf{k},\mbf{q}) \\
&=  \frac{|U|}{2} \frac{1}{\mathcal{N}} \sum_{\mbf{q} \al m'} U^\sigma_{\al m}(\mbf{k})^*  U_{\al m'}^\sigma(\mbf{k}+\mbf{q})  U^\sigma_{\al m'}(\mbf{k}+\mbf{q})^*  U_{\al n}^\sigma(\mbf{k})  \\
&=  \frac{|U|}{2}  \sum_{\al} U^\sigma_{\al m}(\mbf{k})^* \frac{1}{\mathcal{N}} \sum_\mbf{q} P^\sigma_{\al \al}(\mbf{k}+\mbf{q})  U_{\al n}^\sigma(\mbf{k})  \\
\eea
and using the uniform pairing condition \Eq{eq:UPCint}, $\frac{1}{\mathcal{N}} \sum_\mbf{q} P^\sigma_{\al \al}(\mbf{k}+\mbf{q}) = \epsilon$, we have
\bea
\label{eq:Rmn}
R^\sigma_{mn}(\mbf{k})  &=  \frac{|U|}{2}  \sum_{\al} U^\sigma_{\al m}(\mbf{k})^* \eps U_{\al n}^\sigma(\mbf{k}) \\
&=  \frac{\eps |U|}{2}  [U_\sigma^\dag(\mbf{k}) U_\sigma(\mbf{k})]_{mn} \\
&=  \frac{\eps |U|}{2} \delta_{mn} \\
\eea
which is diagonal for all $\mbf{k}$ and has completely flat bands with an energy gap $\eps|U|/2$ above the ground state. \Eq{eq:charge1excit} simply reads
\bea
\label{eq:estateeq}
\null H \gamma^\dag_{\mbf{k},n,\sigma} \ket{GS} &=  \frac{\eps|U|}{2} \gamma^\dag_{\mbf{k},n,\sigma} \ket{GS}
\eea
making use of $H\ket{GS} = 0$. We refer to $\eps|U|/2$ as the single-electron gap, whose analogue in BCS theory is the Bogoliubov quasi-particle gap.

We can repeat this calculation for the hole excitations, defined by
\bea
\null [H, \gamma_{\mbf{k},n,\sigma}] \ket{GS} &=  \sum_m \tilde{R}^\sigma_{nm}(\mbf{k}) \gamma_{\mbf{k},m,\sigma}   \ket{GS} \ .
\eea
Taking the hermitian conjugate of \Eq{eq:formfactorS}, we obtain the requisite formula
\bea
\null [\bar{S}^z_{\mbf{q},\al}, \gamma_{\mbf{k}, n, \sigma}] &=  -\frac{1}{\sqrt{\mathcal{N}}} \sum_m s^z_\sigma \gamma_{\mbf{k}-\mbf{q},m,\sigma} M^{mn}_{\sigma,\al}(\mbf{k},\mbf{q})^* \\
&=  -\frac{1}{\sqrt{\mathcal{N}}} \sum_m M^{nm}_{\sigma,\al}(\mbf{k}+\mbf{q},-\mbf{q}) s^z_\sigma \gamma_{\mbf{k}-\mbf{q},m,\sigma} \ . \\
\eea
All steps in the calculation of $R^\sigma_{mn}(\mbf{k})$ proceed identically for $\tilde{R}^\sigma_{nm}(\mbf{k})$ but with the mapping $M^{mn}_{\sigma,\al}(\mbf{k},\mbf{q}) \to M^{nm}_{\sigma,\al}(\mbf{k}+\mbf{q},-\mbf{q})$, so from \Eq{eq:Rsteps} we find
\bea
\tilde{R}^\sigma_{nm}(\mbf{k}) &= \frac{|U|}{2} \frac{1}{\mathcal{N}} \sum_{\mbf{q} \al m'} M^{m'm}_{\sigma,\al}(\mbf{k},\mbf{q}) M^{nm'}_{\sigma,\al}(\mbf{k}+\mbf{q},-\mbf{q})  \\
&= \frac{|U|}{2} \frac{1}{\mathcal{N}} \sum_{\mbf{q} \al m'} M^{nm'}_{\sigma,\al}(\mbf{k}+\mbf{q},-\mbf{q}) M^{m'm}_{\sigma,\al}(\mbf{k},\mbf{q}) \\
&= R^\sigma_{nm}(\mbf{k}) \\
\eea
and hence single particle and hole excitations have the same energies.  This is not true for systems that do not satisfy the uniform pairing condition.  For example, the flat band systems in magic angle twisted bilayer graphene have different particle and hole excitations \cite{2021PhRvB.103t5415B}.

\subsection{Richardson Criterion}
\label{app:Richcrit}

The Richardson criterion\cite{1964NucPh..52..221R,1977JMP....18.1802R} is an approximation of the optimal binding energy of the Cooper pair, defined by $E_\Delta(N) = E(N+2) - E(N) \, - 2(E(N+1)-E(N))$  where $E(N)$ is the groundstate energy at particle number $N$.  In the flat band case, $E(2n) = 0 ~~\forall n$ due to the eta-pairing symmetry because $\ket{n}$ are zero energy ground states. Thus $E_\Delta(N) = - \eps|U| < 0$ demonstrating perfect pairing at all densities. Indeed, we will see in \App{app:Cooperpairspectrum} that $\eps|U|$ is the maximum binding energy computed from the charge $+2$ Cooper pair excitations.

When weak dispersion (kinetic energy) is added to the model, we will determine which density shows the strongest pairing: that is, the most negative binding energy $E_\Delta(N)$. To evaluate $E(N),E(N+2)$ at $N=2n$ in the eta-pairing groundstates, we need the expectation value of the kinetic energy in the states $\ket{n}$, and for $E(N+1)$ we need to compute the expectation value in the charge +1 states atop the $N = 2n$ ground state, which is a four operator correlator and can be computed with Wick's theorem.

We consider a more general perturbation than discussed in the Main Text. We consider a perturbation to the kinetic energy defined in \Eq{eq:gammasing}, namely
\bea
\tilde{H} + \tilde{H}' &= \sum_{\mbf{k},\al \be,\sigma}c^\dag_{\mbf{k},\al,\sigma} \tilde{h}^\sigma_{\al \be}(\mbf{k}) c_{\mbf{k},\be,\sigma} + \sum_{\mbf{k},\al \be,\sigma\sigma'}c^\dag_{\mbf{k},\al,\sigma} \tilde{h}'_{\al \sigma, \be\sigma'}(\mbf{k}) c_{\mbf{k},\be,\sigma'} \\
\eea
where $\tilde{h}_{\al \be}(\mbf{k})$ is the flat band Hamiltonian and we treat $\tilde{h}'_{\al \sigma,\be \sigma'}(\mbf{k})$ as a small perturbation (additional hoppings, spin-orbit couplings to break $\mathcal{T}$ and $S_z$, etc). When projected into the flat bands (still requiring that the conduction bands are much higher in energy than the scale of $\tilde{h}'$ and $|U|$), we find
\bea
\tilde{H}' &\to \sum_{\mbf{k},\al \be,\sigma\sigma'}\bar{c}^\dag_{\mbf{k},\al,\sigma} \tilde{h}'_{\al \sigma, \be\sigma'}(\mbf{k}) \bar{c}_{\mbf{k},\be,\sigma'}\\
&= \sum_{\mbf{k},\al\al' \be\be',\sigma\sigma'}\bar{c}^\dag_{\mbf{k},\al',\sigma} P^{\sigma}_{\al' \al}(\mbf{k})  \tilde{h}'_{\al \sigma, \be\sigma'}(\mbf{k}) P^{\sigma'}_{\be \be'}(\mbf{k}) \bar{c}_{\mbf{k},\be',\sigma'}\\
&= \sum_{\mbf{k},\al\be,\sigma\sigma',mn}\gamma^\dag_{\mbf{k},m,\sigma} U^{\sigma*}_{\al m}(\mbf{k}) \tilde{h}'_{\al \sigma, \be\sigma'}(\mbf{k}) U^{\sigma'}_{\be n}(\mbf{k}) \gamma_{\mbf{k},n,\sigma'} \\
\eea
using \Eq{eq:bargamma} to write the $\bar{c}$ operators in terms of $\gamma$ operators. Thus the correction to the flat band kinetic energy Hamiltonian takes the form
\bea
\tilde{H}' = \sum_{\mbf{k}mn\sigma\sigma'} \tilde{E}'_{m\sigma,n\sigma'}(\mbf{k}) \gamma^\dag_{\mbf{k},m,\sigma}\gamma_{\mbf{k},n,\sigma'}, \qquad  \tilde{E}'_{m\sigma,n\sigma'}(\mbf{k})  = \sum_{\al\be}U^{\sigma*}_{\al m}(\mbf{k}) \tilde{h}'_{\al \sigma, \be\sigma'}(\mbf{k}) U^{\sigma'}_{\be n}(\mbf{k}) \\
\eea
or in matrix form, recalling $U_\sigma(\mbf{k})$ is an $N_{orb} \times N_{f}$ rectangular matrix,
\bea
\tilde{E}'(\mbf{k}) = \bpm U^\dag_\u(\mbf{k}) & \\
& U^\dag_\d(\mbf{k}) \epm \bpm
\tilde{h}'_{\u\u}(\mbf{k}) & \tilde{h}'_{\u\d}(\mbf{k}) \\
\tilde{h}'_{\d\u}(\mbf{k}) & \tilde{h}'_{\d\d}(\mbf{k}) \\
\epm \bpm U_\u(\mbf{k}) & \\
& U_\d(\mbf{k}) \epm
\eea
in the spin $\otimes$ orbital tensor product basis. We develop an expression for Richardson's criterion in the general case, but it may be helpful to point out the simplest case: dispersion $\tilde{E}'_{m,\sigma}(\mbf{k})$ can be added to the flat bands by taking $ \tilde{E}'_{m\sigma,n\sigma'}(\mbf{k}) = \delta_{nm} \delta_{\sigma \sigma'} \tilde{E}'_{m,\sigma}(\mbf{k})$.

We now need to compute expectation values. The ground states at $N=2n$ and $N+2 = 2(n+1)$ are $\ket{n}, \ket{n+1}$ at zeroth order in $\tilde{E}'$. When $\tilde{H}'$ is added, we determine the shift in energy with degenerate perturbation theory; by computing the matrix elements of ${\tilde H}'$ with respect to the unperturbed states:
\bea
\braket{z|\tilde{H}'|z} &= \sum_{\mbf{k}mn\sigma\sigma'} \tilde{E}'_{m\sigma,n\sigma'}(\mbf{k})  \braket{z|\gamma^\dag_{\mbf{k},m,\sigma}\gamma_{\mbf{k},n,\sigma'} |z} \\
&= \sum_{\mbf{k}mn\sigma\sigma'} \tilde{E}'_{m\sigma,n\sigma'}(\mbf{k}) \delta_{mn} \delta_{\sigma\sigma'} \frac{|z|^2}{1+|z|^2} \braket{z|z} \\
&= |z|^2 (1+|z|^2)^{N_f\mathcal{N}-1} 2N_f \mathcal{N} \tilde{E}', \qquad 2N_f \mathcal{N} \tilde{E}' = \sum_{\mbf{k}m\sigma} \tilde{E}'_{m\sigma,m\sigma}(\mbf{k}) \\
\eea
using \Eq{eq:wickcontractions}. We defined $\tilde{E}' = \frac{1}{2N_f \mathcal{N}} \sum_{\mbf{k}m\sigma} \tilde{E}'_{m\sigma,m\sigma}$ as the average of the entire spectrum, which is finite in the thermodynamic limit. Expanding both sides in powers of $z$, we find
\bea
\label{eq:Richarson1}
\sum_n |z|^{2n} \binom{N_f\mathcal{N}}{n} \braket{n|\tilde{H}'|n} &= \sum_n |z|^{2(n+1)} \binom{N_f\mathcal{N}-1}{n} (2N_f \mathcal{N} \tilde{E}') \\
 \binom{N_f\mathcal{N}}{n} \braket{n|\tilde{H}|n} &= \binom{N_f\mathcal{N}-1}{n-1}(2N_f \mathcal{N} \tilde{E}') \\
\braket{n|\tilde{H}|n} &= \frac{n}{N_f\mathcal{N}} (2N_f \mathcal{N} \tilde{E}') = \nu (2N_f \mathcal{N} \tilde{E}')
\eea
recalling $ \frac{n}{N_f\mathcal{N}} = \nu$ is the filling. Next we need the expectation value of $\tilde{H}'$ in the electron excitations \Eq{eq:estateeq}, but first we must normalize them. We compute
\bea
\label{eq:1pnorm}
\braket{z|\gamma_{\mbf{k},m,\sigma} \gamma^\dag_{\mbf{k},m,\sigma}|z} &= \frac{1}{1+|z|^2} \braket{z|z} \\
\binom{N_f \mathcal{N}}{n} \braket{n|\gamma_{\mbf{k},m,\sigma} \gamma^\dag_{\mbf{k},m,\sigma}|n} &= \binom{N_f \mathcal{N} -1}{n} \\
\braket{n|\gamma_{\mbf{k},m,\sigma} \gamma^\dag_{\mbf{k},m,\sigma}|n} &= 1-\nu \ .\\
\eea
Because all $\gamma^\dag_{\mbf{k},m,\sigma}\ket{n}$ excitations are degenerate, we compute their effective Hamiltonian at first order in degenerate perturbation theory:
\bea
\braket{z|\gamma_{\mbf{k},m',\sigma'} \tilde{H} \gamma^\dag_{\mbf{k},m,\sigma}|z} &= \sum_{\mbf{k}'ijss'} \tilde{E}'_{is,js'}(\mbf{k}')  \braket{z|\gamma_{\mbf{k},m',\sigma'} \gamma^\dag_{\mbf{k}',i,s}\gamma_{\mbf{k}',j,s'} \gamma^\dag_{\mbf{k},m,\sigma}|z} \ . \\
\eea
There are only three contractions:
\bea
 \braket{z|\gamma_{\mbf{k},m',\sigma'} \gamma^\dag_{\mbf{k}',i,s}\gamma_{\mbf{k}',j,s'} \gamma^\dag_{\mbf{k},m,\sigma}|z} &= \frac{1}{(1+|z|^2)^2} \Big( \delta_{\mbf{k}\mbf{k}'} \delta_{im'} \delta_{\sigma' s} \delta_{jm} \delta_{\sigma s'} - |z|^2 \delta_{\mbf{k},-\mbf{k}'} \delta_{m'j} \delta_{im} s_z^{s'} \delta_{\sigma',-s'} s_z^{s} \delta_{\sigma,-s} \\
 & \qquad\qquad\qquad + |z|^2 \delta_{mm'}\delta_{\sigma \sigma'} \delta_{ij} \delta_{ss'}
\Big) \braket{z|z} \ .
\eea
Plugging in, we find
\bea
\braket{z|\gamma_{\mbf{k},m',\sigma'} \tilde{H}' \gamma^\dag_{\mbf{k},m,\sigma}|z} &= (1+|z|^2)^{N_f \mathcal{N}-2} \Big( \tilde{E}'_{m'\sigma',m\sigma}(\mbf{k}) -|z|^2 s_z^{\sigma'} s_z^{\sigma} \tilde{E}'_{m(-\sigma),m'(-\sigma')}(-\mbf{k}) +  |z|^2 \delta_{mm'} \delta_{\sigma \sigma'} 2N_f \mathcal{N} \tilde{E}' \ .
\Big)
\eea
Now expanding in powers of $|z|^2$ and including the normalization \Eq{eq:1pnorm}, we find
\bea
\label{eq:degepertumatr}
\binom{N_f\mathcal{N}}{n} \frac{\braket{n|\gamma_{\mbf{k},m',\sigma'} \tilde{H}' \gamma^\dag_{\mbf{k},m,\sigma}|n} }{1-\nu} &= \frac{1}{1-\nu} \lp \binom{N_f\mathcal{N}-2}{n} \tilde{E}'_{m'\sigma',m\sigma}(\mbf{k}) - \binom{N_f\mathcal{N}-2}{n-1} s_z^{\sigma'} s_z^{\sigma} \tilde{E}'_{m(-\sigma),m'(-\sigma')}(-\mbf{k}) \right. \\
&\left.\qquad + \binom{N_f\mathcal{N}-2}{n-1} \delta_{mm'} \delta_{\sigma \sigma'}2N_f \mathcal{N} \tilde{E}' \rp \ . \\
\eea
Simplifying the binomal coefficients gives
\bea
\frac{\braket{n|\gamma_{\mbf{k},m',\sigma'} \tilde{H}' \gamma^\dag_{\mbf{k},m,\sigma}|n} }{1-\nu} &= \frac{1}{1-\nu} \lp \frac{(N_f\mathcal{N}-n-1)(N_f\mathcal{N}-n)}{N_f \mathcal{N}(N_f\mathcal{N}-1)} \tilde{E}'_{m'\sigma',m\sigma}(\mbf{k}) - \frac{n(N_f\mathcal{N}-n)}{N_f \mathcal{N}(N_f\mathcal{N}-1)} s_z^{\sigma'} s_z^{\sigma} \tilde{E}'_{m(-\sigma),m'(-\sigma')}(-\mbf{k}) \right. \\
&\left.\qquad + \frac{n(N_f\mathcal{N}-n)}{N_f \mathcal{N}(N_f\mathcal{N}-1)}  \delta_{mm'} \delta_{\sigma \sigma'}2N_f \mathcal{N} \tilde{E}' \rp \\
&=  \lp \frac{N_f\mathcal{N}-n-1}{N_f\mathcal{N}-1} \tilde{E}'_{m'\sigma',m\sigma}(\mbf{k}) - \frac{n}{N_f\mathcal{N}-1} s_z^{\sigma'} s_z^{\sigma} \tilde{E}'_{m(-\sigma),m'(-\sigma')}(-\mbf{k})  + \frac{n}{N_f\mathcal{N}-1}  \delta_{mm'} \delta_{\sigma \sigma'}2N_f \mathcal{N} \tilde{E}' \rp \\
\eea
To write the effective Hamiltonian more transparently, we define the matrix notation
\bea
\null [\tilde{E}'(\mbf{k})]_{m'\sigma', m\sigma} &= \tilde{E}'_{m'\sigma',m\sigma}(\mbf{k}) \\
\null [\sigma_y \tilde{E}'^T(-\mbf{k}) \sigma_y]_{m'\sigma', m\sigma} &= s_z^{\sigma'} s_z^{\sigma}\tilde{E}'_{m(-\sigma),m'(-\sigma')}(-\mbf{k}) \\
\eea
where $\sigma_y$ is a Pauli matrix acting only on the spin indices. Because $\tilde{H}$ is Hermitian, $E^T = E^*$. Thus we can write the matrix \Eq{eq:degepertumatr} obtained from degenerate perturbation theory as
\bea
\frac{\braket{n|\gamma_{\mbf{k},m',\sigma'} \tilde{H}' \gamma^\dag_{\mbf{k},m,\sigma}|n} }{1-\nu} &= \left[  \frac{N_f\mathcal{N}-n-1}{N_f\mathcal{N}-1}  \tilde{E}'(\mbf{k}) -  \frac{n}{N_f\mathcal{N}-1}  \sigma_y \tilde{E}'^*(-\mbf{k}) \sigma_y + \frac{n}{N_f\mathcal{N}-1} \mathbb{1} \, 2N_f \mathcal{N} \tilde{E}'\right]_{m'\sigma',m\sigma} \ . \\
\eea
The eigen-energies of this matrix split the electron excitations which were degenerate at energy $\eps|U|/2$, forming a band structure for the charge +1 excitations. Let $\la_{min}[M(\mbf{k})]$ denote the smallest eigenvalue of $M(\mbf{k})$ for all $\mbf{k}$. Then the lowest energy state of the one-particle excitations is
\bea
\label{eq:Richarson2}
E(N+1) = \la_{min}\left[\frac{\eps |U|}{2} \mathbb{1} + \frac{N_f\mathcal{N}-n-1}{N_f\mathcal{N}-1}  \tilde{E}'(\mbf{k}) -  \frac{n}{N_f\mathcal{N}-1}  \sigma_y \tilde{E}'^*(-\mbf{k}) \sigma_y + \frac{n}{N_f\mathcal{N}-1} \mathbb{1} \, 2N_f \mathcal{N} \tilde{E}'\right]
\eea
corresponding to the lowest energy single-particle excitation in perturbation theory.  With \Eqs{eq:Richarson1}{eq:Richarson2}, the Richardson binding energy can be written as
\bea
\label{eq:Delta}
E_\Delta(N) &= E(N+2) + E(N) - 2 E(N+1) \\
&= \frac{n +(n+1)}{N_f\mathcal{N}} (2N_f \mathcal{N} \tilde{E}') \mathbb{1}  - 2 \la_{min} \left[ \frac{\eps|U|}{2}\mathbb{1} + \frac{N_f\mathcal{N}-n-1}{N_f\mathcal{N}-1}  \tilde{E}'(\mbf{k}) -  \frac{n}{N_f\mathcal{N}-1}  \sigma_y \tilde{E}'^*(-\mbf{k}) \sigma_y + \frac{n}{N_f\mathcal{N}-1} \mathbb{1}2 N_f \mathcal{N} \tilde{E}'  \right] \\
&= \lp \frac{2n+1}{N_f\mathcal{N}} -\frac{2n}{N_f\mathcal{N}-1} \rp (2N_f \mathcal{N} \tilde{E}') - \eps|U| -2 \la_{min} \left[ \frac{N_f\mathcal{N}-n-1}{N_f\mathcal{N}-1}  \tilde{E}'(\mbf{k}) -  \frac{n}{N_f\mathcal{N}-1}  \sigma_y \tilde{E}'^*(-\mbf{k}) \sigma_y \right] \\
&= - \eps|U| + 2(1-2\nu) \tilde{E}'  -2 \la_{min} \left[ (1-\nu)  \tilde{E}'(\mbf{k}) -  \nu \sigma_y \tilde{E}'^*(-\mbf{k}) \sigma_y \right] + O(1/\mathcal{N})\\
\eea
where in the last line we took the thermodynamic limit with $\nu = n/N_f\mathcal{N}$. In going from the second to third lines we have pulled out terms proportional to the identity out of $\lambda_\text{min}$: $\lambda_\text{min}[aI + B] = a + \lambda_\text{min}[B]$. To determine at what filling $\nu$ the superconductor is strongest (i.e. has the largest gap), we need to determine when $-E_\Delta(N)$ is largest. In general, this would require diagonalizing $(1-\nu)  \tilde{E}'(\mbf{k}) - \nu \sigma_y \tilde{E}'^*(-\mbf{k}) \sigma_y$ as a function of $\nu$. This is easily done for a specific model since $\tilde{E}'(\mbf{k})$ is an effective single-particle Hamiltonian. Generically, \Eq{eq:Delta} can reveal material-specific effects on superconductivity due spin-orbit coupling, applied magnetic field, and other small perturbations.

However, we are able to make a general statement when time-reversal symmetry is preserved by $\tilde{H}'$, ensuring that $\sigma_y \tilde{E}'^*(-\mbf{k}) \sigma_y = \tilde{E}'(\mbf{k})$. We first take $\tilde{E}'=0$ for convenience; we will restore nonzero $\tilde{E}'$ later. Then noting that $\la_{min}[\tilde{E}'(\mbf{k})] \leq 0, \la_{max}[\tilde{E}'(\mbf{k})] \geq 0$, we find
\bea
E_\Delta(N) &= - \eps|U| -2 \la_{min} \left[ (1-2\nu)  \tilde{E}'(\mbf{k}) \right]  \\
&= - \eps|U| -2 \begin{cases}
(1-2\nu) \la_{min} [\tilde{E}'(\mbf{k})], & \nu < 1/2\\
(1-2\nu)  \la_{max} [\tilde{E}'(\mbf{k})], & \nu > 1/2\\
\end{cases} \\
&= - \eps|U| + 2|1-2\nu| \begin{cases}
 |\la_{min} [\tilde{E}'(\mbf{k})] |, & \nu < 1/2\\
 \la_{max} [\tilde{E}'(\mbf{k})], & \nu > 1/2\\
\end{cases} \\
\eea
so because the second term is non-negative with a minimum at $\nu = 1/2$, we see that $E_\Delta(N)$ is minimized (the binding energy is largest and negative) at $\nu = 1/2$. This holds for arbitrary time-reversal symmetric $\tilde{E}'(\mbf{k})$ and is independent of the position of the peak in the density of states, which is very different from standard BCS theory.

Restoring nonzero $E'$ gives
\bea
E_\Delta(N) &= - \eps|U| +2(1-2\nu) {\tilde E}' -2 \begin{cases}
(1-2\nu) \la_{min} [\tilde{E}'(\mbf{k})], & \nu < 1/2\\
(1-2\nu)  \la_{max} [\tilde{E}'(\mbf{k})], & \nu > 1/2\\
\end{cases} \\
&= - \eps|U| -2 \begin{cases}
(1-2\nu) (\la_{min} [\tilde{E}'(\mbf{k}) -{\tilde E}' I]), & \nu < 1/2\\
(1-2\nu)  (\la_{max} [\tilde{E}'(\mbf{k}) -{\tilde E}' I]), & \nu > 1/2\\
\end{cases}.
\eea  Because we have subtracted the average ${\tilde E}'$ from the matrix ${\tilde E}'(\kk)$, $\la_{min} [\tilde{E}'(\mbf{k}) -{\tilde E}' I] \leq 0$ and $\la_{max} [\tilde{E}'(\mbf{k}) -{\tilde E}' I] \geq 0$.  The energy $E_\Delta(N)$ is still minimized at $\nu = 1/2$.

\subsection{Charge $+2$ Excitations}
\label{app:charge2}

We now study the charge $+2$ excitations. To do so, we will compute the scattering matrix $R^{\sigma\sigma'}_{\mbf{k}'m'n',\mbf{k}mn}(\mbf{p})$ on the complete basis of charge $+2$ excitations at many-body momentum $\mbf{p}$ defined by
\bea
\label{eq:Hcharge2}
\null [H, \gamma^\dag_{\mbf{p}+\mbf{k},m,\sigma}\gamma^\dag_{-\mbf{k},n,\sigma'}] \ket{GS} &=  \sum_{\mbf{k}'} \gamma^\dag_{\mbf{p}+\mbf{k}',m',\sigma}\gamma^\dag_{-\mbf{k}',n',\sigma'}  \ket{GS} [R^{\sigma \sigma'}(\mbf{p})]_{\mbf{k}'m'n',\mbf{k}mn} \ .
\eea
There are three flavors of Cooper pair. The $\sigma = \sigma' = \pm \frac{1}{2}$ cases lead to spin $\pm1$ Cooper pairs, while the $\sigma = - \sigma'$ case leads to a spin 0 Cooper pair. We will find that only the spin 0 Cooper pair has low-energy eigenstates below the two-particle continuum. For instance, we know that there must be a zero-energy excitation $\eta^\dag$ at $\mbf{p}=0$ since $\eta$ is a symmetry.

This calculation is made simpler by using the results from the charge $+1$ electron excitation matrix. Using \Eqs{eq:formfactorS}{eq:SSgamma}, the single electron commutator is
\bea
\null [H, \gamma^\dag_{\mbf{k}, n, \sigma}] &= \frac{\eps |U|}{2} \gamma^\dag_{\mbf{k}, n, \sigma} + \frac{|U|}{2} \sum_{\mbf{q} \al m} 2  \frac{1}{\sqrt{\mathcal{N}}} s^z_\sigma \gamma^\dag_{\mbf{k}+\mbf{q},m,\sigma} M^{mn}_{\sigma,\al}(\mbf{k},\mbf{q}) \bar{S}^z_{-\mbf{q},\al}
\eea
and the second term annihilates the groundstate. Now it is straightforward to compute
\bea
\null [H, \gamma^\dag_{\mbf{p}+\mbf{k},m,\sigma}\gamma^\dag_{-\mbf{k},n,\sigma'}] \ket{GS} &=  \gamma^\dag_{\mbf{p}+\mbf{k},m,\sigma}  [H, \gamma^\dag_{-\mbf{k},n,\sigma'}] \ket{GS}  + [H, \gamma^\dag_{\mbf{p}+\mbf{k},m,\sigma}] \gamma^\dag_{-\mbf{k},n,\sigma'} \ket{GS}  \\
 &= \eps |U| \gamma^\dag_{\mbf{p}+\mbf{k},m,\sigma}  \gamma^\dag_{-\mbf{k},n,\sigma'} \ket{GS}  + |U| \frac{1}{\sqrt{\mathcal{N}}}  \sum_{\mbf{q} \al m'} s^z_\sigma \gamma^\dag_{\mbf{p}+\mbf{q}+\mbf{k},\sigma,m'} M^{m'm}_{\sigma,\al}(\mbf{k}+\mbf{p},\mbf{q}) [\bar{S}^z_{-\mbf{q},\al}, \gamma^\dag_{-\mbf{k},n,\sigma'}] \ket{GS} \\
  &= \eps |U| \gamma^\dag_{\mbf{p}+\mbf{k},m,\sigma}  \gamma^\dag_{-\mbf{k},n,\sigma'} \ket{GS}  \\
  &\qquad + |U| \frac{1}{\mathcal{N}}  s^z_\sigma  s^z_{\sigma'} \sum_{\mbf{q} \al m'} \gamma^\dag_{\mbf{p}+\mbf{q}+\mbf{k},\sigma,m'} \gamma^\dag_{-\mbf{k}-\mbf{q},\sigma',n'} M^{m'm}_{\sigma,\al}(\mbf{k}+\mbf{p},\mbf{q})  M^{n'n}_{\sigma',\al}(-\mbf{k},-\mbf{q})  \ket{GS} \\
\eea
where in the second line we introduced the commutator $[\bar{S}^z_{-\mbf{q},\al}, \gamma^\dag_{-\mbf{k},n,\sigma'}]$ since $\bar{S}^z_{-\mbf{q},\al}$ annihilates the vacuum. Resumming gives
\bea
\null  [H, \gamma^\dag_{\mbf{p}+\mbf{k},m,\sigma}\gamma^\dag_{-\mbf{k},n,\sigma'}] \ket{GS}   &= \eps |U| \gamma^\dag_{\mbf{p}+\mbf{k},m,\sigma}  \gamma^\dag_{-\mbf{k},n,\sigma'} \ket{GS}  \\
&\qquad + |U| s^z_\sigma  s^z_{\sigma'}  \frac{1}{\mathcal{N}} \sum_{\mbf{k}' \al} \gamma^\dag_{\mbf{p}+\mbf{k}',\sigma,m'} \gamma^\dag_{-\mbf{k}',\sigma',n'}  \ket{GS}  M^{m'm}_{\sigma,\al}(\mbf{k}+\mbf{p},\mbf{k}'-\mbf{k})  M^{n'n}_{\sigma',\al}(-\mbf{k},\mbf{k}-\mbf{k}') \\
\eea
and thus we find
\bea
\label{eq:Rscater}
\null [R^{\sigma \sigma'}(\mbf{p})]_{\mbf{k}'m'n',\mbf{k}mn} &= \eps |U| \delta_{\mbf{k}\mbf{k}'} \delta_{mm'} \delta_{nn'} + s^z_\sigma  s^z_{\sigma'} |U| \frac{1}{\mathcal{N}} \sum_\al M^{m'm}_{\sigma,\al}(\mbf{k}+\mbf{p},\mbf{k}'-\mbf{k})  M^{n'n}_{\sigma',\al}(-\mbf{k},\mbf{k}-\mbf{k}') \ .
\eea
The second term describes the nontrivial scattering processes. We observe that
\bea
\label{eq:nontrivtermMM}
\sum_\al M^{m'm}_{\sigma,\al}(\mbf{k}+\mbf{p},\mbf{k}'-\mbf{k})  M^{n'n}_{\sigma',\al}(-\mbf{k},\mbf{k}-\mbf{k}') &= \sum_\al U^{\sigma*}_{m' \al}(\mbf{p}+\mbf{k}') U^{\sigma}_{m \al}(\mbf{p}+\mbf{k}) \, U^{\sigma'*}_{n' \al}(-\mbf{k}') U^{\sigma'}_{n \al}(-\mbf{k}) \\
&= \sum_\al \mathcal{U}^{\sigma \sigma'}_{\mbf{k}'m'n',\al}(\mbf{p}) \mathcal{U}^{\sigma \sigma' *}_{\mbf{k}mn,\al}(\mbf{p}), \quad \mathcal{U}^{\sigma \sigma'}_{\mbf{k}mn,\al}(\mbf{p}) = U^{\sigma *}_{m \al}(\mbf{p}+\mbf{k}) U^{\sigma' *}_{n \al}(-\mbf{k})  \ .
\eea
Let us consider the case where $\sigma = \sigma'$. Note that if $\sigma = \sigma'$, $\mathcal{U}^{\sigma \sigma'}_{\mbf{k}'m'n',\al}(\mbf{p})$ is symmetric under $\mbf{k}' \to -(\mbf{p}+\mbf{k}'), m' \leftrightarrow n'$. But the electron operators $\gamma^\dag_{\mbf{p}+\mbf{k}',m',\sigma}\gamma^\dag_{-\mbf{k}',n',\sigma}$ which contract against $\mathcal{U}^{\sigma \sigma'}_{\mbf{k}'m'n',\al}(\mbf{p})$ in \Eq{eq:Hcharge2} are anti-symmetric under $\mbf{k}' \to -(\mbf{p}+\mbf{k}'), m' \leftrightarrow n'$. Thus many-body states which are \emph{not} in the kernel of \Eq{eq:nontrivtermMM} all vanish by fermion anti-symmetry when $\sigma = \sigma'$. Hence all the spin $\pm 1$ Cooper pairs all have energy $\eps |U|$, which is twice the single-electron gap. This indicates they are effectively unpaired. Going forward, we take $\sigma = \u, \sigma' = \d$.

We can think of $\mathcal{U} \equiv \mathcal{U}^{\u \d}$ as a large rectangular matrix of dimension $\mathcal{N} N_f^2 \times N_{orb}$. (In the bipartite case, the matrix is $\mathcal{N} N_f^2 \times N_{L}$ because $U^\sigma_{m\al}(\mbf{k}) = 0$ for $\al \notin L$.) In matrix notation, we have
\bea
\label{eq:lowrank}
R^{\u \d}(\mbf{p}) &= \eps |U| \mathbb{1} - |U| \frac{1}{\mathcal{N}} \mathcal{U}(\mbf{p}) \mathcal{U}^\dag(\mbf{p}) \ . \\
\eea
Note that $ \mathcal{U}(\mbf{p}) \mathcal{U}^\dag(\mbf{p})$ is positive semi-definite, so the spin 0 Cooper pair spectrum is below $\eps |U|$, indicating attractive pairing. Because $\mathcal{U}$ is rectangular, it has low rank and $\frac{1}{\mathcal{N}} \mathcal{U}\mathcal{U}^\dag$ has many zero eigenvalues. Its nonzero eigenvalues are the same as the eigenvalues of $h(\mbf{p}) = \frac{1}{\mathcal{N}}\mathcal{U}^\dag(\mbf{p}) \mathcal{U}(\mbf{p})$ which is a matrix of dimension $N_{orb} \times N_{orb}$. To see this, define the eigenvector $u$ with energy $\eps(\mbf{p})$ by $h(\mbf{p}) u = \eps(\mbf{p}) u$. Then $\mathcal{U} u$ is an eigenvector of the scattering matrix: $\frac{1}{\mathcal{N}} \mathcal{U}\mathcal{U}^\dag (\mathcal{U} u) =  \mathcal{U} h(\mbf{p}) u = \eps(\mbf{p}) \mathcal{U} u$. This gives $\dim h(\mbf{p}) = N_{orb}$ nontrivial eigenvalues, which are nonnegative because $ \mathcal{U}\mathcal{U}^\dag $ is positive-semi-definite. Note that the rank of $\mathcal{U}\mathcal{U}^\dag$ is at most $N_{orb}$, so all other eigenvalues are zero. Hence we are led to study the spectrum of
\bea
\label{eq:CPham}
h_{\al \be}(\mbf{p}) &= \frac{1}{\mathcal{N}} [\mathcal{U}^\dag(\mbf{p}) \mathcal{U}(\mbf{p})]_{\al \be} \\
&= \frac{1}{\mathcal{N}} \sum_{\mbf{k}mn} U^{\u}_{m \al}(\mbf{p}+\mbf{k}) U^{\d}_{n \al}(-\mbf{k})  U^{\u *}_{m \be}(\mbf{p}+\mbf{k}) U^{\d *}_{n \be}(-\mbf{k})  \\
&= \frac{1}{\mathcal{N}} \sum_{\mbf{k}} P^{\u}_{\al \be}(\mbf{p}+\mbf{k}) P^{\d}_{\al \be}(-\mbf{k}) \\
&=\frac{1}{\mathcal{N}} \sum_{\mbf{k}} P_{\al \be}(\mbf{p}+\mbf{k}) P_{\be \al}(\mbf{k}) \\
\eea
where we used $P^\d(\mbf{k}) = P^\u(-\mbf{k})^*$, and set $P^\u(\mbf{k}) = P(\mbf{k})$.

A number of extremely useful statements will be proven about this effective single-particle Hamiltonian in the next section. For now, we simply remark that the eigenvectors $u^\be_\mu(\mbf{p})$ and eigenvalues $\eps_\mu(\mbf{p})$ are defined as
\bea
\sum_\be h_{\al \be}(\mbf{p}) u^\be_\mu(\mbf{p}) = \eps_\mu(\mbf{p}) u^\al_\mu(\mbf{p}) \ . \\
\eea
We normalize the eigenvectors such that $\sum_\be u^\be_\mu(\mbf{p}) u^{\be *}_\nu(\mbf{p}) = \delta_{\mu \nu}$ where $\mu,\nu \in 0,\dots, N_{orb}-1$ label the boson bands. It is convenient to decreasingly order the eigenvalues of $h(\mbf{p})$ such that $\mu = 0$ corresponds to the largest eigenvalue. The full eigenstates of the scattering matrix \Eq{eq:Rscater} not in the kernel of $\mathcal{U}\mathcal{U}^\dag$ are $\sum_\al \mathcal{U}_{\mbf{k}mn,\al}(\mbf{p}) u^\al_\mu(\mbf{p})$ with eigenvalue
\bea
R^{\u \d}(\mbf{p}) \mathcal{U} u_\mu = |U| (\eps - \eps_\mu(\mbf{p}) )  \mathcal{U} u_\mu
\eea
so the many-body energies are $|U| (\eps - \eps_\mu(\mbf{p}) )$; $\mu=0$ corresponds to the lowest energy band of many-body states.

The eigenvectors $\mathcal{U} u_\mu$ have norm squared $u^\dag_\mu(\mbf{p}) \mathcal{U}^\dag(\mbf{p}) \mathcal{U}(\mbf{p}) u_\mu(\mbf{p}) = \mathcal{N} u^\dag_\mu(\mbf{p}) h(\mbf{p}) u_\mu(\mbf{p}) = \mathcal{N} \eps_\mu(\mbf{p})$. Thus we define the many-body operators
\bea
\label{eq:etap}
\eta^\dag_{\mbf{p},\mu} &= \frac{1}{\sqrt{\mathcal{N} \eps_\mu(\mbf{p})}} \sum_{\mbf{k}\al mn} \gamma^\dag_{\mbf{p}+\mbf{k},m,\u}\gamma^\dag_{-\mbf{k},n,\d} \mathcal{U}_{\mbf{k}mn,\al}(\mbf{p}) u^\al_{\mu}(\mbf{p}) \\
&= \frac{1}{\sqrt{\mathcal{N} \eps_\mu(\mbf{p})}} \sum_{\mbf{k}\al mn} \gamma^\dag_{\mbf{p}+\mbf{k},m,\u}\gamma^\dag_{-\mbf{k},n,\d} U^{\u *}_{m \al}(\mbf{p}+\mbf{k}) U^{\d *}_{n \al}(-\mbf{k})  u^\al_{\mu}(\mbf{p}) \\
&= \frac{1}{\sqrt{\mathcal{N}}} \sum_{\mbf{k}\al} \frac{u^\al_{\mu}(\mbf{p})}{\sqrt{\eps_\mu(\mbf{p})}} \bar{c}^\dag_{\mbf{p}+\mbf{k},\al,\u}\bar{c}^\dag_{-\mbf{k},\al,\d}  \\
&= \frac{1}{\sqrt{\mathcal{N}}} \sum_{\mbf{R}\al} \frac{u^\al_{\mu}(\mbf{p})}{\sqrt{\eps_\mu(\mbf{p})}}  e^{-i \mbf{p} \cdot (\mbf{R}+\mbf{r}_\al)} \bar{c}^\dag_{\mbf{R},\al,\u}  \bar{c}^\dag_{\mbf{R},\al,\d}  \\
\eea
which obey
\bea
\null [H, \eta^\dag_{\mbf{p},\mu}] \ket{GS} &= |U| (\eps - \eps_\mu(\mbf{p}) ) \eta^\dag_{\mbf{p},\mu} \ket{GS} \ . \\
\eea
Comparing with \Eq{eq:etadef}, we see that the Cooper pair operators only show onsite pairing in $\mbf{R},\al$ in the $\bar c$ degrees of freedom, indicating closely bound states. We will show that, up to a normalization, $\eta^\dag_{\mbf{p}=0,\mu=0} \propto \eta^\dag$ and that $\eps_{\mu=0}(\mbf{p}=0) = \eps$. Thus the Cooper pair operators $\eta^\dag_{\mbf{p},\mu}$ are deformations of the commuting $\eta^\dag$ symmetry with the same $\bar{c}^\dag_{\mbf{R},\al,\u}  \bar{c}^\dag_{\mbf{R},\al,\d}$ orbital structure. They are gapless at $\mbf{p}=0$ as a consequence of Goldstone's theorem. The spectrum of the $\eta^\dag_{\mbf{p},\mu}$ operators contains the low lying Cooper pair (which condenses) as well as higher energy Cooper pair excitations (also known as Leggett modes). 

To compute the normalization of the states $\eta^\dag_{\mbf{p},\mu}\ket{n}$, we use Wick's theorem. First we need the correlator
\bea
\label{eq:CPnormcorr}
\bra{z}\gamma_{-\mbf{k}',n',\d} \gamma_{\mbf{q}+\mbf{k}',m',\u} \gamma^\dag_{\mbf{p}+\mbf{k},m,\u}\gamma^\dag_{-\mbf{k},n,\d} \ket{z} &= \frac{1}{(1+|z|^2)^2} \lp |z|^2 \delta_{\mbf{q},0} \delta_{n'm'} \delta_{\mbf{p},0} \delta_{nm} - 0 + \delta_{\mbf{k}\mbf{k}'} \delta_{nn'} \delta_{\mbf{p},\mbf{q}}\delta_{mm'} \rp \braket{z|z} \\
&= \frac{ \delta_{\mbf{p},\mbf{q}}}{(1+|z|^2)^2} \lp |z|^2 \delta_{\mbf{p},0} \delta_{n'm'}  \delta_{nm}  + \delta_{\mbf{k}\mbf{k}'} \delta_{nn'}\delta_{mm'} \rp \braket{z|z} \ . \\
\eea
We will also need the identity (see \Eq{eq:nontrivtermMM})
\bea
\label{eq:calUupc}
\frac{1}{\mathcal{N}}\sum_{\mbf{k}m} \mathcal{U}^{\sigma,-\sigma}_{\mbf{k}mm,\al}(0) &=\frac{1}{\mathcal{N}} \sum_{\mbf{k}m} U^{\sigma*}_{m \al}(\mbf{k}) U^{-\sigma *}_{m \al}(-\mbf{k}) = \frac{1}{\mathcal{N}} \sum_{\mbf{k}m} U^{\sigma*}_{m \al}(\mbf{k}) U^{\sigma}_{m \al}(\mbf{k}) = \frac{1}{\mathcal{N}}\sum_{\mbf{k}} P^\sigma_{\al \al}(\mbf{k}) = \eps
\eea
using the uniform pairing condition. Thus we obtain
\bea
\label{eq:etapnorm}
\bra{z}\eta_{\mbf{q},\nu} \eta^\dag_{\mbf{p},\mu}\ket{z} &= \delta_{\mbf{p},\mbf{q}} \frac{(1+|z|^2)^{N_f \mathcal{N}-2} }{\mathcal{N} \sqrt{\eps_\mu(\mbf{p})\eps_\nu(\mbf{q})}} \sum_{\mbf{k}\mbf{k}', \al \be} \sum_{mn,m'n'} \mathcal{U}^*_{\mbf{k}'m'n',\be}(\mbf{q}) u^{\be*}_\nu(\mbf{q}) \lp |z|^2 \delta_{n'm'} \delta_{\mbf{p},0} \delta_{nm} + \delta_{\mbf{k}\mbf{k}'} \delta_{nn'} \delta_{mm'} \rp \mathcal{U}_{\mbf{k}mn,\al}(\mbf{p}) u^\al_\mu(\mbf{p}) \\
&= \delta_{\mbf{p},\mbf{q}} \frac{(1+|z|^2)^{N_f \mathcal{N}-2} }{\mathcal{N} \sqrt{\eps_\mu(\mbf{p})\eps_\nu(\mbf{p})}} \sum_{\al \be} u^{\be*}_\nu(\mbf{p}) \lp |z|^2 \delta_{\mbf{p},0} \sum_{\mbf{k}'m'}\mathcal{U}^*_{\mbf{k}'m'm',\be}(0) \sum_{\mbf{k}m}\mathcal{U}_{\mbf{k}mm,\al}(0)  + [\mathcal{U}^\dag(\mbf{p})\mathcal{U}(\mbf{p})]_{\be\al} \rp u^\al_\mu(\mbf{p}) \\
&= \delta_{\mbf{p},\mbf{q}} \frac{(1+|z|^2)^{N_f \mathcal{N}-2} }{\sqrt{\eps_\mu(\mbf{p})\eps_\nu(\mbf{p})}} \sum_{\al \be} u^{\be*}_\nu(\mbf{p}) \lp |z|^2 \delta_{\mbf{p},0} \mathcal{N} \eps^2 +h_{\be\al} (\mbf{p})\rp u^\al_\mu(\mbf{p}) \\
&= \delta_{\mbf{p},\mbf{q}} \frac{(1+|z|^2)^{N_f \mathcal{N}-2} }{\sqrt{\eps_\mu(\mbf{p})\eps_\nu(\mbf{p})}} \lp |z|^2 \delta_{\mbf{p},0} \mathcal{N} \eps^2  \sum_{\al \be} u^{\be*}_\nu(0) u^{\al}_\mu(0) +\delta_{\mu \nu} \eps_{\mu}(\mbf{p})\rp \ . \\
\eea
In \App{app:Cooperpairspectrum}, we will prove that $\eps_0(\mbf{p}=0) = \eps$ and $ \sum_{\al} u^{\al}_\mu(0) = \delta_{\mu,0} \sqrt{N_{L}}$ where $N_L$ is the number of orbitals where $P_{\al \be}(\mbf{k})$ is nonzero. With this simple result, we obtain
\bea
\bra{z}\eta_{\mbf{q},\nu} \eta^\dag_{\mbf{p},\mu}\ket{z} &= \delta_{\mbf{p},\mbf{q}} \frac{(1+|z|^2)^{N_f \mathcal{N}-2} }{\sqrt{\eps_\mu(\mbf{p})\eps_\nu(\mbf{p})}} \lp |z|^2 \delta_{\mbf{p},0} \mathcal{N} N_{L}\eps^2 \delta_{\mu,\nu} \delta_{\mu,0}  +\delta_{\mu \nu} \eps_{\mu}(\mbf{p})\rp \\
&= \delta_{\mbf{p},\mbf{q}} \delta_{\mu,\nu} (1+|z|^2)^{N_f \mathcal{N}-2} \lp |z|^2 \delta_{\mbf{p},0} \mathcal{N} N_{L}\eps \delta_{\mu,0} +1\rp \\
&= \delta_{\mbf{p},\mbf{q}} \delta_{\mu,\nu} (1+|z|^2)^{N_f \mathcal{N}-2} \lp |z|^2 \delta_{\mbf{p},0} \delta_{\mu,0} N_f \mathcal{N}  +1\rp \\
\eea
making use of $\eps = N_f/N_{L}$. We see that there is an anomalous term when $\mbf{p}=0,\mu=0$. This is because $\eta^\dag_{\mbf{p},0} \propto \eta^\dag$, which produces a zero-energy eta-pairing state. These states are the ground states we have already computed the normalization for in \Eq{eq:normedetastates}. Expanding in powers of $z$ and picking $\mbf{p}\neq 0, \mu \neq 0$, we obtain
\bea
\sum_n |z|^{2n} \binom{N_f\mathcal{N}}{n} \bra{n}\eta_{\mbf{q},\nu} \eta^\dag_{\mbf{p},\mu}\ket{n} &= \sum_n |z|^{2n} \delta_{\mbf{p},\mbf{q}} \delta_{\mu,\nu}  \binom{N_f\mathcal{N}-2}{n} , \qquad \mbf{p}\neq 0, \mu \neq 0 \\
\bra{n}\eta_{\mbf{q},\nu} \eta^\dag_{\mbf{p},\mu}\ket{n} &= (1-\nu)^2 \delta_{\mbf{p},\mbf{q}} \delta_{\mu,\nu}  \\
\eea
and in the last line we took the thermodynamic limit $\mathcal{N}\to \infty$.

\subsection{Examples of the Cooper Pair Spectrum}
\label{app:examples}

To illustrate the Cooper pair spectrum, we consider three examples: an inversion-protected SSH chain in 1D, an obstructed atomic limit in 2D, and an $S$-matrix construction of fragile bands in 2D. We use the same notation $\tilde{h}(\mbf{k})$ for the single-particle electron Hamiltonian, $P(\mbf{k})$ for the flat band projector, and $h(\mbf{p})$ for the bosonic pairing Hamiltonian for all the models used in this section. We recall that in general, the projection into the flat bands is valid when $W \ll |U| \ll t$ where $W$ is the bandwidth of the flat band ($W=0$ if the band is perfectly flat), $|U|$ is the interaction strength, and $t$ is the single-particle gap above the flat band.  

\subsubsection{SSH Chain}
\label{app:SSHchain}

The 1D Hamiltonian we consider was studied in \Ref{2020Sci...367..794S} as a minimal model of a flat band obstructed atomic insulator. We place $s$ and $p$ orbitals at the 1a position (the center of the unit cell) in the 1D space group $pm$. The electron Hamiltonian is
\bea
\tilde{h}(k) &= \sin k \, \sigma_2 + \cos k \, \sigma_3
\eea
with exactly flat bands at energies $\pm 1$. The eigenstate of the lower band is
\bea
U(k) = \frac{1}{2} \bpm 1 - e^{-ik} \\ 1+ e^{- i k}\epm  \\
\eea
and one can calculate the Berry phase $U^\dag i \del_k U = \frac{1}{2}$, indicating an obstruction to the trivial atomic phase where the Wannier functions are onsite. Indeed, $h(k)$ obeys inversion symmetry with $D[\mathcal{I}] h(k)D^\dag[\mathcal{I}] = h(-k)$ where $D[\mathcal{I}] = \sigma_3$, and real space invariants and/or symmetry eigenvalues diagnose $h(k)$ as being in an obstructed atomic limit \cite{2020Sci...367..794S}.

Projecting into the single flat band, we find that the uniform pairing condition is obeyed:
\bea
\label{eq:SSHproj}
\int \frac{dk}{2\pi} P_{\al\al}(k) = \frac{N_f}{N_{orb}} = \eps= \frac{1}{2} , \qquad P(k) = U(k)U^\dag(k) = \left(
\begin{array}{cc}
 \sin ^2\frac{k}{2} & \frac{i}{2} \sin k \\
 -\frac{i}{2} \sin k & \cos ^2 \frac{k}{2} \\
\end{array}
\right) \ .
\eea
Note that the uniform pairing condition is not enforced by inversion because the $s$ and $p$ orbitals are \emph{different} irreps of the 1a Wyckoff position and hence are not interrelated by crystalline symmetry. In other words, adding $\mathcal{I}$-preserving hoppings could destroy the uniform pairing condition.  However, there may be non-crystallographic symmetries that enforce UPC.  In this model, two anti-commuting symmetries: ${\cal C} = \sigma_1$ and ${\cal Z} = \kk \rightarrow \kk + \pi$, combine into a commuting unitary symmetry ${\cal CZ}$ that enforces UPC:

\begin{align}
{\cal CZ} P(\kk) {\cal CZ} = \sigma_1 \begin{bmatrix}
 \cos ^2\frac{k+\pi}{2} & -\frac{i}{2} \sin (k +\pi) \\
 \frac{i}{2} \sin (k +\pi) & \sin ^2 \frac{k+\pi}{2} \\
\end{bmatrix} \sigma_1 = P(\kk) \\
P_{\alpha\alpha}(\kk) = \sum_\beta [\sigma_{1}]_{\alpha\beta} P_{\beta\beta}(\kk+\pi).
\label{}
\end{align}

The form of the interaction is still density-density, with the spin up $s$-orbital density multiplying the spin down $s$-orbital density and the same for the $p$-orbitals:
\bea
H_\text{int} &= -|U| \sum_{\mbf{R},\al = s,p} \bar{n}_{\mbf{R},\al,\u} \bar{n}_{\mbf{R},\al,\d} .
\eea

Using the formula for the projector in \Eq{eq:SSHproj}, we find the pairing Hamiltonian
\bea
h_{\al \be}(p) = \int \frac{dk}{2\pi} P_{\al \be}(k+p) P_{\be \al}(k)  = \frac{1}{8} \left(
\begin{array}{cc}
2 + \cos p & \cos p \\
\cos p & 2 + \cos p\\
\end{array}
\right)_{\al \be}
\eea
which has eigenvalues $\eps_\mu(p) = \frac{1}{4}(1+\cos p), \frac{1}{4}$. The many-body energies are $E_\mu(p)/|U| = \eps - \eps_\mu(p) = \frac{1}{4}(1- \cos p), \frac{1}{4}$. The unpaired continuum of states appears at $E/|U| = \eps = \frac{1}{2}$. The spectrum is depicted in \Fig{fig:SSHOAL}(a).

As a final comment, we note that in this symmetry group $pm$ , one may imagine deforming the $s$ and $p$ orbitals at 1a into symmetric and antisymmetric combinations of orbitals at non-maximal Wyckoff positions of $s$ orbitals at 2b. This deformation involves a unitary transformation that combines $s$ and $p$ orbitals at position ${\bf R}$ into symmetric and antisymmetric combinations.  While the orbitals now obey the uniform pairing condition, the density-density interactions are not invariant under this transformation.  However, the following interaction
\bea
H &= -\dfrac{|U|}{2} \sum_{\mbf{R},\sigma = \u,\d} [\bar{n}_{\mbf{R},s,\sigma} + \bar{n}_{\mbf{R},p,\sigma}]^2
\eea is invariant, though we are currently unable to treat this interaction with the eta-pairing operator.

\subsubsection{$p3$ Compact Obstructed atomic Insulator}
\label{app:p3OAL}

\begin{figure*}
\includegraphics[height=.3\textwidth]{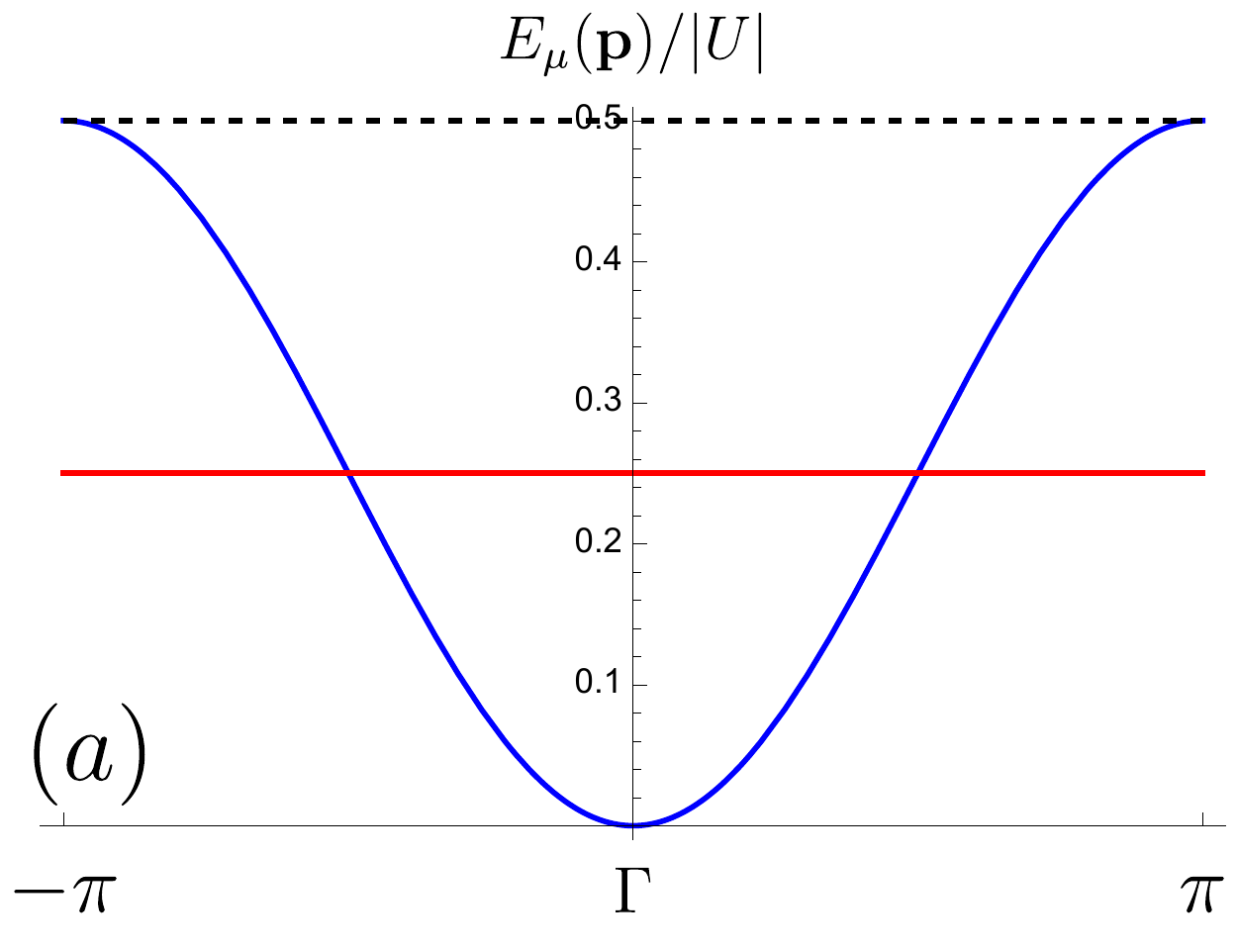}\qquad
\includegraphics[height=.3\textwidth]{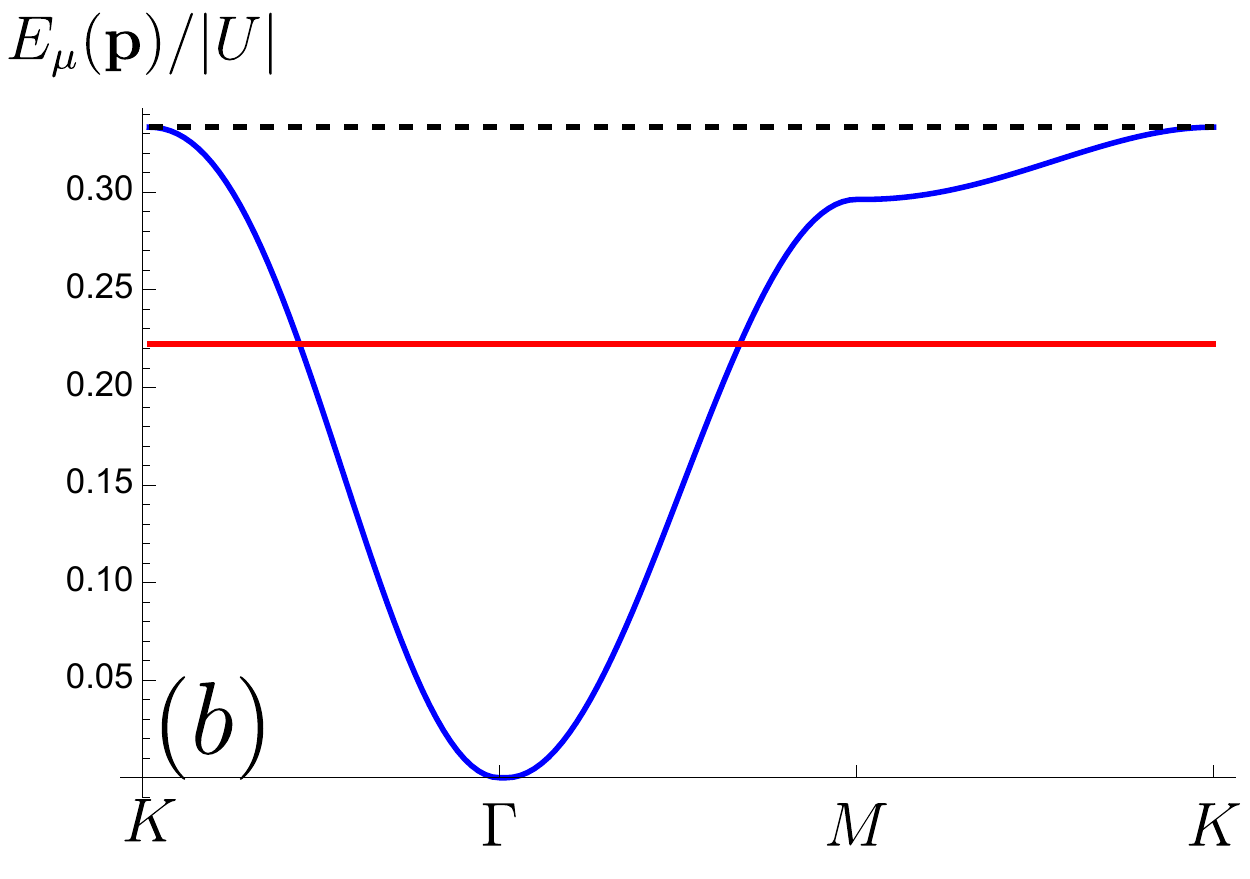}
\caption{The many-body Cooper pair/density excitation spectrum $E_\mu(\mbf{p})= |U|(\eps - \eps_\mu(\mbf{p}))$ is shown for $(a)$ the 1D SSH-like model in \App{app:SSHchain} and $(b)$ the 2D compact obstructed atomic limit in \App{app:p3OAL}. The blue band is the quadratic gapless mode guaranteed by the $su(2)$ eta-pairing symmetry. The red band is perfectly flat for both models and is singly degenerate at $\frac{E}{|U|} = \frac{1}{4}$ in $(a)$ and doubly degenerate at $\frac{E}{|U|} = \frac{2}{9}$ in $(b)$. The dashed black line shows the macroscopically degenerate continuum of unpaired states at $E = \eps |U|$. The flatness of the Leggett modes (in red) is due to the simple, low-harmonic structure of the eigenstates. See the Cooper pair spectrum in the Main Text for nontrivially dispersing Leggett modes. }
\label{fig:SSHOAL}
\end{figure*}

A second compact OAI model with the uniform pairing condition and flat bands was first presented in \Ref{2022PhRvL.128h7002H} in the wallpaper group $p3$. The Hamiltonian is built from $s$ and $p_x, p_y$ orbitals on the 1a position (the unit cell center) which form the $C_3$ irreps $A, {}^1E, {}^2E$. As constructed in \Ref{2022PhRvL.128h7002H}, the single valence band of the Hamiltonian forms a compact obstructed atomic limit, much like the obstructed phase of the SSH chain in \App{app:SSHchain}, and is indicated by real space invariants. Thus $\eps = \frac{N_f}{N_{orb}} = \frac{1}{3}$. The valence band eigenstate is
\bea
U(\mbf{k}) = \frac{1}{3} \bpm 1 \\ 1 \\1 \epm +  \frac{1}{3} \bpm 1 \\ e^{- \frac{2\pi i}{3}} \\ e^{ \frac{2\pi i}{3}} \epm e^{i (k_1 + k_2)} +  \frac{1}{3} \bpm 1 \\ e^{\frac{2\pi i}{3}} \\ e^{-\frac{2\pi i}{3}} \epm e^{i k_2} \ .
\eea
We can calculate the pairing Hamiltonian analytically from $P(\mbf{k}) = U(\mbf{k}) U^\dag(\mbf{k})$:
\bea
h_{\al \be}(\mbf{p}) &= \int \frac{dk_1 dk_2}{(2\pi)^2} P_{\al \be}(\mbf{p}+\mbf{k}) P_{\be \al}(\mbf{k}) = \frac{1}{9} \delta_{\al \be} + \frac{2}{81} (\cos(p_1 + p_2) + \cos p_1 + \cos p_2) \\
h(\mbf{p}) &= \frac{2}{81} (\cos(p_1 + p_2) + \cos p_1 + \cos p_2) \bpm 1 & 1 & 1 \\ 1 & 1 & 1 \\ 1 & 1 & 1 \\  \epm + \frac{1}{9}\bpm 1 & & \\ & 1 & \\
& & 1 \\ \epm
\eea
which has eigenvalues $\eps_\mu(\mbf{p}) = \frac{1}{27}(3 + 2 \cos (p_1+p_2) + 2 \cos p_1 + 2 \cos p_2), \frac{1}{9},  \frac{1}{9}$. The many-body spectrum $E_\mu(\mbf{p})/|U| = \frac{1}{3} - \eps_\mu(\mbf{p})$ is shown in \Fig{fig:SSHOAL}(b) along with the un-paired continuum at $E/|U| = \eps = \frac{1}{3}$.

\subsubsection{$p6mm$ Fragile $S$-matrix Bands}
\label{eq:kagomecooper}

Finally, we construct a perfectly flat fragile band in symmetry group $p6mm$ using the $S$-matrix method introduced in \Ref{2021NatPh..18..185C}. To this end, we choose a 1D irrep (only one to enforce UPC) and place there an $s$-orbital, as we show in App.~\ref{eq:CPsymmetry} the Cooper pair transforms like an electronic $s$-orbital under the crystalline symmetries.  As the set of bands induced from an atomic position is trivial, to get a fragile set of bands one requires the induced band representation decomposes into two sets, one of them fragile.  As shown in \Tab{tab:2Dcooperpairs}, the Cooper pair band structure we desire can be induced by placing any 1D irrep (for instance an $s$ orbital) at the 3c (kagome) positions. The uniform pairing condition is guaranteed because the orbitals form an irrep of a single Wyckoff position. To obtain flat bands, we use the $S$-matrix construction define the two sublattices as $L = A_1^{3c}$ with $N_L = 3$ orbitals per unit cell and  $\tilde{L} = A_1^{1a}$ with $N_{\tilde{L}} = 1$ orbital per unit cell. The $A_1$ irrep denotes the $s$ orbital representation at the 1a and 3c positions (see the Bilbao Crystallographic Server \url{https://www.cryst.ehu.es/cgi-bin/cryst/programs/mbandrep.pl}). The momentum space irreps of the two sublattices are
\bea
\mathcal{B}_L &= A_1^{3c} \uparrow 6mm = \Gamma_1 \oplus \Gamma_5 + K_1 \oplus K_3 + M_1 \oplus M_3 \oplus M_4 \\
\mathcal{B}_{\tilde{L}} &= A_1^{1a} \uparrow 6mm = \Gamma_1 + K_1 + M_1 \ . \\
\eea
In the bipartite $S$-matrix construction, hoppings are only allowed between the $L$ and $\tilde{L}$ sublattices resulting in $N_f = N_L - N_{\tilde{L}} = 3- 1  = 2$ flat bands. The momentum space irreps of the flat bands are\cite{2021NatPh..18..185C}
\bea
\mathcal{B}_L \ominus \mathcal{B}_{\tilde{L}} = \Gamma_5 + K_3  + M_3 \oplus M_4
\eea
which is fragile topological \cite{rsis}. To be concrete, we choose hoppings such that the single-particle electron Hamiltonian is
\bea
\label{eq:tildeh6mm}
\tilde{h}(\mbf{k}) = \bpm 0 & S^\dag(\mbf{k}) \\ S(\mbf{k})& 0_{3\times 3} \epm, \qquad S(\mbf{k}) = t \bpm \cos \mbf{k} \cdot \mbf{a}_1/2 \\ \cos \mbf{k} \cdot (\mbf{a}_1+\mbf{a}_2)/2 \\ \cos \mbf{k} \cdot \mbf{a}_2/2 \epm = t \bpm \cos k_1/2 \\ \cos (k_1+k_2)/2 \\ \cos k_2/2 \epm \\
\eea
where $\mbf{a}_1 = (0,1), \mbf{a}_2 = C_3 \mbf{a}_1$ are the lattice vectors and $\mbf{a}_1/2, (\mbf{a}_1+\mbf{a}_2)/2, \mbf{a}_2/2$ are the sites of the 2c position. The projector into the flat bands of $\tilde{h}(\mbf{k})$ is given by
\bea
\label{eq:projfrag}
P(\mbf{k}) &=  \bpm 0 & 0 & 0 & 0 \\ 0 & 2 + \cos k_2 + \cos (k_1 +k_2) & - 2 \cos \frac{k_1}{2} \cos \frac{k_1+k_2}{2} & - \cos \frac{k_1}{2} \cos \frac{k_2}{2} \\ 0 & - 2 \cos \frac{k_1}{2} \cos \frac{k_1+k_2}{2} & 2 + \cos k_1 + \cos k_2 & - 2 \cos \frac{k_2}{2} \cos \frac{k_1+k_2}{2} \\ 0 & - 2 \cos \frac{k_1}{2} \cos \frac{k_2}{2} & - 2 \cos \frac{k_2}{2} \cos \frac{k_1+k_2}{2} & 2 + \cos k_1 + \cos k_1 + k_2 \\ \epm / (3 + \cos k_1 + \cos (k_1 + k_2) + \cos k_2 )
\eea
which is identically zero on the $1a$ position and obeys
\bea
\int \frac{dk_1 dk_2}{(2\pi)^2} P_{\al \al}(\mbf{k}) = \bpm 0 & & & \\
 & \frac{2}{3} & & \\
 & & \frac{2}{3} &  \\
 & & & \frac{2}{3} \\
\epm
\eea
which shows that the uniform pairing condition is satisfied on the 3c position, as guaranteed by \App{app:UPC}, with $\eps = \frac{N_f}{N_L} = \frac{2}{3}$.

\begin{figure*}
\includegraphics[height=.3\textwidth]{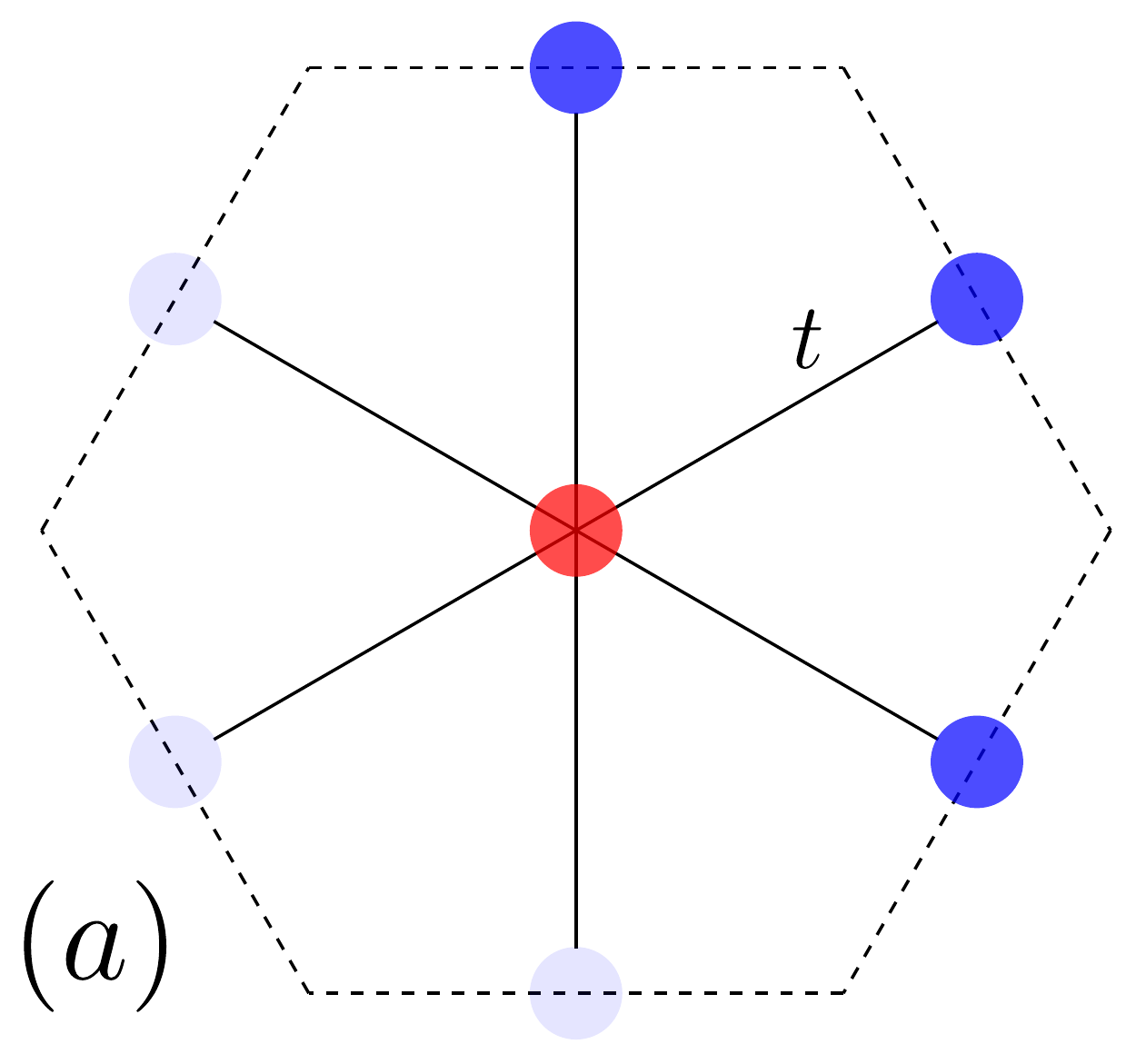}
\includegraphics[height=.3\textwidth]{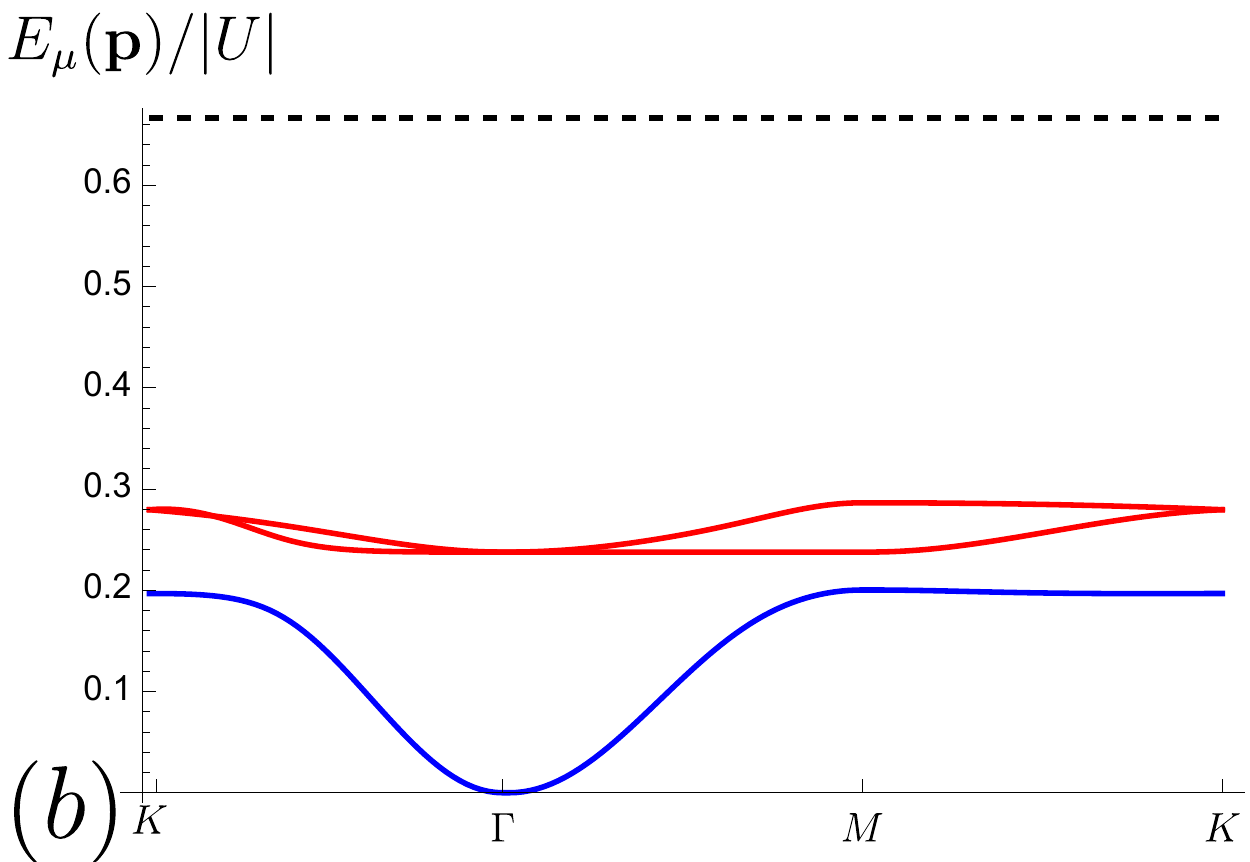}
\caption{$(a)$ Bipartite hoppings $t$ of the single-particle electron Hamiltonian \Eq{eq:tildeh6mm} resulting in two-fold degenerate fragile flat bands. The three $s$ orbitals at the 3c position are blue, and the one $s$ orbital at the 1a position is marked in red. $(b)$ The many-body Cooper pair/Goldstone mode spectrum $E_\mu(\mbf{p})= |U|(\eps - \eps_\mu(\mbf{p}))$ is shown. The blue band is the quadratic gapless mode guaranteed by the $su(2)$ eta-paiting symmetry. The two red bands are fragile topological (see Eq.~\ref{eq:cooperPairP6mmReps}) and are gapped from the low-energy quadratic band. The the dashed black line shows the macroscopically degenerate continuum of unpaired states at $E = \eps |U|$.}
\label{fig:6mmfrag}
\end{figure*}

We are unable to compute $h_{\al \be}(\mbf{p})$ analytically from the projectors in \Eq{eq:projfrag}, but it is very simple to compute the spectrum numerically.  The lattice model and Cooper pair band structure are depicted in Fig.~\ref{fig:6mmfrag}.  The irreps of these bands are precisely those irreps induced by $s$ orbitals at $3c$:

\bea
\mathcal{B}_\text{Cooper} &= A_1^{3c} \uparrow 6mm = \Gamma_1 \oplus \Gamma_5 + K_1 \oplus K_3 + M_1 \oplus M_3 \oplus M_4.
\eea  The Cooper pair band structure is decomposable, and as shown in Fig.~\ref{fig:6mmfrag}(b), the blue band and red bands carry irreps
\bea
\mathcal{B}_1 &= \Gamma_1 + K_1 + M_1 \\
\mathcal{B}_2 &= \Gamma_5 + K_3 + M_3 \oplus M_4 \ . \\
\label{eq:cooperPairP6mmReps}
\eea  The first band carries the $\Gamma_1$ irrep that is the zero-energy Cooper pair, and its global topology is trivial because $\Gamma_1+K_1+M_1$ is the band representation of an $s$ orbital at the 1a position. The second set of bands is fragile topological as predicted from topological quantum chemistry, and cannot induced from local orbitals. 

\subsection{Density Excitations}
\label{app:goldstones}

We now study the (spin) density modes which are in the form
\bea
\label{eq:densitymodes}
\null [H, \gamma^\dag_{\mbf{p}+\mbf{k},m,\sigma}\gamma_{\mbf{k},n,\sigma'}] \ket{GS} &=  \sum_{\mbf{k}'} \gamma^\dag_{\mbf{p}+\mbf{k}',m',\sigma}\gamma_{\mbf{k}',n',\sigma'}  \ket{GS} [\mathcal{R}^{\sigma \sigma'}(\mbf{p})]_{\mbf{k}'m'n',\mbf{k}mn} \ .
\eea
The eigenstates obtained from \Eq{eq:densitymodes} are charge-neutral, so they contribute to the partition function at fixed particle-number.

To compute $\mathcal{R}^{\sigma \sigma'}(\mbf{p})$, we need the identities from \App{app:charge1}:
\bea
\null [\bar{S}^z_{\mbf{q},\al},\gamma_{\mbf{k},n,\sigma}] &= -\frac{1}{\sqrt{\mathcal{N}}} s^z_\sigma \sum_m M^{nm}_{\sigma,\al}(\mbf{k}-\mbf{q},\mbf{q})  \gamma_{\mbf{k} -\mbf{q},m,\sigma} \\
\null [H, \gamma_{\mbf{k}, n, \sigma}] &= \frac{\eps |U|}{2} \gamma_{\mbf{k}, n, \sigma} - \frac{|U|}{2} \sum_{\mbf{q} \al m} 2  \frac{1}{\sqrt{\mathcal{N}}} s^z_\sigma M^{nm}_{\sigma,\al}(\mbf{k}-\mbf{q},\mbf{q})  \gamma_{\mbf{k}-\mbf{q},m,\sigma} \bar{S}^z_{-\mbf{q},\al}
\eea
and as usual, the second term in $[H, \gamma_{\mbf{k}, n, \sigma}]$ annihilates the groundstate. The scattering matrix $\mathcal{R}$ is obtained from a direct calculation:
\bea
\null [H, \gamma^\dag_{\mbf{p}+\mbf{k},m,\sigma}\gamma_{\mbf{k},n,\sigma'}] \ket{GS} &=  \gamma^\dag_{\mbf{p}+\mbf{k},m,\sigma}  [H, \gamma_{\mbf{k},n,\sigma'}] \ket{GS}  + [H, \gamma^\dag_{\mbf{p}+\mbf{k},m,\sigma}] \gamma_{\mbf{k},n,\sigma'} \ket{GS}  \\
 &= \eps |U| \gamma^\dag_{\mbf{p}+\mbf{k},m,\sigma}  \gamma_{\mbf{k},n,\sigma'} \ket{GS}  + |U| \frac{1}{\sqrt{\mathcal{N}}}  \sum_{\mbf{q} \al m'} s^z_\sigma \gamma^\dag_{\mbf{p}+\mbf{q}+\mbf{k},m',\sigma} M^{m'm}_{\sigma,\al}(\mbf{k}+\mbf{p},\mbf{q}) [\bar{S}^z_{-\mbf{q},\al}, \gamma_{\mbf{k},n,\sigma'}] \ket{GS} \\
  &= \eps |U| \gamma^\dag_{\mbf{p}+\mbf{k},m,\sigma}  \gamma_{\mbf{k},n,\sigma'} \ket{GS}  \\
  &\qquad - |U| \frac{1}{\mathcal{N}}  s^z_\sigma  s^z_{\sigma'} \sum_{\mbf{q} \al m' n'} \gamma^\dag_{\mbf{p}+\mbf{q}+\mbf{k},\sigma,m'} \gamma_{\mbf{k}+\mbf{q},\sigma',n'} M^{m'm}_{\sigma,\al}(\mbf{k}+\mbf{p},\mbf{q})  M^{nn'}_{\sigma',\al}(\mbf{k}+\mbf{q},-\mbf{q})  \ket{GS} \\
\eea
from which we identify the scattering matrix
\bea
\null [\mathcal{R}^{\sigma \sigma'}(\mbf{p})]_{\mbf{k}'m'n',\mbf{k}mn} &= \eps |U| \delta_{\mbf{k}\mbf{k}'} \delta_{mm'} \delta_{nn'} - s^z_\sigma  s^z_{\sigma'} |U| \frac{1}{\mathcal{N}} \sum_\al M^{m'm}_{\sigma,\al}(\mbf{k}+\mbf{p},\mbf{k}'-\mbf{k})  M^{nn'}_{\sigma',\al}(\mbf{k}',\mbf{k}-\mbf{k}') \\
&= \eps |U| \delta_{\mbf{k}\mbf{k}'} \delta_{mm'} \delta_{nn'} - s^z_\sigma  s^z_{\sigma'} |U| \frac{1}{\mathcal{N}} \sum_\al U^{\sigma *}_{m' \al}(\mbf{p}+\mbf{k}') U^\sigma_{m \al}(\mbf{p}+\mbf{k}) U^{\sigma' *}_{n \al}(\mbf{k}) U^{\sigma' }_{n' \al}(\mbf{k}') \\
&= \eps |U| \delta_{\mbf{k}\mbf{k}'} \delta_{mm'} \delta_{nn'} - s^z_\sigma  s^z_{\sigma'} |U| \frac{1}{\mathcal{N}} \sum_\al U^{\sigma *}_{m' \al}(\mbf{p}+\mbf{k}') U^{\sigma' }_{n' \al}(\mbf{k}') \ U^\sigma_{m \al}(\mbf{p}+\mbf{k}) U^{\sigma' *}_{n \al}(\mbf{k}) \ . \\
\eea
Let us recall the definition from \Eq{eq:nontrivtermMM},
\bea
\mathcal{U}^{\sigma \sigma'}_{\mbf{k}mn,\al}(\mbf{p}) &= U^{\sigma *}_{m \al}(\mbf{p}+\mbf{k}) U^{\sigma' *}_{n \al}(-\mbf{k}) \ . 
\eea
In the Cooper pair case (see discussion after \Eq{eq:nontrivtermMM}), the anti-commutation of the fermions, schematically $\gamma^\dag_{i\u} \gamma^\dag_{j\u} + \gamma^\dag_{j\u} \gamma^\dag_{i\u} =0$, prevented nontrivial excitations of nonzero spin. In the density excitation case, $\gamma^\dag_{i\u} \gamma_{i \d} + \gamma^\dag_{j\u} \gamma_{i \u} \neq 0$, so nontrivial spin $\pm1$ excitations are allowed. Enumerating the possible choices of $\sigma, \sigma'$ in \Eq{eq:densitymodes} gives the effective Hamiltonians
\bea
\gamma^\dag_{\mbf{p}+\mbf{k},m,\u}\gamma_{\mbf{k},n,\u}: \qquad \mathcal{R}^{\u \u}(\mbf{p}) &= \eps |U| \mathbb{1} - |U| \frac{1}{\mathcal{N}} \mathcal{U}_{\u \d}(\mbf{p}) \mathcal{U}_{\u \d}^\dag(\mbf{p}) \\
\gamma^\dag_{\mbf{p}+\mbf{k},m,\d}\gamma_{\mbf{k},n,\d}: \qquad \mathcal{R}^{\d \d}(\mbf{p}) &= \eps |U| \mathbb{1} - |U| \frac{1}{\mathcal{N}} \mathcal{U}_{\d \u}(\mbf{p}) \mathcal{U}_{\d \u}^\dag(\mbf{p}) \\
\gamma^\dag_{\mbf{p}+\mbf{k},m,\u}\gamma_{\mbf{k},n,\d}: \qquad \mathcal{R}^{\u \d}(\mbf{p}) &= \eps |U| \mathbb{1} + |U| \frac{1}{\mathcal{N}} \mathcal{U}_{\u \u}(\mbf{p}) \mathcal{U}_{\u \u}^\dag(\mbf{p}) \\
\gamma^\dag_{\mbf{p}+\mbf{k},m,\d}\gamma_{\mbf{k},n,\u}: \qquad \mathcal{R}^{\d \u}(\mbf{p}) &= \eps |U| \mathbb{1} + |U| \frac{1}{\mathcal{N}} \mathcal{U}_{\d \d}(\mbf{p}) \mathcal{U}_{\d \d}^\dag(\mbf{p}) \ . \\
\eea
Note that the first two lines are spin 0 operators and the $\mathcal{U}\mathcal{U}^\dag$ term comes with an overall minus, while the last two lines are spin $\pm 1$ and the $\mathcal{U}\mathcal{U}^\dag$ term comes with an overall plus. Because $\mathcal{U}\mathcal{U}^\dag$ is positive semi-definite, we learn that the spin $0$ excitations are always at or below the two-electron gap $\eps |U|$, whereas the spin $\pm 1$ excitations are at or above the $\eps |U|$. This leads to a simple phenomenological prediction: the spin susceptibility is exponentially suppressed at temperatures below the spin gap $\eps|U|$.

By the same argument in \Eq{eq:lowrank}, the nonzero eigenvalues of $\mathcal{U}\mathcal{U}^\dag$ are determined by the effective single-particle Hamiltonians
\bea
\label{eq:harrows}
\mathcal{R}^{\u \u}: \quad h_{\u \u}(\mbf{p}) &= h(\mbf{p}) \\
\mathcal{R}^{\d \d}: \quad h_{\d \d}(\mbf{p}) &= \frac{1}{\mathcal{N}} \sum_{\mbf{k}} P^{\d}_{\al \be}(\mbf{p}+\mbf{k}) P^{\u}_{\al \be}(-\mbf{k}) =  \frac{1}{\mathcal{N}} \sum_{\mbf{k}} P^{\u}_{\al \be}(\mbf{p}+\mbf{k}) P^{\d}_{\al \be}(-\mbf{k}) \\
&= h(\mbf{p}) \\
\mathcal{R}^{\u \d}: \quad h_{\u \d}(\mbf{p}) &= \frac{1}{\mathcal{N}} \sum_{\mbf{k}mn} U^{\u}_{m \al}(\mbf{p}+\mbf{k}) U^{\u}_{n \al}(-\mbf{k})  U^{\u *}_{m \be}(\mbf{p}+\mbf{k}) U^{\u *}_{n \be}(-\mbf{k})  =\frac{1}{\mathcal{N}} \sum_{\mbf{k}} P^{\u}_{\al \be}(\mbf{p}+\mbf{k})  P^{\d}_{\be \al}(\mbf{k})  \\
&=  \frac{1}{\mathcal{N}} \sum_{\mbf{k}} P_{\al \be}(\mbf{p}+\mbf{k}) P_{\be\al}(-\mbf{k})^*  \\
\mathcal{R}^{\d \u}: \quad h_{\d \u}(\mbf{p}) &= \frac{1}{\mathcal{N}} \sum_{\mbf{k}} P^{\d}_{\al \be}(\mbf{p}+\mbf{k})  P^{\d}_{\al \be}(-\mbf{k}) = h_{\u\d}(-\mbf{p})^* \\
\eea
so we see that the two branches of spin $0$ Goldstone modes have the same spectrum as the spin 0 Cooper pair, $h(\mbf{p})$ in \Eq{eq:CPham}. This is evidence of an enlarged symmetry group, which we will explain in forthcoming work. In \Eq{eq:harrows}, the Hamiltonians $h^{\u\d}(\mbf{p})$ of the spin $\pm1$ density excitations are related by $\mathcal{T}$ which flips the spins. If the single-particle bands have $SU(2)$ spin symmetry, then $P(\mbf{k}) = P(-\mbf{k})^*$ and $h_{\u \d}(\mbf{p}) = h(\mbf{p})$. This symmetry will manifest itself in the excitation spectrum, where there will be particle-hole symmetric excitations about the flat band continuum at $E = \eps |U|$.  Otherwise, the spectrum of $h_{\u \d}(\mbf{p})$ is generically different from that of $h(\mbf{p})$. Because our interest in this work is the low-energy excitations, we postpone a detailed study of $h_{\u \d}(\mbf{p})$ to future work.

We now write down the spin-0 density operators as in \Eq{eq:etap}:
\bea
\rho_{\mbf{p},\mu,\sigma} &= \frac{1}{\sqrt{\mathcal{N} \eps_\mu(\mbf{p})}} \sum_{\mbf{k}\al mn} \gamma^\dag_{\mbf{p}+\mbf{k},m,\sigma}\gamma_{\mbf{k},n,\sigma} \mathcal{U}^{\sigma,-\sigma}_{\mbf{k}mn,\al}(\mbf{p}) u^\al_{\mu}(\mbf{p}) \\
&= \frac{1}{\sqrt{\mathcal{N} \eps_\mu(\mbf{p})}} \sum_{\mbf{k}\al mn} \gamma^\dag_{\mbf{p}+\mbf{k},m,\sigma}\gamma_{\mbf{k},n,\sigma} U^{\sigma *}_{m \al}(\mbf{p}+\mbf{k}) U^{-\sigma, *}_{n \al}(-\mbf{k})  u^\al_{\mu}(\mbf{p}) \\
&= \frac{1}{\sqrt{\mathcal{N}}} \sum_{\mbf{k}\al} \frac{u^\al_{\mu}(\mbf{p})}{\sqrt{\eps_\mu(\mbf{p})}} \bar{c}^\dag_{\mbf{p}+\mbf{k},\al,\sigma}\bar{c}_{\mbf{k},\al,\sigma}  \\
&= \frac{1}{\sqrt{\mathcal{N}}} \sum_{\mbf{R}\al} \frac{u^\al_{\mu}(\mbf{p})}{\sqrt{\eps_\mu(\mbf{p})}}  e^{-i \mbf{p} \cdot (\mbf{R}+\mbf{r}_\al)} \bar{n}_{\mbf{R},\al,\sigma} \ . \\
\eea
To compute the normalized states, we use Wick's theorem. To start, we require the correlator
\bea
\braket{z| \gamma^\dag_{\mbf{k}',n',\sigma} \gamma_{\mbf{q}+\mbf{k}',m',\sigma} \gamma^\dag_{\mbf{p}+\mbf{k},m,\sigma}\gamma_{\mbf{k},n,\sigma}|z} &= \frac{\delta_{\mbf{p}\mbf{q}}}{(1+|z|^2)^2} \lp |z|^4 \delta_{\mbf{p},0} \delta_{n'm'} \delta_{nm} - 0 + |z|^2 \delta_{\mbf{k},\mbf{k}'} \delta_{nn'} \delta_{mm'} \rp \braket{z|z} \ . \\
\eea
The presence of the $\delta_{\mbf{p},0}$ term is similar to \Eq{eq:CPnormcorr}. Following the same steps as in \Eq{eq:etapnorm} and using \Eq{eq:calUupc}, we find
\bea
\label{eq:etarhonorm}
\bra{z}\rho^\dag_{\mbf{q},\nu} \rho_{\mbf{p},\mu}\ket{z} &= \delta_{\mbf{p},\mbf{q}} \frac{(1+|z|^2)^{N_f \mathcal{N}-2} }{\mathcal{N} \sqrt{\eps_\mu(\mbf{p})\eps_\nu(\mbf{q})}} \sum_{\mbf{k}\mbf{k}', \al \be} \mathcal{U}^{\sigma,-\sigma}_{\mbf{k}'m'n',\be}(\mbf{q})^* u^{\be*}_\nu(\mbf{q}) \lp  |z|^4 \delta_{n'm'} \delta_{\mbf{p},0} \delta_{nm} + |z|^2 \delta_{\mbf{k}\mbf{k}'} \delta_{nn'} \delta_{mm'} \rp \mathcal{U}^{\sigma,-\sigma}_{\mbf{k}mn,\al}(\mbf{p}) u^\al_\mu(\mbf{p}) \\
&= \delta_{\mbf{p},\mbf{q}} \frac{(1+|z|^2)^{N_f \mathcal{N}-2} }{\sqrt{\eps_\mu(\mbf{p})\eps_\nu(\mbf{p})}} \sum_{\al \be} u^{\be*}_\nu(\mbf{p}) \lp |z|^4 \delta_{\mbf{p},0} \mathcal{N} \eps^2 +|z|^2 h_{\be\al} (\mbf{p})\rp u^\al_\mu(\mbf{p}) \\
&=  \delta_{\mbf{p},\mbf{q}} \delta_{\mu,\nu} (1+|z|^2)^{N_f \mathcal{N}-2} \lp |z|^4 \delta_{\mbf{p},0} \delta_{\mu,0} N_f \mathcal{N}  + |z|^2 \rp\ . \\
\eea

The case of $\mbf{p}=0,\mu=0$ again corresponds to the symmetry $\bar{N}_\u$ and $\bar{N}_\d$ operators at zero energy. Otherwise, we compute
\bea
\sum_n |z|^{2n} \binom{N_f\mathcal{N}}{n} \bra{n}\rho^\dag_{\mbf{q},\nu} \rho_{\mbf{p},\mu}\ket{n} &= \sum_n |z|^{2n+2} \delta_{\mbf{p},\mbf{q}} \delta_{\mu,\nu}  \binom{N_f\mathcal{N}-2}{n} , \qquad \mbf{p}\neq 0, \mu \neq 0 \\
\bra{n} \rho^\dag_{\mbf{q},\nu} \rho_{\mbf{p},\mu} \ket{n} &= \nu (1-\nu) \delta_{\mbf{p},\mbf{q}} \delta_{\mu,\nu}  \\
\eea
which fixes the normalization of the states.

\section{Spectrum and Topology of the Cooper Pairs}
\label{app:Cooperpairspectrum}

In this section, we study the effective single-particle boson Hamiltonian $h(\mbf{p})$ (\Eq{eq:hpairing} of the Main Text) in detail, first showing clearly the invariance of all many-body observables under the momentum space gauge freedom corresponding to the orbital positions (\App{eq:conventons}). We then prove a number of generic features of the spectrum of $h(\mbf{p})$ (\App{app:spectrumbounds}), demonstrating a unique low-energy branch corresponding to the $s$-wave Cooper pairing channel. Its effective mass is determined by the minimal quantum metric (\App{app:quantummetric}). We then prove the symmetry properties of $h(\mbf{p})$ and determine the many-body symmetry transformations of the Cooper pair (\App{eq:CPsymmetry}). Finally, we use topological quantum chemistry to enumerate the possible band connectivities and discover fragile topological Cooper pair bands (\App{app:bosontopology}).

\subsection{Momentum Space Conventions}
\label{eq:conventons}

Before discussing the spectrum of $h(\mbf{p})$, we discuss the dependence of $h(\mbf{p})$ on the choice of momentum space operator convention in \Eq{eq:cRck}. We start by recalling the single-particle Hamiltonian
\bea
\tilde{H} &= \sum_{\mbf{R}\mbf{R}',\al\be,\sigma} c^\dag_{\mbf{R},\al,\sigma} t^\sigma_{\al \be}(\mbf{R} - \mbf{R}') c_{\mbf{R}',\be,\sigma} = \sum_{\mbf{k}\al \be,\sigma} c^\dag_{\mbf{k},\al,\sigma} \tilde{h}^\sigma_{\al \be}(\mbf{k}) c_{\mbf{k},\be,\sigma} \\
\eea
where the momentum operators are $c^\dag_{\mbf{k},\al,\sigma} = \frac{1}{\sqrt{\mathcal{N}}} \sum_\mbf{R} e^{- i \mbf{k} \cdot (\mbf{R} + \mbf{r}_\al)} c^\dag_{\mbf{R},\al,\sigma}$. The choice of the $\mbf{r}_\al$ term in the momentum space operators is arbitrary but convenient for representations of the space group symmetries \cite{2018Sci...361..246W,2022PhRvL.128h7002H}. We now show that the spectrum of $h(\mbf{p})$ and the many-body operators $\eta^\dag_{\mbf{p},\mu}$ are all invariant under $\mbf{r}_\al \to \mbf{r}_\al + \mbf{x}_\al$. However, under $\mbf{r}_\al \to \mbf{r}_\al + \mbf{x}_\al$, we see nontrivial transformations
\bea
c^\dag_{\mbf{k},\al,\sigma} &\to e^{- i \mbf{k} \cdot \mbf{x}_\al} c^\dag_{\mbf{k},\al,\sigma} \\
\tilde{h}^\sigma_{\al \be}(\mbf{k})  &\to e^{i \mbf{k} \cdot \mbf{x}_\al} \tilde{h}^\sigma_{\al \be}(\mbf{k}) e^{-i \mbf{k} \cdot \mbf{x}_\be} \ .
\eea
If we define the local embedding matrix $[V_\mbf{x}(\mbf{k})]_{\al \be} = e^{-i \mbf{k} \cdot \mbf{x}_\al} \delta_{\al \be}$, then the transformation can be written $\tilde{h}(\mbf{k}) \to V_\mbf{x}^\dag(\mbf{k}) \tilde{h}(\mbf{k}) V_\mbf{x}(\mbf{k})$. Thus the electron eigenvectors $U_\sigma(\mbf{k})$ obey $U_\sigma(\mbf{k}) \to V_\mbf{x}^\dag(\mbf{k}) U_\sigma(\mbf{k})$ and the projectors transform like $\tilde{h}$: $P_\sigma(\mbf{k}) \to V^\dag_\mbf{x}(\mbf{k}) P_\sigma(\mbf{k})V_\mbf{x}(\mbf{k})$. As such, we have
\bea
P^\sigma_{\al \be}(\mbf{k}) \to e^{i \mbf{k} \cdot (\mbf{x}_\al-\mbf{x}_\be)} P^\sigma_{\al \be}(\mbf{k}) \ . \\
\eea
We now consider the transformation of the Cooper pair Hamiltonian:
\bea
h_{\al \be}(\mbf{p}) &\to \frac{1}{\mathcal{N}} \sum_\mbf{k} e^{i (\mbf{p}+\mbf{k}) \cdot (\mbf{x}_\al-\mbf{x}_\be)} e^{i \mbf{k} \cdot (\mbf{x}_\be-\mbf{x}_\al)} P_{\al \be}(\mbf{p}+\mbf{k}) P_{\be \al}(\mbf{k}) \\
&= e^{i \mbf{p} \cdot (\mbf{x}_\al-\mbf{x}_\be)} \frac{1}{\mathcal{N}} \sum_\mbf{k} P_{\al \be}(\mbf{p}+\mbf{k}) P_{\be \al}(\mbf{k}) \\
&= [V^\dag_\mbf{x}(\mbf{p}) h(\mbf{p}) V_\mbf{x}(\mbf{p})]_{\al \be} \ .
\eea
Hence the Cooper pair eigenvectors transform just like the electron eigenvectors: $u_\mu(\mbf{p}) \to V^\dag_\mbf{x}(\mbf{p}) u_\mu(\mbf{p})$. We see that the Cooper pair operators are invariant under the embedding:
\bea
\eta^\dag_{\mbf{p},\mu} &= \frac{1}{\sqrt{\mathcal{N}}} \sum_{\mbf{k}\al} \frac{u^\al_{\mu}(\mbf{p})}{\sqrt{\eps_\mu(\mbf{p})}} \bar{c}^\dag_{\mbf{p}+\mbf{k},\al,\u}\bar{c}^\dag_{-\mbf{k},\al,\d} \\
&\to \frac{1}{\sqrt{\mathcal{N}}} \sum_{\mbf{k}\al} \frac{u^\al_{\mu}(\mbf{p}) e^{i \mbf{p}\cdot \mbf{x}_\al}}{\sqrt{\eps_\mu(\mbf{p})}} e^{-i (\mbf{k}+\mbf{p})\cdot\mbf{x}_\al} \bar{c}^\dag_{\mbf{p}+\mbf{k},\al,\u} e^{i \mbf{k}\cdot\mbf{x}_\al} \bar{c}^\dag_{-\mbf{k},\al,\d} \\
&= \eta^\dag_{\mbf{p},\mu} \\
\eea
It is also obvious that the spectrum of the Cooper pairs, $\eps_\mu(\mbf{p})$, is invariant because the $V_\mbf{x}(\mbf{k})$ conjugate the Cooper pair Hamiltonian. Hence, as expected, all of the many-body observables are invariant under the arbitrary choice of $\mbf{x}_\al$.

\subsection{Cooper Pair Hamiltonian and Spectrum}
\label{app:spectrumbounds}

The fact that the Cooper pair and density excitations are determined by an effective single-particle Hamiltonian $h(\mbf{p})$ is very evocative. It shows that the tightly-bound Cooper pairs behave much like a free boson whose hoppings arise from the quantum geometry, encoded in $P(\mbf{k})$, of the flat bands. We will study the effective single-particle boson Hamiltonian and prove a number of properties about its spectrum and topology.

Let us restate the effective Hamiltonian derived in \Eq{eq:CPham},
\bea
h_{\al \be}(\mbf{p}) = \frac{1}{\mathcal{N}} \sum_\mbf{k} P_{\al \be}(\mbf{p}+\mbf{k}) P_{\be \al}(\mbf{k})
\eea
where $P(\mbf{k})= U(\mbf{k})U^\dag(\mbf{k})$ is the Hermitian projector into the spin-$\u$ electronic flat bands. The $\al,\be$ orbital indices range from $1,\dots, N_{orb}$. In the bipartite crystalline lattice construction, $N_{orb} \to N_L$ is understood to be the number of orbitals in the $L$ sublattice. ($P_{\al \be}(\mbf{k}) = 0$ for $\al$ or $\be \notin L$.)

First we show $h(\mbf{p})$ is Hermitian:
\bea
h^*_{\be \al}(\mbf{p}) = \frac{1}{\mathcal{N}} \sum_\mbf{k} P^*_{\be \al}(\mbf{p}+\mbf{k}) P^*_{\al \be}(\mbf{k}) = \frac{1}{\mathcal{N}} \sum_\mbf{k} P_{\al \be}(\mbf{p}+\mbf{k}) P_{\be \al}(\mbf{k}) = h_{\al \be}(\mbf{p}) \ .
\eea
The eigenvalues of $h(\mbf{p})$ are denoted by $\eps_\mu(\mbf{p})$ where $\mu = 0, \dots, N_{orb} - 1$ will denote the boson band index. We will now show that the spectrum of $h(\mbf{p})$ is bounded by $0 \leq \eps_{\mu}(\mbf{p}) \leq \eps$. Importantly, the many-body energies (see \Eq{eq:lowrank}) of the boson excitations are given by $|U| (\eps - \eps_\mu(\mbf{p}) )$ so the bounds on the eigenvalues of $h(\mbf{p})$ show that the many-body spectrum of the boson bands is between $0$ and $\eps |U|$. Note that the largest eigenvalue of $h(\mbf{p})$ corresponds to the lowest energy in the many-body spectrum. Thus we can think of $|U|\eps_\mu(\mbf{p})$ as the binding energy gained by the pairing relative to the unpaired continuum of states at energy $\eps |U|$.

To show $h(\mbf{p})$ is positive semi-definite, we observe that for an arbitrary vector $v$,
\bea
\sum_{\al \be} v^*_\al h_{\al \be}(\mbf{p}) v_\be &= \frac{1}{\mathcal{N}} \sum_\mbf{k} \sum_{mn,\al\be} v^*_\al U_{\al m}(\mbf{p}+\mbf{k}) U^*_{\be m}(\mbf{p}+\mbf{k})  U_{\be n}(\mbf{k}) U^*_{\al n}(\mbf{k})v_\be \\
&= \frac{1}{\mathcal{N}} \sum_\mbf{k} \sum_{mn}  [\sum_\al U^*_{\al m}(\mbf{p}+\mbf{k})  v_\al U_{\al n}(\mbf{k})]^* \sum_\be U^*_{\be m}(\mbf{p}+\mbf{k}) v_\be U_{\be n}(\mbf{k})  \\
&= \frac{1}{\mathcal{N}} \sum_\mbf{k} \sum_{mn} V^*_{mn}  V_{mn}  \\
&= \frac{1}{\mathcal{N}} \sum_\mbf{k} ||V||_F^2 \geq 0 \ .
\eea
where we defined the $N_f \times N_f$ matrix $V_{mn} = \sum_\be U^*_{\be m}(\mbf{p}+\mbf{k}) v_\be U_{\be n}(\mbf{k})$ and used the Frobenius norm $||V||_F^2 = \sum_{mn} |V_{mn}|^2$. This establishes $\eps_\mu(\mbf{p}) \geq 0$. Alternatively, we observe that $P_{\al \be}(\mbf{p}+\mbf{k}) P_{\be \al}(\mbf{k}) = [P(\mbf{p}+\mbf{k}) \circ P^T(\mbf{k})]_{\al \be}$ is the Hadamard product of two positive semi-definite matrices $P(\mbf{p}+\mbf{k})$ and $P^T(\mbf{k})$. By the Schur product theorem, $P_{\al \be}(\mbf{p}+\mbf{k}) P_{\be \al}(\mbf{k})$ is positive semi-definite, and so $h_{\al \be}(\mbf{p})$ is the sum of positive semi-definite matrices with positive coefficients, and hence is positive-semi definite itself.

To prove $\eps_\mu(\mbf{p}) \leq \eps$, we show that $\eps \mathbb{1} - h(\mbf{p})$ is positive semi-definite. Using the uniform pairing condition, note that
\bea
\label{eq:zerobound}
\null [\eps \mathbb{1} - h(\mbf{p})]_{\al \be} &= \eps \delta_{\al \be} \,- h_{\al \be}(\mbf{p})  = \frac{1}{\mathcal{N}} \sum_\mbf{k} P_{\al \al}(\mbf{k}) \delta_{\be\al} \ - h_{\al \be}(\mbf{p}) \\
&= \frac{1}{\mathcal{N}} \sum_\mbf{k} P_{\al \be}(\mbf{p}+\mbf{k}) \delta_{\be\al} \ - h_{\al \be}(\mbf{p}) \\
&= \frac{1}{\mathcal{N}} \sum_\mbf{k} P_{\al \be}(\mbf{p}+\mbf{k}) (\delta_{\be\al} - P_{\be \al}(\mbf{k})) \\
&= \frac{1}{\mathcal{N}} \sum_\mbf{k} P_{\al \be}(\mbf{p}+\mbf{k}) Q_{\be \al}(\mbf{k}) \\
\eea
where we defined $Q(\mbf{k}) = \mathbb{1} - P(\mbf{k})$. Note that $Q(\mbf{k})$ is also a positive semi-definite matrix (it is a projector), and so $\eps \mathbb{1}- h(\mbf{p})$ is positive semi-definite by the Schur product theorem.

We have shown that $\eps_\mu(\mbf{p}) \leq \eps$. We now prove that this bound is tight by exhibiting the corresponding eigenvector at $\mbf{p}=0$. Claim: there is an eigenvector $u_0^\al(\mbf{p}=0) = 1/\sqrt{N_{orb}}$ with eigenvalue $\eps$. This is simply the normalized uniform eigenvector corresponding to the eta-pairing symmetry $\eta^\dag$. The claim is proven using the uniform pairing condition:
\bea
\label{eq:u0mathcaln}
\null [h(0) u_0]_\al &= \frac{1}{\mathcal{N}} \sum_\mbf{k} \sum_\be P_{\al \be}(\mbf{k}) P_{\be \al}(\mbf{k}) u_0^\be(0) \\
&= \frac{1}{\sqrt{N_{orb}}} \frac{1}{\mathcal{N}} \sum_\mbf{k} \sum_\be P_{\al \be}(\mbf{k}) P_{\be \al}(\mbf{k}) \\
&= \frac{1}{\sqrt{N_{orb}}} \frac{1}{\mathcal{N}} \sum_\mbf{k} [P(\mbf{k})^2]_{\al \al} \\
&= \frac{1}{\sqrt{N_{orb}}} \frac{1}{\mathcal{N}} \sum_\mbf{k} P_{\al \al}(\mbf{k}) \\
&= \eps u^\al_0 \ . \\
\eea
The existence of this state is tied to the $\eta^\dag$ symmetry. The $\mbf{p}=0$ Cooper pair excitation spectrum includes the $\eta^\dag$ operator in its Hilbert space, and  $\eta^\dag$ commutes with the Hamiltonian and therefore has zero energy. Because the many-body energy is $|U|(\eps - \eps_\mu(\mbf{p}))$, there must be a band with $\eps_\mu(0) = \eps$, and as such this excitation is gapless. We refer to this as the $\mu=0$ band. Indeed, the excitation operator defined in \Eq{eq:etap} yields
\bea
\eta^\dag_{\mbf{p}=0,\mu=0} &= \frac{1}{\sqrt{\mathcal{N}}} \sum_{\mbf{k}\al} \frac{1/\sqrt{N_{orb}}}{\sqrt{\eps}} \bar{c}^\dag_{\mbf{k},\al,\u}\bar{c}^\dag_{-\mbf{k},\al,\d} = \frac{1}{\sqrt{N_f\mathcal{N}}} \sum_{\mbf{k}\al} \bar{c}^\dag_{\mbf{k},\al,\u}\bar{c}^\dag_{-\mbf{k},\al,\d} = \frac{1}{\sqrt{N_f \mathcal{N}}} \eta^\dag
\eea
using $\eps = N_f/N_{orb}$. (In the bipartite case, $N_{orb} \to N_L$.)

We now show that the gapless Cooper pair band is unique, i.e. that there are no other zero-energy excitations for all $\mbf{p}$. To do so, we first Fourier transform into real space using \Eq{eq:projdef}, which reads
\bea
P_{\al \be}(\mbf{k}) = \sum_{\mbf{R}} e^{i \mbf{k} \cdot (\mbf{R}+\mbf{r}_\al-\mbf{r}_\be)} p_{\al \be}(\mbf{R}) \ .
\eea
A direct calculation gives
\bea
\label{eq:hprealspace}
h_{\al \be}(\mbf{p}) &= \sum_{\mbf{R}\mbf{R}'} \frac{1}{\mathcal{N}} \sum_\mbf{k} e^{i (\mbf{p}+\mbf{k}) \cdot (\mbf{R}+\mbf{r}_\al-\mbf{r}_\be) + i \mbf{k} \cdot (\mbf{R}'+\mbf{r}_\be-\mbf{r}_\al)} p_{\al \be}(\mbf{R}) p_{\be \al}(\mbf{R}') \\
&= \sum_{\mbf{R}} e^{i \mbf{p} \cdot (\mbf{R}+\mbf{r}_\al-\mbf{r}_\be)} p_{\al \be}(\mbf{R}) p_{\be \al}(-\mbf{R}) \\
&= \sum_{\mbf{R}} e^{i \mbf{p} \cdot (\mbf{R}+\mbf{r}_\al-\mbf{r}_\be)} |p_{\al \be}(\mbf{R})|^2 \\
\eea
using $p^\dag(\mbf{R}) = p(-\mbf{R})$. We recognize the last line of \Eq{eq:hprealspace} as an expression for the matrix elements of the real space hopping matrix $t_{\al \be}(\mbf{R}-\mbf{R}') = |p_{\al \be}(\mbf{R}-\mbf{R}')|^2$. This shows that the effective hoppings felt by the bosons are due to the correlation function $p(\mbf{R})$ of the flat bands. Note that $\text{spec } t_{\al \be}(\mbf{R}-\mbf{R}') = \text{spec }|p_{\al \be}(\mbf{R}-\mbf{R}')|^2 = \{\text{spec } h(\mbf{p}) \forall \mbf{p} \}$. To show that $\eps_0(\mbf{p}=0)$ is the unique largest eigenvalue of the whole band structure of $h(\mbf{p})$, the crucial feature is that the hopping elements are non-negative: $ |p_{\al \be}(\mbf{R}-\mbf{R}')|^2 \geq 0$. Hence we can apply the Perron-Frobenius theorem, which states than a non-negative irreducible matrix has a \emph{unique} largest positive eigenvalue, which we have shown explicitly is $\eps$. Because $|p_{\al \be}(\mbf{R}-\mbf{R}')|^2$ is non-negative, we must also check that $t_{\al\be}(\mbf{R}-\mbf{R}')$ is irreducible to apply the theorem. An irreducible matrix $A$ is one for which $A e \notin e \forall e$ where $e$ is a basis of any coordinate subspace. Physically, irreducibility is equivalent to the lattice being strongly connected, meaning any two orbitals are connected by a sequence of hoppings $|p_{\al \be}(\mbf{R}-\mbf{R}')|^2$. This is \emph{not} the case in a trivial atomic limit where $p_{\al \be}(\mbf{R}-\mbf{R}') \propto \delta_{\mbf{R}\mbf{R}'}$. \Ref{2022PhRvL.128h7002H} proved nonzero lower bounds on $||p(\mbf{R}\neq0)||^2$ in terms of real space invariants \cite{rsis} which are the indices diagnosing obstructed atomic insulators and fragile phases. Excluding certain pathological cases (like a 2D system formed from decoupled 1D chains, or an occupied trivial atomic limit band added with a direct sum into the projector $P(\mbf{k})$), the results of \Ref{2022PhRvL.128h7002H} show that $t_{\al \be}(\mbf{R}-\mbf{R}')$ is strongly connected if the flat bands are in an obstructed, fragile, or stable phase.

We prove a further result on the spectrum of $h(\mbf{p})$. So far we have shown that $0 \leq \eps_\mu(\mbf{p}) \leq \eps$. We now show that the spectral average of the band structure is $\eps^2$:
\bea
\frac{1}{\mathcal{N}N_{L}} \sum_{\mbf{p}\mu} \eps_\mu(\mbf{p}) = \frac{1}{\mathcal{N}N_{L}} \sum_{\mbf{p}} \Tr h(\mbf{p}) = \frac{1}{\mathcal{N}N_{L}} \sum_{\mbf{p} \al} h_{\al \al}(\mbf{p}) = \frac{1}{\mathcal{N}^2N_{L}} \sum_{\mbf{p}\mbf{k} \al} P_{\al \al}(\mbf{p}+\mbf{k})  P_{\al \al}(\mbf{k}) =  \frac{1}{N_{L}} \sum_{\al}  \eps^2 = \eps^2 \ .  \\
\eea
Recalling the many-body energies $E_\mu(\mbf{p}) = |U|(\eps - \eps_\mu(\mbf{p}))$, the many-body density of states $g(E)$ obeys
\bea
\label{eq:dosint}
\int_{0}^{\eps|U|} E g(E) dE = \frac{1}{\mathcal{N}} \sum_{\mbf{p}\mu} E_\mu(\mbf{p})  = |U| N_L (\eps - \eps^2) =  |U| N_f (1 - \eps) \ . \\
\eea
The upper bound of integral in \Eq{eq:dosint} is understood to be just below $\eps|U|$, where there is a thermodynamically large degeneracy of flat density excitation bands (analogous to the unpaired continuum in the Cooper pair sector).

Our last result concerns a simple transformation of the model under $P(\mbf{k}) \to \mathbb{1}-P(\mbf{k})$, which is equivalent to exchanging the projected and unprojected bands. For simplicity, we assume that all orbitals form an irrep of a single Wyckoff position, e.g. we do not use the bipartite construction in what follows. Note that under $P(\mbf{k}) \to \mathbb{1}-P(\mbf{k})$, we have $N_f \to N_{orb} - N_f$ and hence $\eps \to 1-\eps$. Now we compute
\bea
h_{\al \be}(\mbf{p}) &\to \frac{1}{\mathcal{N}} \sum_\mbf{k} (\delta_{\al \be} - P_{\al \be}(\mbf{k}+\mbf{p}))(\delta_{\al \be} - P_{\be \al}(\mbf{k})) \\
&=\delta_{\al \be} - \delta_{\al \be}  \frac{1}{\mathcal{N}} \sum_\mbf{k} P_{\al\al}(\mbf{k}+\mbf{p}) - \delta_{\al \be}  \frac{1}{\mathcal{N}} \sum_\mbf{k} P_{\al\al}(\mbf{k}) + h_{\al \be}(\mbf{p}) \\
&=\delta_{\al \be} - 2 \eps \delta_{\al \be}+ h_{\al \be}(\mbf{p}) \ . \\
\eea
Thus the many-body energies transform as
\bea
E_\mu(\mbf{p}) = |U|(\eps - \eps_\mu(\mbf{p})) \to  |U|\big(1 - \eps - (1 - 2 \eps +\eps_\mu(\mbf{p})) \big) = |U|(\eps - \eps_\mu(\mbf{p})) \\
\eea
and hence are invariant. The energy of the unpaired continuum states does change, however, $\eps|U| \to (1-\eps) |U|$. Because we proved above (\Eq{eq:zerobound}) that the bound state energies $E_\mu(\mbf{p}) \leq \eps |U|$ are below the unpaired continuum, we see in fact that $E_\mu(\mbf{p}) \leq \min \{\eps |U|, (1-\eps) |U|\}$. This follows by taking $P(\mbf{k}) \to \mathbb{1}-P(\mbf{k})$ and using $E_\mu(\mbf{p}) \to E_\mu(\mbf{p})$, $\eps \to 1-\eps$.

\subsection{Cooper Pair mass and Minimal Quantum Metric}
\label{app:quantummetric}

We proved in \App{app:spectrumbounds} that the many-body spectrum contains a gapless mode $E_{\mu=0}(\mbf{p}=0) = 0$. In this section, we use perturbation theory to compute the small $\mbf{p}$ expansion around this point, yielding the low energy behavior of the bosonic excitations. The result we will prove is $\eps_0(\mbf{p}) = \eps - \frac{g_{ij}}{N_{L}} p_i p_j$ where $g_{ij}$ is the minimal quantum metric (introduced in \Ref{2022arXiv220311133H} and explicitly constructed below) and $N_L$ is the number of orbitals where $P(\mbf{k})$ is nonzero. We sum over repeated spatial indices $i,j$ in this section. As a result, the many-body energy is
\bea
E_0(\mbf{p}) = |U|(\eps - \eps_0(\mbf{p})) = |U| \frac{g_{ij}}{N_{L}} p_i p_j + \dots
\eea
which shows the low-lying Goldstone and Cooper pair modes are quadratic. This amounts to a many-body calculation of the Cooper pair mass, which has been previously approximated from the two-body problem\cite{2018PhRvB..98v0511T} and is directly related to the mean-field superfluid weight \cite{2015NatCo...6.8944P,2021arXiv211100807T}.

Because we have proven that $\eps_0(\mbf{p}=0)$ is the unique maximal eigenvalue in \App{app:spectrumbounds}, its small $\mbf{p}$ corrections can be obtained using non-degenerate perturbation theory. At zeroth order, $h(0)$ has eigenvectors $u_\mu$, with $u_0 = 1/\sqrt{N_{L}}$ corresponding to the maximal eigenvalue $\eps_\mu = \eps$. We determine the corrections to $h(0)$ by expanding in $\mbf{p}$:
\bea
h_{\al \be}(\mbf{p}) &= \frac{1}{\mathcal{N}} \sum_\mbf{k} P_{\al \be}(\mbf{k}) P_{\be \al}(\mbf{k}) + p_i \frac{1}{\mathcal{N}} \sum_\mbf{k} \del_i P_{\al \be}(\mbf{k}) P_{\be \al}(\mbf{k}) + \frac{1}{2} p_i p_j \frac{1}{\mathcal{N}} \sum_\mbf{k} \del_{ij} P_{\al \be}(\mbf{k}) P_{\be \al}(\mbf{k}) + \dots  \ . \\
\eea
Denote $h^i_{\al \be} = \del_i h_{\al \be} |_{\mbf{p}=0}$. By time-reversal $h^*_{\al \be}(\mbf{p}) = h_{\al \be}(-\mbf{p})$, we expand to first order and find
\bea
(p_i h^i_{\al \be} )^* &= - p_i h^i_{\al \be}, \\
\eea
so $h^i_{\al \be} = - h^i_{\be \al}$ is anti-symmetric because $(h^i_{\al \be})^* = h^i_{\be \al}$ using Hermiticity. This can also be seen with an integration by parts. This immediately shows that the first order correction to $\eps_0(\mbf{p})$ vanishes because
\bea
u^\dag_0 h^i u_0 &= \frac{1}{N_L} \sum_{\al \be} h^i_{\al \be} = 0 \ . \\
\eea
We now consider the second order correction. There are contributions to the second order correction:
\bea
\label{eq:pertubtwoerms}
\frac{1}{2} p_ip_j \del_{ij} \eps_0(\mbf{p}) &= \frac{1}{2} p_ip_j \lp  2 \Re \sum_{\mu \neq 0} \frac{u^\dag_0 h^i u_\mu u_\mu^\dag h^j u_0}{\eps - \eps_\mu(0)} + \frac{1}{\mathcal{N}} \sum_{\mbf{k} \al \be} \frac{1}{N_L}  \del_{ij} P_{\al \be}(\mbf{k}) P_{\be \al}(\mbf{k}) \rp \equiv \frac{1}{2} p_i p_j (A_{ij} + B_{ij}) \ .
\eea
The first term $A_{ij}$ is the usual second order correction from a linear perturbation $p_i h^i$, and the second term $B_{ij}$ is simply the expectation value of the second order correction in $u_0$. The second term is related to the quantum metric after an integration by parts:
\bea
B_{ij} &= \frac{1}{N_{L}} \frac{1}{\mathcal{N}} \sum_{\mbf{k}\al\be} \del_{ij} P_{\al \be}(\mbf{k}) P_{\be \al}(\mbf{k}) =  -\frac{1}{N_{L}} \frac{1}{\mathcal{N}} \sum_\mbf{k}  \Tr \del_i P(\mbf{k}) \del_j P(\mbf{k}) = - \frac{2}{N_{L}} g_{ij} \preceq 0,  \\
g_{ij} &= \frac{1}{\mathcal{N}} \sum_\mbf{k} g_{ij}(\mbf{k}) , \qquad  g_{ij}(\mbf{k}) = \frac{1}{2} \Tr \del_i P(\mbf{k})\del_j P(\mbf{k})
\eea
using the convention $\Tr \mathcal{G}_{ij}(\mbf{k}) = \Tr P(\mbf{k})\del_i P(\mbf{k}) \del_j P(\mbf{k}) = g_{ij}(\mbf{k}) - \frac{i}{2} f_{ij}(\mbf{k})$. Note that \Ref{2022PhRvL.128h7002H} proves $\Tr P(\mbf{k}) \{\del_{i} P(\mbf{k}) , \del_{j} P(\mbf{k}) \} = \Tr \del_i P(\mbf{k})\del_j P(\mbf{k})$. Next, we observe that $A_{ij} \succeq 0$ because $\eps-\eps_\mu(0) \geq 0$. Thus $A$ and and $B$ have competing contributions to $\del_{ij} \eps_0(\mbf{p})$.

Because the spectrum of $h(\mbf{p})$ is invariant under invariant under $\mbf{r}_\al \to \mbf{r}_\al+ \mbf{x}_\al$,  $\frac{1}{2} p_ip_j \del_{ij} \eps_0(\mbf{p})$ is invariant. However, $A_{ij}$ and $B_{ij}$ are not individually invariant. We now show that there is a unique choice of $\mbf{x}_\al$, up to a constant, where $A_{ij} = 0$ and the quantum metric alone determines $\del_{ij} \eps_0(\mbf{p})$. Because $A_{ij} + B_{ij}$ is independent of $\mbf{x}_\al$, taking $A_{ij} \to 0$ corresponds to the maximal $B_{ij}$ and hence the minimal $g_{ij}$ for all $\mbf{x}_\al$.

First we show that there exists $\mbf{x}_\al$ where $A_{ij} = 0$, and then we show such a choice respects the symmetries of the system. Note that we keep $\mbf{r}_\al$, the physical location of the Wyckoff position, constant.

Returning to \Eq{eq:pertubtwoerms}, we recall that $\eps - \eps_\mu > 0$ and $u^\dag_0 p_i h^i u_\mu u_\mu^\dag p_j h^j u_0$ is nonnegative. Thus $A_{ij} = 0$ iff $u_\mu^\dag h^j u_0 = 0$ for all $\mu \neq 0$. Because the $u_\mu$ eigenvectors form a complete and orthogonal basis, $u_\mu^\dag h^j u_0 = 0$ iff $ h^j u_0 \propto u_0$. Thus $u_0$ must be an eigenvector of $h^i$. For a given choice of orbital locations, this is not guaranteed. However, $h^i$ is not invariant under shifts of orbital location $\mbf{r}_\al \to \mbf{r}_\al + \mbf{x}_\al$. Using $P_{\al \be}(\mbf{k}) \to e^{i \mbf{k} \cdot (\mbf{x}_\al-\mbf{x}_\be)} P_{\al \be}(\mbf{k})$, the transformation is
\bea
h^i_{\al \be} =  \frac{1}{\mathcal{N}} \sum_\mbf{k} \del_i P_{\al \be}(\mbf{k}) P_{\be \al}(\mbf{k}) &\to \frac{1}{\mathcal{N}} \sum_\mbf{k} \del_i P_{\al \be}(\mbf{k}) P_{\be \al}(\mbf{k}) + i (x^i_\al - x^i_\be)  \frac{1}{\mathcal{N}} \sum_\mbf{k} P_{\al \be}(\mbf{k}) P_{\be \al}(\mbf{k}) \\
&= h^i_{\al\be} + i  (x^i_\al - x^i_\be) h_{\al \be}(0)
\eea
and $x^i_\al$ denotes the $i$th spatial component of $\mbf{x}_\al$. Note that $h_{\al \be}(0)$ is Hermitian and real, so $ i (x^i_\al - x^i_\be) h_{\al \be}(0)$ is Hermitian and anti-symmetric, as is $h^i_{\al\be}$. If $\mbf{x}_\al = \mbf{x}$ is a constant shift, then $h^i_{\al\be} \to h^i_{\al\be}$.

Our task is to look for $\mbf{x}_\al$ such that $u_0$ is an eigenvector of $h^i$ (which guarantees that $A_{ij}=0$). The eigenvalue equation is
\bea
\label{eq:eigevalcond}
\sum_\be ( h^i_{\al\be} + i  (x^i_\al - x^i_\be) h_{\al \be}(0) ) u^\be_0 = \la u^\al_0 \\
\eea
where $\la$ is the eigenvalue. First we show that if \Eq{eq:eigevalcond} condition holds, then $\la = 0$. Contracting both sides with $u^\al_0 = 1/\sqrt{N_L}$ gives
\bea
\la &= \frac{1}{N_{L}} \sum_{\al \be} h^i_{\al\be} + i (x^i_\al - x^i_\be) h_{\al \be}(0) = 0 \\
\eea
because of anti-symmetry in $\al\be$. Then \Eq{eq:eigevalcond} reads
\bea
0 &= \sum_\be h^j_{\al\be} + i  \sum_\be (x^j_\al - x^j_\be) h_{\al \be}(0) \\
0 &= \sum_\be h^j_{\al\be} + i x^j_\al \sum_\be h_{\al \be}(0) - i  \sum_\be  h_{\al \be}(0)  x^j_\be \\
\sum_\be  h_{\al \be}(0)  x^j_\be &= -i \sum_\be h^j_{\al\be} + \eps x^j_\al  \\
\null [ (h(0) -\eps) x^j]_\al &= -i \sum_\be h^j_{\al\be}
\eea
where we used the uniform pairing condition $\sum_\be h_{\al \be}(0) = \eps$. We now define the vector $-i \sum_\be h^j_{\al\be}  = h^j_\al$ which is real and obeys $\sum_\al h^j_\al = 0$, so it is orthogonal to the uniform eigenvector $u_0$. Because $h(0)$ has a unique eigenvector $u_0$ with eigenvalue $\eps$ (see \Eq{eq:u0mathcaln}), $(h(0) -\eps)$ is invertible up to the $u_0$ eigenspace. This means $x^j_\al$ is uniquely determined up to constant shifts because $u_0^\al \propto 1$. Thus we obtain $x^j_\al$ up to constant shifts by applying the pseudo-inverse\cite{prasolov1994problems} $(h(0) -\eps)^+$:
\bea
\label{eq:xalphafixed}
x^j_\al &= [-i (h(0) -\eps)^+ h^j u_0\sqrt{N_L}]_\al, \qquad [h^j u_0\sqrt{N_L}]_\al = \sum_\be h^j_{\al \be} \\
\mbf{x}_\al &= [i (\eps-h(0))^+ (\pmb{\nabla}h(\mbf{p})|_{\mbf{p}=0}) u_0\sqrt{N_L}]_\al  \ . \\
\eea
The pseudo-inverse of a Hermitian matrix is very simple. For a hermitian matrix $h$ with eigenvalues $\eps_\mu$ and eigen-decomposition $h = \sum_\mu \eps_\mu p_\mu$, the pseudo-inverse is $h^+ = \sum_\mu \eps^+_\mu p_\mu$ where $p_\mu$ is the projector onto the $\eps_\mu$ eigenspace, $\eps^+ = 1/\eps$ if $\eps \neq 0$, and $\eps^+ = 0$ otherwise. Hence $hh^+ = h^+h$ acts as the identity on the nonzero eigenspaces and 0 otherwise.

Having proven in \Eq{eq:xalphafixed} that orbitals shifts $\mbf{x}_\al$ always exist for which $A_{ij}=0$, we now show that $\mbf{x}_\al$ preserve the space group symmetries $g \in G$. In particular, this will show that orbitals at high-symmetry Wyckoff positions, whose location is fixed by symmetry, are necessarily the choices where $A_{ij} = 0$. In this case, the minimal quantum metric is the quantum metric evaluated in the $\mbf{r}_\al$ convention \Eq{eq:cRck}.

\begin{figure*}
\includegraphics[height=.3\textwidth]{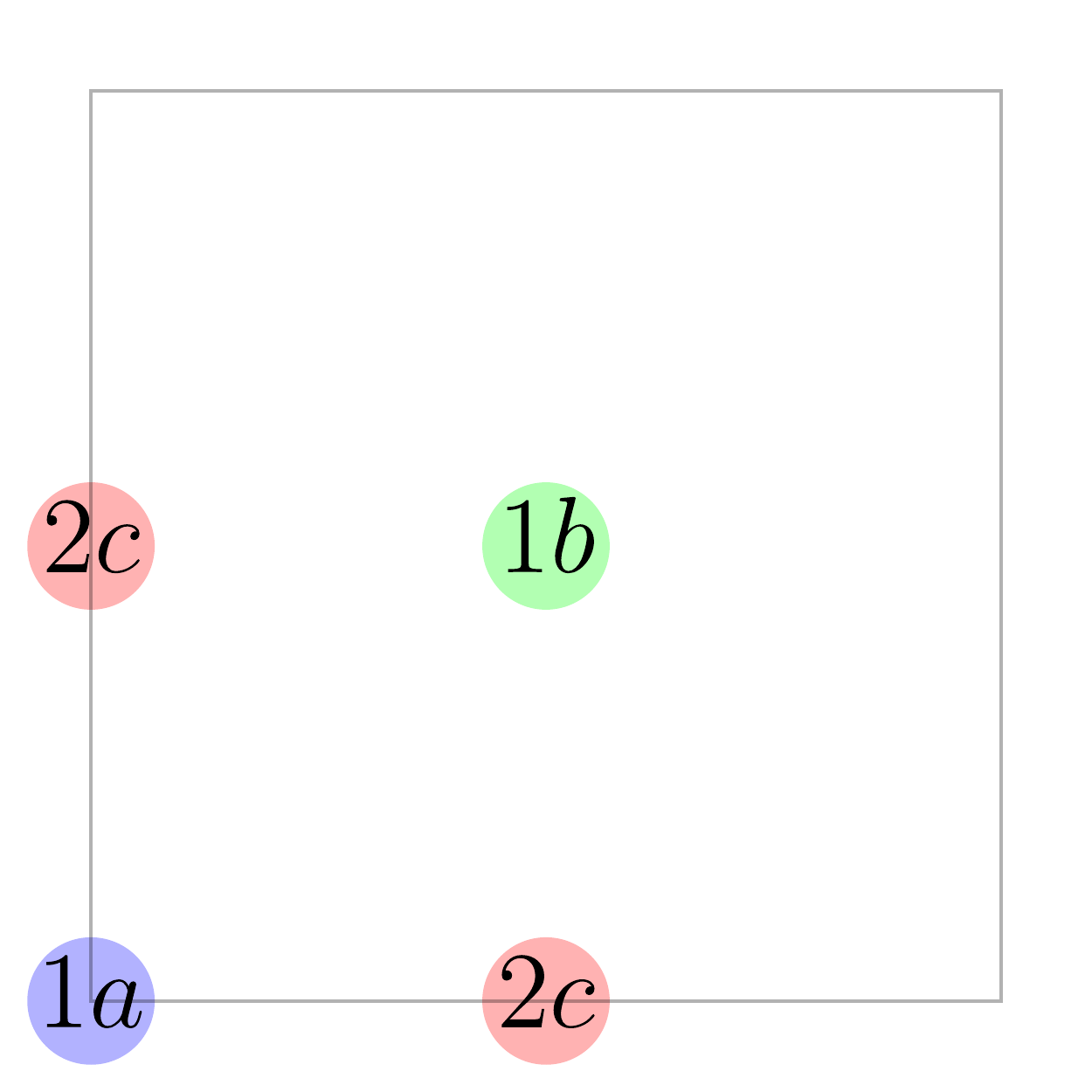}
\caption{Unit cell of the wallpaper group $p4mm$ generated by $C_4$ and translations (we take $\mbf{a}_1 = (1,0),\mbf{a}_2 = (0,1)$ for ease). There are three maximal Wyckoff positons 1a = $(0,0)$,1b= $(1/2,1/2)$,2c= $\{(0,1/2),(1/2,0)\}$. The site-symmetry groups are $G_{1a} = 4mm, G_{1b} = 4mm, G_{2c} = 2mm$. In all cases, the rotations of the site-symmetry groups are enough to pin $\mbf{x}_\al = 0$. }
\label{fig:wyckoff4}
\end{figure*}

To prove this result, we need the action of the space group symmetries on $h(\mbf{p})$. We prove in \App{eq:CPsymmetry} that
\bea
\Gamma[g] h(\mbf{p}) \Gamma^\dag[g] = h(g^{-1} \mbf{p}), \qquad g \in G
\eea
where the representation $\Gamma_{\al\be}[g]=|D_{\al\be}[g]|$ is a permutation matrix. It follows that 
\bea
\label{eq:ghp}
\Gamma[g] \pmb{\nabla}  h(\mbf{p}) \Gamma^\dag[g] = \pmb{\nabla}  h(g^{-1} \mbf{p}) = g^{-1} \pmb{\nabla}h|_{g^{-1} \mbf{p}} \ .
\eea
Evaluating \Eq{eq:ghp} at $\mbf{p} = 0$ gives $\Gamma[g] \pmb{\nabla}  h|_{\mbf{p}=0}\Gamma^\dag[g]  = g^{-1} \pmb{\nabla}h|_{\mbf{p}=0}$. Then
\bea
\label{eq:gxgG}
g^{-1} \mbf{x}_\al &= [i (\eps-h(0))^+ (g^{-1}\pmb{\nabla}h(\mbf{p})|_{\mbf{p}=0}) u_0]_\al \\
&= [i (\eps-h(0))^+ \Gamma[g] (\pmb{\nabla}h(\mbf{p})|_{\mbf{p}=0}) \Gamma^\dag[g] u_0]_\al \\
&= \sum_\be \Gamma_{\al \be}[g] [i (\eps-h(0))^+  (\pmb{\nabla}h(\mbf{p})|_{\mbf{p}=0}) [g] u_0]_\be \\
&= \sum_\be \Gamma_{\al \be}[g] \mbf{x}_\be \\
\eea
where we used $\Gamma[g] h(0) \Gamma^\dag[g] = h(0)$ and $\Gamma[g] u_0=u_0$ because $\Gamma[g]$ is a permutation matrix and $u_0$ is the uniform eigenvector. Thus if two orbitals are related by a symmetry $g$, their $\mbf{x}_\al$ that guarantee the minimal metric are also related by $g$. With \Eq{eq:gxgG}, we can prove that $\mbf{x}_\al =0$ if the site symmetry group $G_x$ of the Wyckoff position $x$ contains a symmetry $h$ such that $h \mbf{v} \neq \mbf{v}$ for all $\mbf{v}\neq0$ and $\Gamma[h] = \mathbb{1}$. As an example, $h$ can be any nontrivial rotation at a maximal Wyckoff position. In this case
\bea
h \mbf{x}_\al =  \sum_\be \Gamma_{\al \be}[h] \mbf{x}_\be = \mbf{x}_\al,
\eea
but by assumption, the only solution to $h \mbf{x}_\al = \mbf{x}_\al$ is $\mbf{x}_\al = 0$. We give two examples in space group $p4mm$ (see \Fig{fig:wyckoff4}). (1) Consider an $s$ (or $p$) orbital at the 2c $= \{\frac{1}{2} \hat{x},\frac{1}{2} \hat{y}\}$ position. The site symmetry group is $G_{2c} = 2mm$ consisting of $C_2$ reflection and mirrors. The $s$ orbitals transform in a 1D irrep, so $\Gamma[C_2] = 1$. Thus $-\mbf{x}_\al = C_2 \mbf{x}_{\al} = \mbf{x}_\al$, so $\mbf{x}_\al = 0$. (2) Consider $p_x,p_y$ orbitals at the 1b $=\frac{1}{2} \hat{x}+\frac{1}{2} \hat{y}$ position. The representation of $C_4$ is $D[C_4] = i \sigma_2$ (the $p_x,p_y$ orbitals are rotated into each other), so $\Gamma[C_4]= \sigma_1$. Then $\Gamma[C_2]= \mathbb{1}$ and $-\mbf{x}_\al = C_2 \mbf{x}_\al = \mbf{x}_\al$ as before. We remark that if the orbitals are not at a maximal Wyckoff position (such as the 6j position in \App{app:Aaron}), the symmetries allow $\mbf{x}_\al$ to be nonzero. 

\subsection{Space Group Symmetry and Representations of the Boson Bands}
\label{eq:CPsymmetry}

Because $h(\mbf{p})$ is a single-particle Hamiltonian, its bands can display nontrivial topology. In fact, we will show that topological quantum chemistry can be applied to $h(\mbf{p})$ to prove a number of nontrivial statements about the possible band decompositions and symmetry representations. To bring this formalism to bear, we must first work out the symmetries of $h(\mbf{p})$. We denote the $N_L\times N_L$ representations of the symmetries of $h_{\al\be}(\mbf{p})$ by $\Gamma[g]$.

First we show that $h(\mbf{p})$ has a bosonic time-reversal symmetry with representation $\Gamma[\mathcal{T}] = K$. We check
\bea
\label{eq:trsboson}
h^*_{\al \be}(\mbf{p}) = \frac{1}{\mathcal{N}} \sum_\mbf{k} P^*_{\al \be}(\mbf{p}+\mbf{k}) P^*_{\be \al}(\mbf{k}) = \frac{1}{\mathcal{N}} \sum_\mbf{k} P_{\al \be}(\mbf{k})  P_{\be \al}(\mbf{p}+\mbf{k}) = \frac{1}{\mathcal{N}} \sum_\mbf{k} P_{\al \be}(\mbf{k}-\mbf{p})  P_{\be \al}(\mbf{k})  = h_{\al \be}(-\mbf{p}) \ .
\eea
We also need to check the embedding matrices. Using $P(\mbf{k}+\mbf{G}) = V[\mbf{G}] P(\mbf{k}) V^\dag[\mbf{G}]$ where $V_{\al\be}[\mbf{G}] = \delta_{\al \be} e^{ - i \mbf{r}_\al \cdot \mbf{G}}$ is the embedding matrix and $\mbf{G}$ is a reciprocal lattice vector (see the Appendix of \Ref{2022PhRvL.128h7002H}), we compute
\bea
\label{eq:orbitallocation}
h_{\al \be}(\mbf{p}+\mbf{G}) &= \frac{1}{\mathcal{N}} \sum_\mbf{k} P_{\al \be}(\mbf{p}+\mbf{G}+\mbf{k}) P_{\be \al}(\mbf{k}) = \frac{1}{\mathcal{N}} \sum_\mbf{k} e^{-i \mbf{G} \cdot \mbf{r}_\al} P_{\al \be}(\mbf{p}+\mbf{k}) e^{i \mbf{G} \cdot \mbf{r}_\be}P_{\be \al}(\mbf{k}) = [V(\mbf{G}) h(\mbf{p}) V^\dag(\mbf{G})]_{\al \be} \\
\eea
which is the same as the transformation of the electron Hamiltonian. The embedding matrix contains the location $\mbf{r}_\al$ of the orbital basis in the lattice. Since the embedding matrix $V[\mbf{G}]$ of $h(\mbf{p})$ is the same as the initial electron Hamiltonian (see \Eq{eq:pRandembeddingV}), we find that the orbitals of the effective boson tight-binding model are located at the same position as the electron orbitals. The representations of the orbitals, however, may be different.

To study the spatial symmetries, we recall that $D[g] P(\mbf{k}) D^\dag[g] = P(g\mbf{k})$ (see \Eq{eq:simhtilde}). In indices, we have
\bea
\label{eq:Hreps1}
h_{\al \be}(g\mbf{p}) &= \frac{1}{\mathcal{N}} \sum_\mbf{k} P_{\al \be}(g\mbf{p}+\mbf{k}) P_{\be \al}(\mbf{k}) = \frac{1}{\mathcal{N}} \sum_\mbf{k} P_{\al \be}(g\mbf{p}+g\mbf{k}) P_{\be \al}(g\mbf{k}) \\
&= \frac{1}{\mathcal{N}} \sum_{\mbf{k}, \kappa\kappa'\la\la'} D_{\al \kappa}[g] P_{\kappa \kappa'}(\mbf{p}+\mbf{k})  D_{\be \kappa'}^*[g] D_{\be \la}[g] P_{\la \la'}(\mbf{k})  D_{\al \la'}^*[g] \\
&= \frac{1}{\mathcal{N}} \sum_{\mbf{k}, \kappa\kappa'\la\la'} D_{\al \kappa}[g] D_{\al \la'}^*[g] \ P_{\kappa \kappa'}(\mbf{p}+\mbf{k}) P_{\la \la'}(\mbf{k}) \  D_{\be \kappa'}^*[g] D_{\be \la}[g] . \\
\eea
We now need the following fact which we prove momentarily: in every point group, there exists a choice of basis where
\bea
\label{eq:DDdD}
D_{\al \kappa}[g] D_{\al \la'}^*[g] = \delta_{\la \kappa}|D_{\al \la}[g]|^2, \qquad g \in G \ .
\eea
To understand \Eq{eq:DDdD}, we see that each column of $D[g]$ can only have a single non-zero element. Because $D[g]$ is unitary, this element must be a complex phase. Thus \Eq{eq:DDdD} is equivalent to the statement that $D[g]$ is a complex permutation matrix. This is trivially satisfied if there is a single orbital per unit cell. More generally, a group with this property, that there exists a basis where all irreps are complex permutation matrices, is called a monomial group. It was recently proven that all spinless point groups in 2D and 3D are monomial \cite{PhysRevB.102.115117}. Because $D[g]$ are spinless representations of the point group (or little group) of the $\Gamma$ point $\mbf{p}=0$, \Eq{eq:DDdD} holds. Conveniently, all the representation matrices given on the Bilbao crystallographic server \url{https://www.cryst.ehu.es/cgi-bin/cryst/programs/representations_point.pl?tipogrupo=spg} are in the complex permutation matrix form so there is only one nonzero entry per row. However, recall that our convention in \App{app:spin} is to take $D[\mathcal{T}]$ to act as the identity on the orbital index. It is now convenient to define
\bea
\label{eq:defgamma}
D_{\al \kappa}[g] D_{\al \la'}^*[g] = \delta_{\la \kappa}|D_{\al \la}[g]|^2 = \delta_{\la \kappa}|D_{\al \la}[g]| \equiv \delta_{\la \kappa} \Gamma_{\al \la}[g], \qquad \Gamma_{\al \be}[g] = |D_{\al \be}[g]|
\eea
where $\Gamma[g]$, which are real permutation matrices, are the representations of $g\in G$ on $h(\mbf{p})$. This follows from \Eqs{eq:Hreps1}{eq:defgamma}:
\bea
\label{eq:gammag}
h_{\al \be}(g\mbf{p}) &= \frac{1}{\mathcal{N}} \sum_{\mbf{k}, \la\la'} D_{\al \la}[g]D_{\al \la}^*[g]  P_{\la \la'}(\mbf{p}+\mbf{k}) P_{\la' \la}(\mbf{k}) D_{\be \la'}^*[g] D_{\be \la'}[g] \\
&= \frac{1}{\mathcal{N}} \sum_{\mbf{k}, \la\la'} \Gamma_{\al \la}[g]  P_{\la \la'}(\mbf{p}+\mbf{k}) P_{\la' \la}(\mbf{k}) \Gamma_{\be \la'}[g]\\
&= [\, \Gamma[g] h(\mbf{p}) \Gamma^\dag[g] \,]_{\al \be} .
\eea
In this way, we have shown that $h(\mbf{p})$ possess all the space group symmetries of the electron Hamiltonian, but in a ``bosonized" form. Because $\Gamma[g]$ are all real permutation matrices, we can think of the boson orbitals as those obtained from replacing all electron orbitals with $s$ orbitals. We emphasize that this result holds on the UPC lattices we have constructed, where the orbitals form an irrep of a single Wyckoff position (\App{app:UPC}). 

Because $\Gamma[g]$ is real, it is obvious that $\Gamma[\mathcal{T}]$ commutes with all spatial operators, so the topology of the Cooper pair bound states is classified by time-reversal symmetric space groups with $\Gamma[\mathcal{T}]^2 = +1$. There is an additional constraint on the spectrum: we proved that $u_0^\al(\mbf{p}=0) = 1/\sqrt{N_L}$. But $\Gamma[g]$ is a permutation matrix, so $\Gamma[g] u_0(\mbf{p}=0)  = u_0(\mbf{p}=0)$ and hence the irrep of the $\eps_0(\mbf{p})$ band at $\mbf{p}=0$ is always the trivial irrep. We now show that in the higher Cooper pair bands and at nonzero momenta, the Cooper pair wavefunction can (and will) exhibit other pairing symmetries. 

Using the representation matrices $\Gamma[g]$, we can compute the action of $g$ on the many-body Cooper pair creation operators. From \Eq{eq:etap}, we first derive
\bea
g \eta^\dag_{\mbf{p},\mu} g^\dag &= \frac{1}{\sqrt{\mathcal{N}}} \sum_{\mbf{k}\al\al'\al''} \frac{u^\al_{\mu}(\mbf{p})}{\sqrt{\eps_\mu(\mbf{p})}} \bar{c}^\dag_{g\mbf{p}+g\mbf{k},\al',\u} D^\u_{\al' \al}[g] \bar{c}^\dag_{-g\mbf{k},\al'',\d}D^\d_{\al'' \al}[g] \\
&= \frac{1}{\sqrt{\mathcal{N}}} \sum_{\mbf{k}\al\al'\al''} \frac{1}{\sqrt{\eps_\mu(\mbf{p})}} \bar{c}^\dag_{g\mbf{p}+\mbf{k},\al',\u} D_{\al' \al}[g]  D^*_{\al'' \al}[g] u^\al_{\mu}(\mbf{p}) \bar{c}^\dag_{-\mbf{k},\al'',\d} \\
&= \frac{1}{\sqrt{\mathcal{N}}} \sum_{\mbf{k}\al\al'\al''} \frac{1}{\sqrt{\eps_\mu(g\mbf{p})}} \bar{c}^\dag_{g\mbf{p}+\mbf{k},\al',\u} \delta_{\al' \al''} \Gamma_{\al' \al}[g] u^\al_{\mu}(\mbf{p}) \bar{c}^\dag_{-\mbf{k},\al'',\d} \\
\eea
where we used that $D[g] = D^\u[g] = D^\d[g]^*$ by $\mathcal{T}$ and $S_z$. Now we use the space group symmetries of $h(\mbf{p})$ such that $\eps_\mu(g\mbf{p}) = \eps_\mu(\mbf{p})$ and $\sum_\al \Gamma_{\al' \al}[g] u^\al_{\mu}(\mbf{p}) = \sum_{\mu'} u^{\al'}_{\mu'}(g\mbf{p}) B_{\mu' \mu}^g(\mbf{p})$ where $[B^g(\mbf{p})]_{\mu'\mu} = u^\dag_{\mu'}(g\mbf{p}) \Gamma[g] u_\mu(\mbf{p})$ is the sewing matrix of the bands \cite{andreibook}. If $\eps_\mu(\mbf{p}) \neq \eps_{\mu'}(\mbf{p})$, then $B^g_{\mu' \mu}(\mbf{p}) = 0$ and thus
\bea
\label{eq:etagtrans}
g \eta^\dag_{\mbf{p},\mu} g^\dag &= \frac{1}{\sqrt{\mathcal{N}}} \sum_{\mbf{k}\al \mu'} \frac{u^{\al}_{\mu'}(g\mbf{p})}{\sqrt{\eps_\mu(g\mbf{p})}} B^g_{\mu' \mu }(\mbf{p}) \bar{c}^\dag_{g\mbf{p}+\mbf{k},\al,\u}  \bar{c}^\dag_{-\mbf{k},\al,\d} = \sum_{\mu'} \eta^\dag_{g\mbf{p},\mu'} B^g_{\mu' \mu}(\mbf{p}) \ . \\
\eea
At the high symmetry points, $B_g(\mbf{p})$ is the representation matrix of the bands of $h(\mbf{p})$, and hence is determined by the irreps of the Cooper pair bands. Hence, the symmetry and topology of the single-particle boson Hamiltonian $h(\mbf{p})$ is detectable through $B_g(\mbf{p})$ in the many-body states. We emphasize that although the bound states Cooper pairs are always spin-singlets, they can have nontrivial symmetry representations ($p$-wave, $d$-wave, etc) due to their multi-orbital nature. This is perhaps unfamiliar because \emph{with only spin indices}, a Cooper pair wavefunction which is odd under $\mbf{p} \to - \mbf{p}$ must be a spin triplet because the spin indices are anti-symmetrized, for example. In contrast, we have shown here that the low-energy Cooper pair bound states, which are spin-singlets, can have be odd under $\mbf{p} \to - \mbf{p}$ since their orbital indices are anti-symmetrized even though their spin indices are symmetric. 

\subsection{Topology of the Boson Bands}
\label{app:bosontopology}

Let us review the constraints placed on the bands of $h(\mbf{p})$. First, we chose the electron orbitals to be irreps of a single Wyckoff position in order to guarantee the UPC (\App{app:UPC}), and we showed that the orbitals of the boson Hamiltonian $h(\mbf{p})$ are at the same Wyckoff position determined by $V[\mbf{G}]$ in \Eq{eq:orbitallocation}. (Note that in the bipartite crystalline lattice construction, the Wyckoff position of the boson Hamiltonian is that of the (larger) $L$ sublattice.) However, the electron orbitals transform in the representation $D[g]$ whereas the boson orbitals transform in the representation $\Gamma[g]$ (\Eq{eq:gammag}). Since $\Gamma_{\al \be}[g] = |D_{\al \be}[g]|$, the orbitals of $h(\mbf{p})$ are effectively $s$ orbitals. $h(\mbf{p})$ also possesses spinless time-reversal (see \Eq{eq:trsboson}). Lastly, we proved that the lowest energy state at the gamma point $\mbf{p} = 0$ is always the trivial ($s$) irrep. 

We now use Topological Quantum Chemistry \cite{2017Natur.547..298B} to determine the momentum space irreps which are induced from the boson orbitals satisfying the constraints in the previous paragraph. Because the symmetry of the Cooper pair wavefunction at a given $\mbf{p}$ is determined by the induced momentum space irreps, we can completely classify the possible pairing symmetries and Cooper pair band connectivities in every space group. 

\begin{figure*}
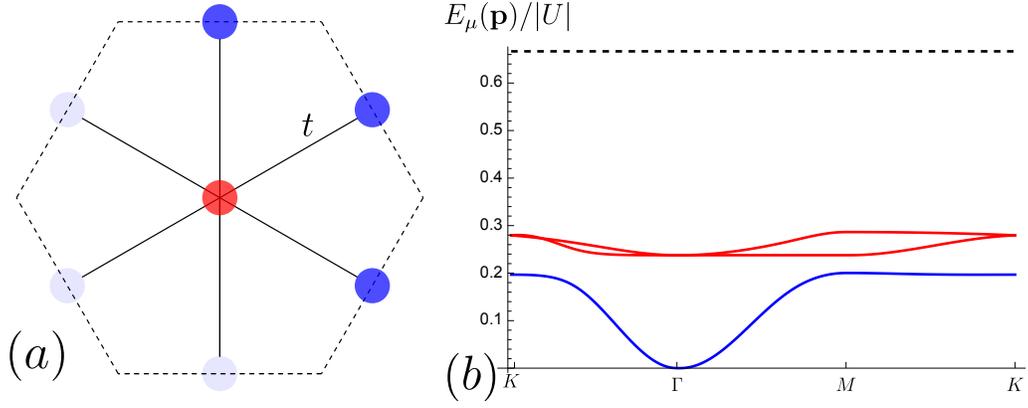

\includegraphics[height=.3\textwidth]{p6mm_hoppings}
\includegraphics[height=.3\textwidth]{6mm_fragile}
\caption{Cooper pair band structure of the Hamiltonian \Eq{eq:tildeh6mm} on the kagome sites. ($(a)$) in \App{eq:kagomecooper}. $(b)$ The band representation computed in \App{eq:kagomecooper} is $\mathcal{B}_1 = \Gamma_1 + K_1 + M_1$ for the low-energy blue band, and $\mathcal{B}_2 = \Gamma_5 + K_3 + M_3\oplus M_4$ for the higher energy pair of red bands.}
\label{fig:6mmfrag2}
\end{figure*}

First we make the momentum space irreps explicit. Consider a high symmetry momentum $\mbf{p}$ satisfying $g \mbf{p} = \mbf{p} + \mbf{G}$ where $\mbf{G}$ is a reciprocal lattice vector. Then  $h(\mbf{p})$ commutes with $V^\dag[\mbf{G}] \Gamma[g]$ and the Cooper pair wavefunctions at $\mbf{p}$ transform in an irrep of the little group operators $V^\dag[\mbf{G}] \Gamma[g]$ (see the Appendix of \Ref{2022PhRvL.128h7002H}). By computing the irreps $V^\dag[\mbf{G}] \Gamma[g]$ for a given model, one can determine the possible pairing symmetries. The Bilbao Crystallographic server has tabulated all such irreps for every Wyckoff position and space group. For instance, in the kagome model in \App{eq:kagomecooper} (see \Fig{fig:6mmfrag2}), we computed the irreps at the gamma point to be $\Gamma_1$ (a $s$-wave 1D irrep) and $\Gamma_5$ (a 2D $d$-wave irrep, angular momentum $\pm2$). This is because the little group of the gamma point contains the operator $\Gamma[C_6]$ which is a $3\times3$ permutation matrix with eigenvalues $1, e^{\pm \frac{2pi i}{3}}$. From the Bilbao Crystallographic server, we observe that the $K$ point has irreps $K_1$ ($s$-wave) and $K_3$ ($d$-wave), and the $M$ point has irreps $M_1$ ($s$-wave) and $M_3 \oplus M_4$ ($p$-wave). Note that the $s$-wave irrep at $\Gamma$ is at zero energy, so physically this is the mode that condenses, leading to the eta-pairing groundstates over zero momentum and zero angular momentum. However at higher temperature, the higher energy branch is also populated. 

The global structure of the Cooper pair bands in momentum space is also determined by topological quantum chemistry. In general, a set of bands in momentum space can only be gapped if the irreps obey a set of compatibility relations \cite{PhysRevB.97.035138}. To apply these compatibility relations to the Cooper pair bands, we recall that the Cooper pair orbitals are induced from a single Wyckoff position, the Cooper bands form an elementary band representation (EBR). If the Wyckoff position is non-maximal, then in general the Cooper pair bands will be trivially decomposable: they can be separated into gapped atomic bands. If the Wyckoff position is maximal (like the $3c$ kagome position in $G = p6mm$), then there are two possibilities: the EBR could be indecomposable meaning that all Cooper pair bands are connected, or the EBR could be decomposable. If the EBR is decomposable, it can be split into disconnected groups of bands, and necessarily at least one band in this decomposition is topological \cite{2018PhRvL.120z6401C,2018arXiv180709729B}. 
 The Bilbao crystallographic server tabulates these properties for all EBRs, and explicitly writes the possible irreps of the split EBR. Finally, there is an additional constraint due to time-reversal symmetry which prevents stable topological bands \cite{PhysRevLett.98.046402} in one, two, and three dimensions. Thus if a branch of a split EBR has a nonzero symmetry indicator (which would diagnose stable topology if the band were gapped everywhere), there must be a Weyl node in the spectrum which prevents the bands from being gapped everywhere \cite{2011PhRvB..83x5132H,2020arXiv201000598E,2017Natur.547..298B}. However, the Cooper pair bands can be gapped with nontrivial topology if the split EBR has a decomposition into fragile topological bands. 
 
 Using the Bilbao Crystallographic server, we exhaustively tabulate every EBR in the 230 space groups which is decomposable and obeys the constraints in the first paragraph of \App{app:bosontopology} (notably, that there are $s$ orbitals on the Wyckoff position). We leave a complete study of the magnetic space groups to other work.

\begin{figure*}
\includegraphics[height=.3\textwidth]{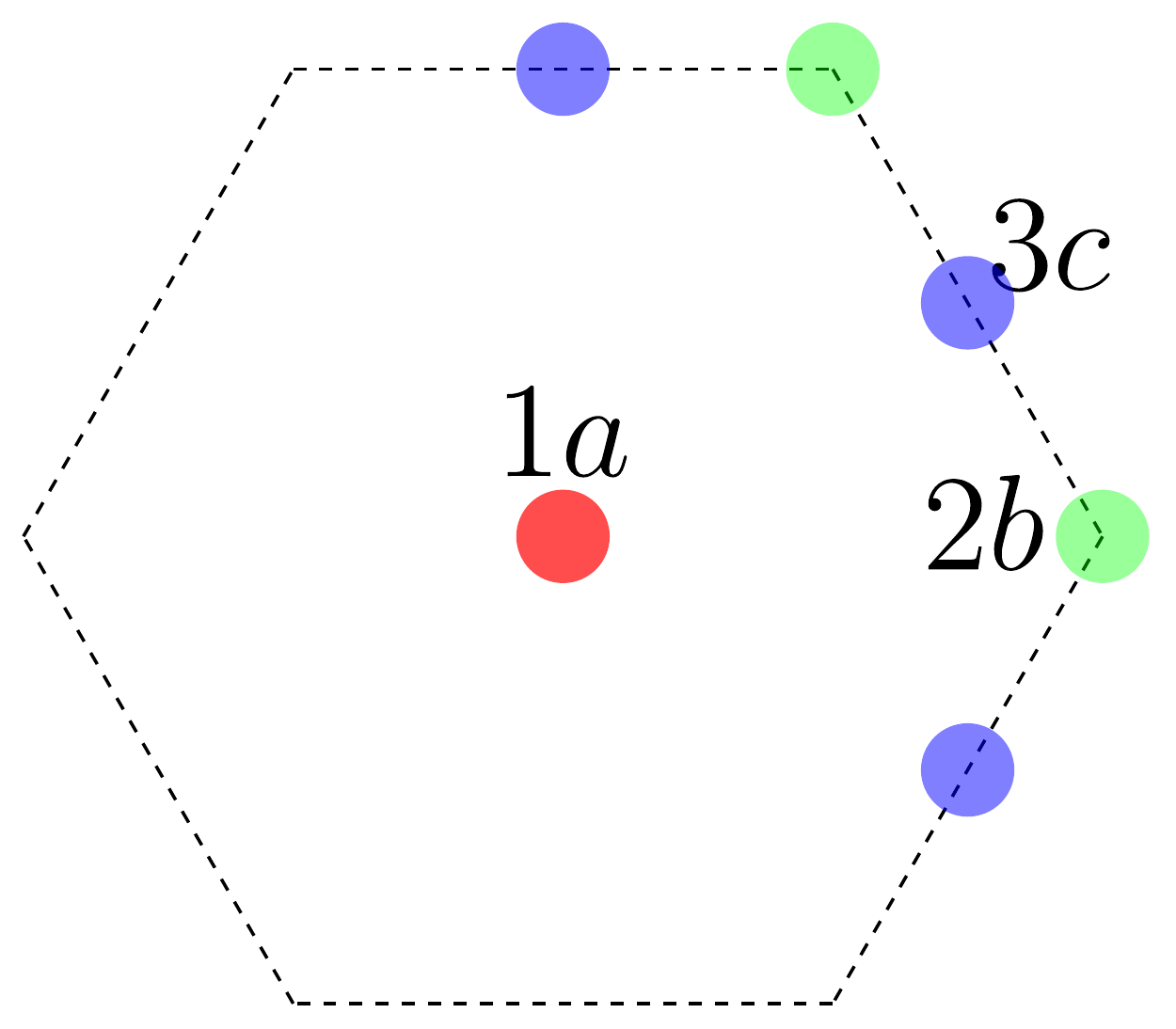}
\caption{We show the maximal Wyckoff positions in $p6$ and $p6mm$. The 1a position forms a triangular lattice, the 2b position forms a hexagonal lattice, and the 3c position forms a kagome lattice. The site symmetry groups of the positions are $G_{1a} = 6, G_{2b} = 3, G_{3c} = 2$. }
\label{fig:wykhex}
\end{figure*}

To illustrate the method, we first consider the 2D wallpaper groups with spinless time-reversal symmetry. Almost every Wyckoff position is indecomposable, meaning that the bands are all connected through high symmetry points in the BZ. Of the 17 spinless wallpaper groups, only $p61'$ and $p6mm1'$ have decomposable bands induced from the trivial irrep at a maximal Wyckoff position. In both cases, this Wyckoff position is the 3c position (see \Fig{fig:wykhex}) which has site symmetry group $G_{3c} = 2 = \{1,C_2\}$. Putting $s$ or $p$ electron orbitals at the 3c position yields the Cooper pair representations $\Gamma[C_2] = \mathbb{1}$ and $\Gamma[C_6] = P_3$ where $P_3$ is the permutation matrix exchanging the 3 sites. This is the representation of an $s$-orbital at the 3c position, denoted $A_{3c} \uparrow G$.  From the Bilbao crystallographic server we find that in $G=p61'$ (the $1'$ denotes time-reversal), there are two possible band decompositions
\bea
\label{eq:banddecomp}
A_{3c} \uparrow p61' &= \Gamma_1 \oplus \Gamma_3\Gamma_5 + K_1 \oplus K_2K_3 + M_1 \oplus M_2 \oplus M_2 \\
&= \begin{cases}
 (\Gamma_1 + K_1 + M_1) +  (\Gamma_3\Gamma_5 + K_2K_3 + 2M_2) & \\
  (\Gamma_1 + K_1 + M_2) +  (\Gamma_3\Gamma_5 + K_2K_3 + M_1\oplus M_2) & \\
  \end{cases}
\eea
where the parentheses denote which bands are connected. Both decompositions contain a single band with a $\Gamma_1$ irrep (the trivial) irreps, which is the quadratic band at low energy. The two higher bands are connected and gapped from the low energy band. The momentum space irreps are defined
\bea
\begin{array}{c|ccc}
61' &1& C_6   \\
\hline
\Gamma_1 &1&1 \\
\Gamma_3\Gamma_5 &2 & -1   \\
\end{array}, \qquad
\begin{array}{r|rr}
3 &1& C_3 \\
\hline
K_1 & 1 & 1 \\
K_2 K_3 & 2 & -1 \\
\end{array},\qquad
 \begin{array}{c|cc}
m&1& M_x \\
\hline
M_1 &1&1\\
M_2 &1 & -1 \\
\end{array} \ .
\eea
We now consider the topology of the two branches in \Eq{eq:banddecomp}. We observe that the $\Gamma_1 + K_1 + M_1$ band is $A_{1a} \uparrow p61'$ and hence is a trivial atomic band. Its complement $\Gamma_3\Gamma_5 + K_2K_3 + 2M_2 = A_{3c} \uparrow G \ominus A_{1a} \uparrow G$ is fragile topological. This is an instance of the general result that gapping band induced from a maximal Wyckoff position must result in topology \cite{2018PhRvL.120z6401C,2018arXiv180709729B}: since the lower band is trivial, the higher two bands must be topological. The other allowed decomposition has a $\Gamma_1 + K_1 + M_2$ band. Using the Smith normal form \cite{2020PhRvB.102c5110E}, we compute the symmetry indicator $\th = m(\Gamma_2) + m(M_2) \mod 2$, which is the number of odd inversion eigenvalues mod 2. We observe that $\th = 1$ for $\Gamma_1 + K_1 + M_2$, indicating is must have Weyl nodes connecting it to the other bands.

We systematize this information in the following table. W.P. stands for the Wyckoff position of the orbitals. The $\eta$ irreps column denotes the orbital of $h(\mbf{p})$ of the site symmetry group of the Wyckoff position (physically it is always an $s$ orbital, but is denoted with different symbols in different space groups), and $e^-$ irrep denotes the possible electron orbitals that yield the bosonic orbital through $\Gamma[g] = |D_{\al \be}[g]|$. $\#$ numbers the possible decompositions and $\th$ is the symmetry indicator, which results in Weyl nodes due to $\Gamma[\mathcal{T}]$. Our classification enumerates the topology of the Cooper pair bands (the first entry corresponds to the gapless quadratic band). When the band is trivial, its atomic representation is given. Otherwise fragile topology or topologically protected Weyl nodes are indicated.

\begingroup
\label{tab:2Dcooperpairs}
\renewcommand\arraystretch{1.3}
\begin{longtable}{|c|c|c|c|c|c|c|c|}
\hline
$G$ & \text{name} & W.P. &\text{$\eta^\dag$ irreps ($e^-$ irreps)} & \# &$\theta$ &\text{Classification}  \\
\hline
16 & $p6$ &  & & &$\th = m(\Gamma_2) + m(M_2) \mod 2 \in \mathds{Z}_2  $ & \\
\hline
& & 3c & $A \  (A,B) $ & 1 &$0,0 $ & $A^{1a} \uparrow G$, fragile \\
& & 3c&  $A \  (A,B) $& 2 &$1,1$ & Weyl, Weyl\\
\cline{3-7}
\hline
17 & $p6mm$ &  & & & none & \\
\hline
& & 3c & $A_1 \  (A_1,A_2,B_1,B_2) $ & 1 & & $A_1^{1a} \uparrow G$, fragile \\
\hline
\end{longtable}
\endgroup
We see that in 2D, the only cases with decomposable bands at maximal Wyckoff positions are in $p6$ and $p6mm$. In all other groups and maximal Wyckoff positions, the Cooper pair bands are connected. In the decomposable cases listed above, only the higher energy bands can be fragile topological.

Performing an exhaustive search 3D, we find 22 space groups containing decomposable bands with a trivial irrep at the $\Gamma$ point. In the case where the symmetry indicator is zero, we check whether the band is fragile or atomic. They are enumerated in the following table. There are many cases where the low-lying Cooper pair band has topologically protected Weyl nodes. We also highlight $G = Fd\bar{3}c$, space group 228, where the only allowed decomposition is into fragile bands.

%
%
%


\begingroup
\renewcommand\arraystretch{1.3}
\begin{longtable}{|c|c|c|c|c|c|c|c|}
\hline
$G$ & \text{name} & W.P. &\text{$\eta$ irreps ($e^-$ irreps)} & \# &$\theta$ &\text{Classification}  \\
\hline
147 & $P\bar{3}$ &  & & &$(\mathds{Z}_2, \mathds{Z}_4) $ & \\
\hline
& &  \multirow{4}{*}{3e} &  \multirow{4}{*}{$A_g \  (A_g,A_u) $}  & 1 &$(0,0), (0,0) $ & $A^{1a}_g \uparrow G$, fragile\\
& & &  & 2 &$(0,1), (0,3) $ & Weyl, Weyl\\
& & &  & 3 &$(1,3), (1,1) $ & Weyl, Weyl\\
& & &  & 4 &$(1,0), (1,0) $ & Weyl, Weyl\\
\cline{3-7}
& & \multirow{4}{*}{3f} & \multirow{4}{*}{$A_g \  (A_g,A_u) $}  & 1 &$(0,3), (0,1) $ & Weyl, Weyl\\
& & &  & 2 &$(0,0), (0,0) $ & $A^{1b}_g \uparrow G$, fragile\\
& & &  & 3 &$(1,2), (1,2) $ & Weyl, Weyl\\
& & &  & 4 &$(1,3), (1,1) $ & Weyl, Weyl\\
\hline
148 & $R\bar{3}$ &  & & &$(\mathds{Z}_2, \mathds{Z}_4) $ & \\
\hline
& & \multirow{4}{*}{9d} &  \multirow{4}{*}{$A_g \  (A_g,A_u) $}  & 1 &$(1,3), (1,1) $ & Weyl, Weyl\\
& & &  & 2 &$(0,0), (0,0) $ & $A^{3b}_g \uparrow G$, fragile\\
& & &  & 3 &$(1,2), (1,2) $ & Weyl, Weyl\\
& & &  & 4 &$(0,3), (0,3) $ & Weyl, Weyl\\
\cline{3-7}
& & \multirow{4}{*}{9e} &  \multirow{4}{*}{$A_g \  (A_g,A_u) $} & 1 &$(0,0), (0,0) $ & $A^{3a}_g \uparrow G$, fragile\\
& & &  & 2 &$(1,1), (1,3) $ & Weyl, Weyl\\
& & &  & 3 &$(1,3), (1,1) $ & Weyl, Weyl\\
& & &  & 4 &$(1,0), (1,0) $ & Weyl, Weyl\\
\hline
162 & $P\bar{3}1m$ &  & &  & $\mathds{Z}_2$  & \\
\hline
& & 3f & $A_g \  (A_g,A_u,B_g,B_u) $ & 1 & 0 & $A_{1g}^{1a} \uparrow G$, fragile \\
\cline{3-7}
& & 3g & $A_g \  (A_g,A_u,B_g,B_u) $ & 1 & 0 & $A_{1g}^{1b} \uparrow G$, fragile \\
\hline
163 & $P\bar{3}1c$ &  & &  & $\mathds{Z}_2$   & \\
\hline
& & \multirow{4}{*}{6g} &  \multirow{4}{*}{$A_g \  (A_g,A_u) $} & 1 & $1,1$ & Weyl, Weyl \\
& &  &  & 2 & $0,0$ & fragile, fragile \\
& &  & & 3 &$1,1$ & Weyl, Weyl \\
& &  &  & 4 &$0,0$ & $A_g^{2b} \uparrow G$, fragile \\
\hline
164 & $P\bar{3}m1$ &  & &  & $\mathds{Z}_2$ & \\
\hline
& & 3e & $A_g \  (A_g,A_u, B_g, B_u) $ & 1 & 0,0 & $A_{1g}^{1a} \uparrow G$, fragile \\
\cline{3-7}
& & 3f & $A_g \  (A_g,A_u, B_g, B_u) $ & 1 & 0,0 & $A_{1g}^{1b} \uparrow G$, fragile \\
\hline
165 & $P\bar{3}c1$ &  & &  & $\mathds{Z}_2$ & \\
\hline
& & \multirow{2}{*}{6e} &  \multirow{2}{*}{$A_g \  (A_g,A_u) $}  & 1 & $1,1$ & Weyl, Weyl \\
& &  & & 2 & $0,0$ & $A_{g}^{2b} \uparrow G$, fragile \\
\hline
166 & $R\bar{3}m$ &  & &  & $\mathds{Z}_2$ & \\
\hline
& & 9d & $A_g \  (A_g,A_u,B_g,B_u) $ & 1 & $0,0$ & $A_{1g}^{3b} \uparrow G$, fragile  \\
\cline{3-7}
& & 9e & $A_g \  (A_g,A_u,B_g,B_u) $ & 1 & $0,0$ & $A_{1g}^{3a} \uparrow G$, fragile \\
\hline
167 & $R\bar{3}c$ &  & &  & $\mathds{Z}_2$ & \\
\hline
& &  \multirow{2}{*}{18d} &  \multirow{2}{*}{$A_g \  (A_g,A_u) $}  & 1 & $1,1$ & Weyl, Weyl \\
& &  &  & 2 & $0,0$ & $A_g^{6b} \uparrow G$, fragile\\
\hline
168 & $P6$ &  & &  & $\mathds{Z}_2$ & \\
\hline
& &  \multirow{2}{*}{3c} &  \multirow{2}{*}{$A \  (A,B) $}  & 1 & $0,0$ & $A^{1a}\uparrow G$, fragile \\
& &  & & 2 & $1,1$ & Weyl, Weyl \\
\hline
175 & $P6/m$ &  & &  & $(\mathds{Z}_2,\mathds{Z}_2,\mathds{Z}_2)$ & \\
\hline
& &  \multirow{2}{*}{3f} & \multirow{2}{*}{$A_g \  (A_g,A_u,B_g,B_u)$} & 1 & $(0,0,0),(0,0,0)$ & $A_g^{1a}\uparrow G$, fragile \\
& &  & & 2 & $(0,1,1),(0,1,1)$ & Weyl, Weyl \\
\cline{3-7}
& &  \multirow{2}{*}{3g} & \multirow{2}{*}{$A_g \  (A_g,A_u,B_g,B_u)$}  & 1 & $(0,0,0),(0,0,0)$ & $A_g^{1b} \uparrow G$, fragile \\
& &  &  & 2 & $(0,1,1),(0,1,1)$ & Weyl, Weyl \\
\hline
176 & $P6_3/m$ &  & &  & $\mathds{Z}_2$ & \\
\hline
& &  \multirow{2}{*}{6g} & \multirow{2}{*}{$A_g \  (A_g,A_u) $}   & 1 & $1,1$ & Weyl, Weyl \\
& &  & & 1 & $0,0$ & $A^{2b}_g \uparrow G$, fragile \\
\hline
177 & $P622$ &  & &  & none & \\
\hline
& & 3f & $A_1 \  (A_1,B_1,B_2,B_3)$ & 1 &  & $A_1^{1a}\uparrow G$, fragile \\
\cline{3-7}
& & 3g & $A_1 \  (A_1,B_1,B_2,B_3)$ & 1 &  & $A_1^{1b} \uparrow G$, fragile \\
\hline
183 & $P6mm$ &  & &  & none & \\
\hline
& & 3c & $A_1 \  (A_1,A_2,B_1,B_2) $ & 1 &  & $A_1^{1a}\uparrow G$, fragile \\
\hline
184 & $P6cc$ &  & &  & $\mathds{Z}_2$ & \\
\hline
& &  \multirow{2}{*}{3c} & \multirow{2}{*}{$A \  (A,B) $}  & 1 & 0,0 & $A_1^{2a}\uparrow G$, fragile \\
& &  &  & 2 & 1,1 & Weyl, Weyl \\
\hline
191 & $P6/mmm$ &  & &  & none & \\
\hline
& & 3f & $A_g \  (A_g,A_u,B_{1g},B_{1u},B_{2g},B_{2u}) $ & 1 &  & $A_{1g}^{1a}\uparrow G$, fragile \\
\cline{3-7}
& & 3g & $A_g \ (A_g,A_u,B_{1g},B_{1u},B_{2g},B_{2u}) $ & 1 &  & $A_{1g}^{1b}\uparrow G$, fragile \\
\hline
192 & $P6/mcc$ &  & &  & $\mathds{Z}_2$ & \\
\hline
& & 6f & $A_1 \  (A_1, B_1, B_2, B_3) $ & 1 & 0,0 & $A_{1}^{2a}\uparrow G$, fragile \\
\cline{3-7}
& &  \multirow{2}{*}{6g} & \multirow{2}{*}{$A_g \ (A_g, A_u, B_g, B_u) $}  & 1 &  0,0 & $A_{1}^{2a}\uparrow G$, fragile \\
& &  &  & 2 & $1,1$ & Weyl, Weyl \\
\hline
193 & $P6_3/mcm$ &  & &  & none & \\
\hline
& & 6f & $A_g \  (A_g, A_u, B_g, B_u) $ & 1 &  & $A_{1g}^{2b}\uparrow G$, fragile \\
\hline
194 & $P6_3/mmc$ &  & & & none & \\
\hline
& & 6g & $A_g \  (A_g, A_u, B_g, B_u) $ & 1 & & $A_{1g}^{2a}\uparrow G$, fragile \\
\hline
202 & $Fm\bar{3}$ &  & &  &  none& \\
\hline
& &  \multirow{2}{*}{24d} & \multirow{2}{*}{$A_g \  (A_g, A_u) $ } & 1 & & $A_{g}^{4a}\uparrow G$, fragile \\
& &  &  & 2 & & $A_{g}^{4b}\uparrow G$, fragile \\
\hline
209 & $F432$ &  & &  &  none& \\
\hline
& &  \multirow{2}{*}{24d} & \multirow{2}{*}{$A_1 \  (A_1, B_1) $} & 1 & & $A_{1}^{4a}\uparrow G$, fragile, $T_2^{4b} \uparrow G$ \\
& &  &  & 2 & & $A_{1}^{4b}\uparrow G$, fragile, $T_2^{4a} \uparrow G$  \\
\hline
225 & $Fm\bar{3}m$ &  & &  & none& \\
\hline
& &  \multirow{2}{*}{24d} & \multirow{2}{*}{$A_g \  (A_g,A_u, B_{1g},B_{1u}) $}   & 1 & & $A_{1g}^{4a}\uparrow G$, fragile \\
& &  &  & 2 & & $A_{1g}^{4b}\uparrow G$, fragile \\
\hline
228 & $Fd\bar{3}c$ &  & &  & none& \\
\hline
& & 48d & $A \  (A,B) $ & 1 & & fragile, fragile \\
\hline

\end{longtable}
\endgroup

We use the explicit formulae in \Ref{2022arXiv220410556G} for the symmetry indicators \cite{2020JPCM...32z3001P,2017NatCo...8...50P,2017PhRvX...7d1069K} in the above groups using the Smith normal form\cite{2020PhRvB.102c5110E}. The results are below. The notation is the same as on the \href{https://www.cryst.ehu.es/cgi-bin/cryst/programs/bandrep.pl}{Bilbao Crystallographic Server}.
\begingroup
\renewcommand\arraystretch{1.3}
\begin{longtable}{|c|c|c|}
\hline
$G$ & \text{name} & Symmetry Indicator $\th$  \\
\hline
\multirow{3}{*}{147} & \multirow{3}{*}{$P\bar{3}$}  & $(\mathds{Z}_2, \mathds{Z}_4) $ \\
\cline{3-3}
&& $m(M_1^+)+m(\Gamma_1^+) \mod 2 $ \\
&& $m(A_1^+)-2m(A_2^+A_3^+)-m(L_1^+)+m(M_1^+)-m(\Gamma_1^+)+2m(\Gamma_2^+\Gamma_3^+) \mod 4 $ \\
\hline
\multirow{3}{*}{148} & \multirow{3}{*}{$R\bar{3}$}  & $(\mathds{Z}_2, \mathds{Z}_4) $ \\
\cline{3-3}
&& $m(T_1^+)+m(L_1^+) \mod 2 $ \\
&& $m(F_1^+)-m(L_1^+)+m(T_1^+)+2m(T_2^+T_3^+)-m(\Gamma_1^+) + 2m(\Gamma_2^+ \Gamma_3^+) \mod 4 $ \\
\hline
\multirow{2}{*}{162} & \multirow{2}{*}{$P\bar{3}1m$}  & $\mathds{Z}_2$ \\
\cline{3-3}
&& $ m(A_1^-) + m(A_3^-) + m(\Gamma_1^-) + m(\Gamma_3^-) + m(L_2^+) + m(M_2^+) \mod 2 $ \\
\hline
\multirow{2}{*}{163} & \multirow{2}{*}{$P\bar{3}1c$}  & $\mathds{Z}_2$ \\
\cline{3-3}
&& $m(A_3) +m(\Gamma_1^-) + m(\Gamma_3^+) +m(M_2^+) \mod 2 $ \\
\hline
\multirow{2}{*}{164} & \multirow{2}{*}{$P\bar{3}m1$}  & $\mathds{Z}_2$ \\
\cline{3-3}
&& $m(A_1^-)+m(A_3^-) +m(\Gamma_1^-) + m(\Gamma_3^+) +m(L_2^+)+m(M_2^+) \mod 2 $ \\
\hline
\multirow{2}{*}{165} & \multirow{2}{*}{$P\bar{3}c1$}  & $\mathds{Z}_2$ \\
\cline{3-3}
&& $m(A_3)+m(\Gamma_1^-)+m(\Gamma_3^+) + m(M_2^+) \mod 2 $ \\
\hline
\multirow{2}{*}{166} & \multirow{2}{*}{$R\bar{3}m$}  & $\mathds{Z}_2$ \\
\cline{3-3}
&& $m(F_2^+)+m(L_2^+) +m(T_1^-) + m(T_3^+) +m(\Gamma_1^-)+m(\Gamma_3^+) \mod 2 $ \\
\hline
\multirow{2}{*}{167} & \multirow{2}{*}{$R\bar{3}c$}  & $\mathds{Z}_2$ \\
\cline{3-3}
&& $m(F_2^+)+m(\Gamma_2^+)+m(\Gamma_3^-) \mod 2 $ \\
\hline
\multirow{2}{*}{168} & \multirow{2}{*}{$P6$}  & $\mathds{Z}_2$ \\
\cline{3-3}
&& $m(A_1)+m(M_1) \mod 2 $ \\
\hline
\multirow{4}{*}{175} & \multirow{4}{*}{$P6/m$}  & $(\mathds{Z}_2, \mathds{Z}_2,\mathds{Z}_2) $ \\
\cline{3-3}
&& $m(A_1^+)+m(A_1^-)+m(\Gamma_1^+)+m(M_1^-) \mod 2 $ \\
&& $m(A_1^+)+m(A_1^-)+m(M_1^+)+m(M_1^-) \mod 2 $ \\
&& $m(A_1^+)+m(A_1^-)+m(A_2^+)+m(\Gamma_1^+)+m(L_2^+)+m(M_1^-) \mod 2 $ \\
\hline
\multirow{2}{*}{176} & \multirow{2}{*}{$P6_3/m$}  & $\mathds{Z}_2$ \\
\cline{3-3}
&& $m(M_1^+)+m(\Gamma_1^+) \mod 2 $ \\
\hline
\multirow{1}{*}{177} & \multirow{1}{*}{$P622$}  & none \\
\hline
\multirow{1}{*}{183} & \multirow{1}{*}{$P6mm$}  & none \\
\hline
\multirow{2}{*}{184} & \multirow{2}{*}{$P6cc$}  & $\mathds{Z}_2$  \\
\cline{3-3}
&& $m(A_3A_4)+m(M_3) \mod 2 $ \\
\hline
\multirow{1}{*}{191} & \multirow{1}{*}{$P6/mmm$}  & none \\
\hline
\multirow{2}{*}{192} & \multirow{2}{*}{$P6/mcc$}  & $\mathds{Z}_2$  \\
\cline{3-3}
&& $m(A_5) +m(\Gamma_1^-)+m(\Gamma_3^+)+m(\Gamma_5^-)+m(\Gamma_6^-)+m(M_2^+)+m(M_3^+) \mod 2 $ \\
\hline
\multirow{1}{*}{193} & \multirow{1}{*}{$P6_3/mcm$}  & none \\
\hline
\multirow{1}{*}{194} & \multirow{1}{*}{$P6_3/mmc$}  & none \\
\hline
\multirow{1}{*}{202} & \multirow{1}{*}{$Fm\bar{3}$}  & none \\
\hline
\multirow{1}{*}{209} & \multirow{1}{*}{$F432$}  & none \\
\hline
\multirow{1}{*}{225} & \multirow{1}{*}{$Fm\bar{3}m$}  & none \\
\hline
\multirow{1}{*}{228} & \multirow{1}{*}{$Fd\bar{3}c$}  & none \\
\hline
\end{longtable}
\endgroup

\section{Ten Band Model for Twisted Bilayer Graphene}
\label{app:Aaron}

In this section, we put forward a variation of the ten band model discussed in Ref.~\onlinecite{2018arXiv180802482P}. Our ten band model captures the flat band irreps of twisted bilayer graphene via an $S$-matrix construction consisting of $6$ orbitals in the $L$ sublattice and $4$ orbitals in the $\tilde L$ sublattice, yielding two flat bands. However, we compute the Wilson loop\cite{2018arXiv180710676S,Alexandradinata:2012sp} of the flat bands in our model and find it has winding number $2$, instead of winding number $1$ as in the Bistritzer-MacDonald Hamiltonian \cite{2018arXiv180710676S,2018arXiv180802482P,2018arXiv180409719B}. In contrast, the model of \Ref{2018arXiv180802482P}, however, in the larger lattice has orbitals at both the $1a$ and $3f$ positions, and thus does not satisfy uniform pairing because the orbitals at the different Wyckoff positions are not related by symmetry.

\begin{figure*}
\includegraphics[width=.95\textwidth]{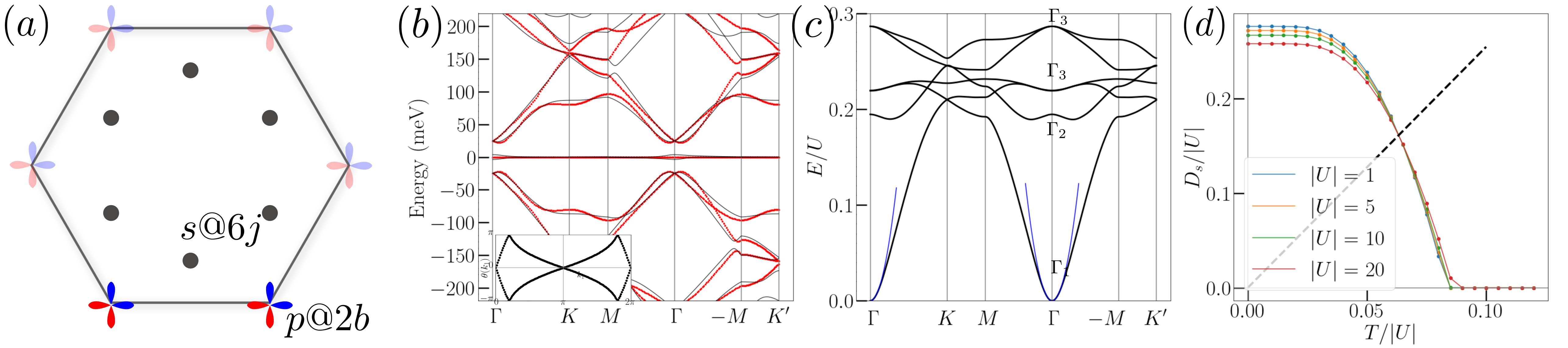}
\caption{We summarize the 10-band UPC model of TBG. $(a)$ shows the electron orbitals of the model in the unit cell, $(b)$ shows the single-particle bands with the Wilson loop over the flat bands inset, $(c)$ shows the Cooper pair spectrum, and $(d)$ shows the mean-field superfluid weight at finite temperature.}
\label{fig:TBGUPC}
\end{figure*}

To remedy this issue we propose a modified $S$-matrix which instead has in the $L$ lattice $s$ orbitals in the 6j position, which are related by $C_{6z} \mathcal{T}$ and hence have uniform pairing.  The $\tilde L$ lattice possesses $p_x, p_y$ orbitals at the $2c$ positions (the hexagon).  In SSG $177.151$, the irreps induced by the two sublattices read
\begin{align}
E_{2c} \uparrow G &= 2 \Gamma_3 + (2K_1 \oplus K_2K_3) + (2M_1 \oplus 2M_2) \\
A_{6j} \uparrow G &= (\Gamma_1 \oplus \Gamma_2 \oplus 2\Gamma_3) + (2K_1 \oplus 2K_2K_3) + (3M_1 \oplus 3M_2) \\
{\cal B}_{\text{FB}} &= (\Gamma_1 \oplus \Gamma_2) + K_2K_3 + (M_1 \oplus M_2).
\label{}
\end{align}  The irreps of the flat band correctly match the irreps carried by the flat bands in twisted bilayer graphene, and thus these two bands must be topological.  Note that as this is a lattice model, one cannot correctly implement the particle-hole symmetry that protects the strong topology in TBG \cite{2021PhRvB.103t5412S}.  The real-space Hamiltonian involves only NN hopping and NNN hopping between $s$ and $p$ orbitals, and there are only four complex independent parameters:

\begin{align}
H_\text{kin}^{\nu,\sigma} &= \sum_{\RR} t_{NN,x}^{\nu,\sigma} c^\dagger_{\RR,1,\sigma,\nu} d_{\RR,p_x,1,\sigma,\nu} + t_{NN,y}^{\nu,\sigma} c^\dagger_{\RR,1,\sigma,\nu} d_{\RR,p_y,1,\sigma,\nu} \nonumber \\
&+ t_{NNN,x}^{\nu,\sigma} c^\dagger_{\RR,1,\sigma,\nu} d_{\RR-\aaa_2,p_x,2,\sigma,\nu} + t_{NNN,y}^{\nu,\sigma} c^\dagger_{\RR,1,\sigma,\nu} d_{\RR-\aaa_2,p_y,2,\sigma,\nu} + (p6') + (H.c.)
\label{}
\end{align} where $\sigma$ denotes the spin and $\nu$ denotes the valley, giving four identical copies of the kinetic Hamiltonian.  Here the $c$ orbitals belong to the $L$ sublattice, and are labeled $\alpha = 1 \dots 6$, labeling each of the $6j$ positions, while the $d$ orbitals belong to the $\tilde L$ sublattice and are labeled $\beta = \{p_x, 1\}, \{p_y, 1\}, \{p_x, 2\}, \{p_y, 2\}$, labeling the $2c$ positions.  In addition to the spin degree of freedom, the TBG model has a valley degree of freedom, $\nu$, from the modes arising from the Dirac cones from the $K$ and $K'$ points in the BZ.  $p6'$ indicates that the appropriate symmetry-related terms are added to yield the full tight-binding Hamiltonians.  We find that the choice of parameters
\begin{align}
t_{NN,x}^{\nu=+1,\sigma=\u} &= 76.50+2.69i, t_{NN,y}^{\nu=+1,\sigma=\u} = -14.64+35.81i, \nonumber \\
t_{NNN,x}^{\nu=+1,\sigma=\u} &= 12.17-7.44i, t_{NNN,y}^{\nu=+1,\sigma=\u} = 5.47-5.45i
\label{}
\end{align} yields an excellent fit of the low-lying bands to the BM model, as depicted in Fig.~\ref{fig:TBGUPC}(b).  The lattice vectors are $\aaa_1 = (-\frac{\sqrt{3}}{2},-\frac{1}{2})a_0, \aaa_2 = (\frac{\sqrt{3}}{2},-\frac{1}{2})a_0$, where $a_0$ is the unit cell length.

The main difference between this ten-band model and previous models arises from the addition of the valley degree of freedom.  The most natural Hubbard-type interaction would be one that includes all spin and valleys summed together in the density:

\begin{align}
H = \dfrac{|U|}{2} \sum_{\RR,\alpha} \left(\sum_{\nu,\sigma} {\bar n}_{\RR,\alpha,\sigma,\nu} \right)^2.
\label{}
\end{align}

However, we find that this interaction is not tractable with our eta-pairing formalism.  Instead, however, the slightly different interaction
\begin{align}
H = \frac{|U|}{2} \sum_{\RR,\alpha} ({\bar S}^z_{\RR, \alpha, \nu\sigma = +1})^2 + ({\bar S}^z_{\RR, \alpha, \nu\sigma = -1})^2,
\label{}
\end{align} where the spin-valley operators ${\bar S}^z_{\RR, \alpha, \nu\sigma}$ are defined as
\begin{align}
{\bar S}^z_{\RR, \alpha, \nu\sigma = +1} &= {\bar n}_{\RR, \alpha, \nu = +1, \sigma = \uparrow} - {\bar n}_{\RR, \alpha, \nu = -1, \sigma = \downarrow} \\
{\bar S}^z_{\RR, \alpha, \nu\sigma = -1} &= {\bar n}_{\RR, \alpha, \nu = +1, \sigma = \downarrow} - {\bar n}_{\RR, \alpha, \nu = -1, \sigma = \uparrow},
\label{eq:TBG_interaction}
\end{align}  can be recast into an eta-pairing problem.  So long as $U(1)$ valley, $U(1)$ spin, and $\cal T$ are preserved, there are two eta-pairing operators that commute with the interaction Hamiltonian
\begin{align}
\eta_{+1}^\dagger = \sum_{\kk,m} \gamma^\dagger_{\kk,m,\nu = +1, \sigma = \uparrow} \gamma^\dagger_{-\kk,m,\nu = -1, \sigma = \downarrow} \\
\eta_{-1}^\dagger = \sum_{\kk,m} \gamma^\dagger_{\kk,m,\nu = +1, \sigma = \downarrow} \gamma^\dagger_{-\kk,m,\nu = -1, \sigma = \uparrow}.
\label{}
\end{align} This spin-valley-polarized model splits into two eta-pairing models, each giving the Cooper pair spectrum shown in Fig.~\ref{figure1}(c).  The two Hamiltonians are labeled by spin-valley, $\nu \sigma = +1$ and $\nu \sigma = -1$, and are
\begin{align}
H^{\nu\sigma = +1} &= \sum_{\sigma\nu = +1} H^{\nu,\sigma}_\text{kin} + \frac{|U|}{2} \sum_{\RR,\alpha} ({\bar S}^z_{\RR, \alpha, \nu\sigma = +1})^2 \\
H^{\nu\sigma = -1} &= \sum_{\sigma\nu = -1} H^{\nu,\sigma}_\text{kin} + \frac{|U|}{2} \sum_{\RR,\alpha} ({\bar S}^z_{\RR, \alpha, \nu\sigma = -1})^2.
\label{}
\end{align}

We now deduce the irreducible representations of the Cooper pair bands.  Let us focus on the $\nu\sigma = +1$ sector.  As argued and rigorously proved in App.~\ref{app:Cooperpairspectrum}, the Cooper pairs themselves transform like $s$ orbitals induced from the larger sublattice $L$, which in this case is the 6j position.  We offer here an intuitive argument for why this must be so.  Consider an \emph{unprojected} Cooper pair operator at site $\RR$ and flavor $\alpha$, in the larger sublattice $L$:
\begin{align}
\eta_{\RR, \alpha}^\dagger = c^\dagger_{\RR,\alpha,\uparrow} c^\dagger_{\RR,\alpha,\downarrow}.
\label{}
\end{align}  Let us assume there is one orbital per position (at least in the larger sublattice $L$), so that the crystalline symmetry $g$ is a permutation matrix up to a phase: the orbital $\alpha$ is located at position ${\bf x}_\alpha$, and under $g$ this is mapped to orbital $\beta$ located at position ${\bf x}_\beta = g {\bf x}_\alpha$.  Under a crystalline symmetry $g$, this operator transforms as
\begin{align}
g \eta_{\RR, \alpha}^\dagger g^{-1} &= g c^\dagger_{\RR,\alpha,\uparrow} c^\dagger_{\RR,\alpha,\downarrow} g^{-1} \\
&= \sum_{\beta,\beta'} D[g]_{\alpha\beta} D^*[g]_{\alpha\beta'} c^\dagger_{g\RR,\beta,\uparrow} c^\dagger_{g\RR,\beta',\downarrow} \\
&= \eta_{g\RR, \beta}^\dagger,~~{\bf x}_\beta = g {\bf x}_\alpha.
\label{}
\end{align}  Any phases introduced in the permutation matrix $D$ is canceled by the complex conjugate from the time-reversed $D$ arising from the opposite spin.  Thus, these $\eta$ operators transform as $s$ orbitals.  When projecting into the flat bands, the orbitals belonging to the smaller sublattice $\tilde{L}$ possess no wavefunction weight, and so those bands do not participate in the Cooper pair Hamiltonian, leaving $s$ orbitals in the larger $L$ sublattice.

In our TBG model, these irreps read
\begin{align}
A_{6j} \uparrow G &= (\Gamma_1 \oplus \Gamma_2 \oplus 2\Gamma_3) + (2K_1 \oplus 2K_2K_3) + (3M_1 \oplus 3M_2).
\label{}
\end{align}  
The $6j$ position is non-maximal, so the bands form a composite representation which can in principle be separated into gapped bands. The $\Gamma_1$ irrep is the zero-energy Cooper pair which has $s$-wave symmetry. The $\Gamma_2$ irrep is odd under $C_{2x}$ and is of $p_z$ orbital character. The $\Gamma_3$ irrep is formed from angular momentum $\pm 1$ states under the $C_{3z}$ operator, and so is of $p_x,p_y$ orbital character (see \Tab{tab:32irreps} below). 
\begin{center}
\label{tab:32irreps}
\begin{table}
\begin{tabular}{c | c c c }
 & $1$ & $C_{3z}$ & $C_{2x}$ \\ 
 \hline
$\Gamma_1$ & $1$ & $1$ & $1$ \\ 
$\Gamma_2$ & $1$ & $1$ & $-1$ \\ 
$\Gamma_3$ & $2$ & $-1$ & $0$ \\ 
\end{tabular}
\caption{Irreps of the group $32$ which is the unitary subgroup of $6'22'$, the little group of the Gamma point in twisted bilayer graphene.}
\end{table}
\end{center}

\section{Mean-field superfluid weight in degenerate flat bands}
\label{app:meanfield}

In this section, we derive the mean-field superfluid weight of a set of $N_f$ degenerate isolated flat bands with uniform pairing. We take $e,\hbar,k_B=1$. In this section only, we use bra-ket notation for the single-particle eigenstates since we are working at the mean-field level. Our calculation is straightforward generalization of the $N_f = 1$ case demonstrated in \Ref{2022arXiv220311133H}. The result agrees with prior work (\Ref{2020PhRvL.124p7002X}) on degenerate flat bands, but correctly accounts for the minimal quantum metric in the self-consistent mean-field solution \cite{2022arXiv220311133H}. 

We describe our system with the attractive mean-field approximated Hubbard model
\begin{align}
  H &= \sum_{\mbf{k}} \crea{\vec{c}_{\mbf{k}}} H_{\rm BdG}(\mbf{k})
  \ani{\vec{c}_{\mbf{k}}} + \sum_{\mbf{k}} {\rm
    Tr }\, \tilde{h}^{\d}(\mbf{k}) - N_{orb} \mathcal{N}\mu -
  \mathcal{N}\sum_{\alpha}\frac{|\Delta_{\alpha}|^2}{U}, \\
  H_{\rm BdG}(\mbf{k}) &= \begin{pmatrix}
    \tilde{h}^{\u}(\mbf{k}) - \mu \mathbb{1}_{N_{orb}} & \vec{\Delta}\\
    \vec{\Delta}^{\dag} & -\tilde{h}^{\d}(\mbf{k})+\mu \mathbb{1}_{N_{orb}}
  \end{pmatrix}, \label{eq.hbdg}
\end{align}
where $\mu$ is the chemical potential, $\tilde{h}^{\sigma}$ is the
single-particle hopping matrix for spin $\sigma$ and
$\ani{\vec{c}_{\mbf{k}}} =
(\ani{c_{\mbf{k},\alpha=1,\u}},\ldots,\ani{c_{\mbf{k},\alpha=N_{orb},\u}},
\crea{c_{-\mbf{k},\alpha=1,\d}},\ldots,\crea{c_{-\mbf{k},\alpha=N_{orb},\d}})$. The
matrix $\vec{\Delta} = {\rm diag}(\Delta_1,\ldots,\Delta_{N_{orb}})$
contains the mean-field order parameters $\Delta_{\alpha} =
\Delta_{i\alpha} =
U\ave{\ani{c_{\mbf{R}\alpha\d}}\ani{c_{\mbf{R}\alpha\u}}}$, where
$\ani{c_{\mbf{R}\alpha\sigma}}$ destroys a particle with spin $\sigma$ at
orbital $\alpha$ in the unit cell $\mbf{R}$. We assume the order parameters
are independent of the unit cell. Since the we have shown in \App{app:excit} that the lowest energy gap is $s$-wave, zero momentum, and has the wavefunction $\propto \sum_{\mbf{R},\al} \bar{c}^\dag_{\mbf{R},\al,\u}\bar{c}^\dag_{\mbf{R},\al,\d}$, we expect our mean-field ansatz to accurately capture this Cooper pair branch.

The superfluid weight for this Hamiltonian is given
by~\cite{2022arXiv220311133H} (where $\partial_{i}=\partial/\partial k_{i}$)
\begin{align}
  [D_s]_{ij} = \frac{1}{\mathcal{N}V_c}
  \sum_{\mbf{k},ab}\frac{n_F(E_a)-n_F(E_b)}{E_b-E_a}\big[
    &\bra{\psi_a}\partial_{i}\widetilde{H}_{\mbf{k}} \ket{\psi_b}
    \bra{\psi_b} \partial_{j}\widetilde{H}_{\mbf{k}} \ket{\psi_a} \nonumber \\
    -&\bra{\psi_a}(\partial_{i}\widetilde{H}_{\mbf{k}}\gamma^z+\delta_{i}\Delta)
    \ket{\psi_b} \bra{\psi_b}
    (\partial_{j}\widetilde{H}_{\mbf{k}}\gamma^z
    +\delta_{j}\Delta)
    \ket{\psi_a}
    \big] - \frac{1}{V_c}C_{ij}, \label{eq.linrespresult}
\end{align}
and
\bea
  \partial_{i}\widetilde{H_{\mbf{k}}} &= \begin{pmatrix}
    \frac{\partial \tilde{h}^{\u}(\mbf{k}')}{\partial
      k_{i}'}\bigg|_{\mbf{k}'=\mbf{k}} & 0 \\
    0 & \frac{\partial \tilde{h}^{\d}(\mbf{k}')^*}{\partial
      k_{i}'}\bigg|_{\mbf{k}'=-\mbf{k}}
  \end{pmatrix}, \
  \delta_{i}\Delta = \begin{pmatrix}
    0 & \frac{{\rm d}\vec{\Delta}}{{\rm d}q_{i}}\bigg|_{\mbf{q}=\vec{0}} \\
    \frac{{\rm d}\vec{\Delta}^{\dag}}{{\rm d}q_{i}}\bigg|_{\mbf{q}=\vec{0}} & 0
  \end{pmatrix}, \
  C_{ij} = \frac{1}{U}\sum_{\alpha} \frac{{\rm
      d}\Delta_{\alpha}}{{\rm d}q_{i}}\frac{{\rm
      d}\Delta_{\alpha}^*}{{\rm d}q_{j}}\bigg|_{\mbf{q}=\vec{0}} +
  {\rm H.c.}
\eea
The vector $\mbf{q}$ in the derivatives ${\rm d}\Delta_{\alpha}/{\rm
  d}q_{i}$ corresponds to the insertion of an electromagnetic field with $\mbf{A}\sim\mbf{q}$, which changes the self-consistent order parameters as  $\Delta_{i\alpha}\to
\Delta_{i\alpha}e^{2i\mbf{q}\cdot (\mbf{R}+\mbf{r}_\al)}$. The Fermi distribution is $n_F(E)=1/(1+e^{\beta
  E})$, where $\beta=1/T$ is the inverse temperature. The eigenvalues
and eigenvectors of $H_{\rm BdG}$ are $E_a$ and $\ket{\psi_a}$
respectively, and $\gamma_z = \sigma_z\otimes \mathbb{1}_{N_{orb}}$,
where $\sigma_i$ are the Pauli matrices. The coefficient
$[n_F(E_a)-n_F(E_b)]/(E_b-E_a)$ should be understood as $-\partial
n_F(E)/\partial E$ when $E_a=E_b$. The total volume of the system is
$\mathcal{N}V_c$ where $V_c$ is the volume of a unit cell.

We will focus on the superfluid weight for a set of
degenerate flat bands $\mathcal{B}$ with energy
$\epsilon_{\overline{m}}$ ($\bar{m}$ denotes a flat band) in a system with time-reversal symmetry
($\tilde{h}^{\u}(\mbf{k}) = [\tilde{h}^{\d}(-\mbf{k})]^*$)
and uniform pairing. Due to time-reversal symmetry, there always
exists a choice of intra-unit-cell orbital positions which guarantees
that the derivatives of the order parameters at $\mbf{q}=\vec{0}$
vanish~\cite{2022arXiv220311133H}. With the uniform pairing condition, these positions are given by \Eq{eq:xalphafixed}. Let us assume that this is the
chosen set of positions, so that the superfluid weight of
Eq.~\ref{eq.linrespresult} simplifies to
\begin{equation}
  [D_s]_{ij} = \frac{1}{V_c\mathcal{N}}
  \sum_{\mbf{k},ab}\frac{n_F(E_a)-n_F(E_b)}{E_b-E_a}\big[
    \bra{\psi_a}\partial_{i}\widetilde{H}_{\mbf{k}} \ket{\psi_b}
    \bra{\psi_b} \partial_{j}\widetilde{H}_{\mbf{k}} \ket{\psi_a}
    -\bra{\psi_a}\partial_{i}\widetilde{H}_{\mbf{k}}\gamma^z
    \ket{\psi_b} \bra{\psi_b}
    \partial_{j}\widetilde{H}_{\mbf{k}}\gamma^z
    \ket{\psi_a}
    \big], \label{eq.simpl_lin_resp}
\end{equation}
This can be expressed in terms of the dispersion relations and Bloch
functions of the bands by expressing the eigenvectors of $H_{\rm BdG}$
as $\ket{\psi_a} =
\sum_{m=1}^{N_{orb}}(w_{+,am}\ket{+}\otimes\ket{m_{\mbf{k}}} +
w_{-,am}\ket{-}\otimes\ket{m_{\mbf{k}}})$,
where $\ket{m_{\mbf{k}}}$ is the eigenvector of
$\tilde{h}^{\u}(\mbf{k})$ with eigenvalue $\epsilon_{m,\mbf{k}}$. We
denote the eigenvectors of $\sigma_z$ with eigenvalues $\pm 1$ by
$\ket{\pm}$. With these definitions, Eq.~\eqref{eq.simpl_lin_resp} can
be rewritten as
\begin{align}
  [D_s]_{ij} &= \frac{1}{V_c\mathcal{N}}\sum_{\mbf{k}}\sum_{mn} C_{pq}^{mn}
  [j_{i}(\mbf{k})]_{mn}[j_{j}(\mbf{k})]_{pq}, \\
  C_{pq}^{mn} &= 4\sum_{ab} \frac{n_F(E_a)-n_F(E_b)}{E_b-E_a}
  w_{+,am}^*w_{+,bn}w_{-,bp}^*w_{-,aq},\\
  [j_{i}]_{mn} &=
  \bra{m_{\sigma,\mbf{k}}}\partial_{i}\tilde{h}^{\sigma}(\mbf{k})\ket{n_{\sigma,\mbf{k}}}\nonumber
  \\
  &= \delta_{mn}\partial_{i}\epsilon_{m,\mbf{k}} +
  (\epsilon_{m,\mbf{k}}-\epsilon_{n,\mbf{k}})\brakett{\partial_{i}m_{\mbf{k}}}{n_{\mbf{k}}} . \label{eq:currentmatrix}
\end{align}

In a system with uniform pairing, we can make the ansatz
$\boldsymbol{\Delta} = \Delta\mathbb{1}$,
and $H_{\rm BdG} =
\sum_{\mbf{k}}\sum_{m=1}^{N_{orb}}[(\epsilon_m-\mu)^2\sigma^z+\Delta^2\sigma^x]\otimes
\ket{m}\bra{m}$, where the momentum dependence of $\eps_m(\mbf{k})$ and $\ket{m_\mbf{k}}$ is suppressed. The
eigenvalues of $H_{BdG}$ are then $\pm E_m = \pm
\sqrt{(\epsilon_m-\mu)^2+\Delta^2}$, and the corresponding
eigenfunctions are $\ket{\psi_m^{+}} = (u_m\ket{+}+v_m\ket{-})\otimes
\ket{m}$ and $\ket{\psi_m^{-}} = (-v_m\ket{+} + u_m\ket{-})\otimes
\ket{m}$, where (again suppressing the momentum dependence)
\begin{equation}
\label{eq:vnok}
  u_m = \frac{1}{\sqrt{2}}\sqrt{1 + \frac{\epsilon_{m}-\mu}{E_m}}, \:\:
  v_m
  = \frac{1}{\sqrt{2}}\sqrt{1-\frac{\epsilon_{m}-\mu}{E_n}}.
\end{equation}
As Eq.~\ref{eq:currentmatrix} shows, all contributions to the superfluid weight which involve
$[j_{i}^{\sigma}]_{mn}$ with $m,n\in\mathcal{B}$ vanish (because $\eps_m(\mbf{k})$=$\eps_n(\mbf{k})$=$\eps_{\overline{m}}$ for the degenerate flat bands) vanish. Thus the
superfluid weight reduces to the geometric contribution
\begin{align}
  [D_s]_{ij} &=
  \frac{2|\Delta|^2}{V_c\mathcal{N}}\sum_{\mbf{k}}\sum_{m\in\mathcal{B}}\sum_{n\notin\mathcal{B}}
  \left[ \frac{{\rm tanh}(\beta E_{\overline{m}}/2)}{E_{\overline{m}}}
    - \frac{{\rm tanh}(\beta E_n/2)}{E_n}\right]
  \frac{\epsilon_n-\epsilon_{\overline{m}}}{\epsilon_n+\epsilon_{\overline{m}}
  - 2\mu} \left(
  \brakett{\partial_{i}m}{n}\brakett{n}{\partial_{j}m} + {\rm H.c.}
  \right) \nonumber \\
  &+ \frac{|\Delta|^2}{V_c\mathcal{N}}\sum_{\mbf{k}}\sum_{\substack{m\neq
      n\\m,n\notin\mathcal{B}}} \left[ \frac{{\rm tanh}(\beta
      E_m/2)}{E_m} - \frac{{\rm tanh}(\beta E_n/2)}{E_n} \right]
  \frac{\epsilon_n-\epsilon_m}{\epsilon_n+\epsilon_m-2\mu}\left(
  \brakett{\partial_{i}m}{n}\brakett{n}{\partial_{j}m} + {\rm H.c.}
  \right).
\end{align}
Let $W_n$ be the band gap between the bands in $\mathcal{B}$ and the
$n$th band. We now assume that $W_n\gg |\epsilon_{\overline{m}}-\mu|$
and $W_n\gg |\Delta|$ for all $n\notin\mathcal{B}$, and the superfluid
weight reduces to~\cite{Liang2017}
\begin{equation}
  [D_s]_{ij} = \frac{4\Delta^2}{V_c} \frac{{\rm tanh}(\beta
    E_{\overline{m}}/2)}{E_{\overline{m}}}g_{ij},
\end{equation}
where $g_{ij} =
(1/\mathcal{N})\sum_{\mbf{k}}\sum_{m\in\mathcal{B}}\sum_{n\notin\mathcal{B}}{\rm
  Re}\brakett{\partial_{i}m}{n}\brakett{n}{\partial_{j}m}$ is the
integral of the quantum metric of the set of bands $\mathcal{B}$. Since we assumed
above that the derivatives of the order parameters are zero, this
is in fact the minimal quantum metric~\cite{2022arXiv220311133H}.

At $T=0$, the filling and the pairing gap are given by
\begin{align}
  N_f\mathcal{N}\nu &= \sum_{\mbf{k}}v^2{\rm Tr} \, P_{\mbf{k}} \, =
  N_f\mathcal{N}v^2 \\
  \Delta_{\alpha} = \Delta &=
  \frac{|U|}{\mathcal{N}}\sum_{\mbf{k}}u v
       [P_{\mbf{k}}]_{\alpha\alpha} = \eps|U|\sqrt{\nu(1-\nu)},
\end{align}
with $\eps=N_f/N_L$, where $N_L$ is the number of orbitals where the flat band
states have a nonzero weight. We have used that $v = v_{\bar{m}},
u = u_{\bar{m}}$ are independent of $\mbf{k}$ and $\bar{m}$ for a flat band (see Eq. \Eq{eq:vnok}) and that
${\rm Tr}\, P = N_f$. Noticing that
$|\Delta|/E_{\overline{m}}=2u v=2\sqrt{\nu(1-\nu)}$,
we get the zero-temperature superfluid weight
\begin{equation}
  [D_s]_{ij} = 8\frac{|U|\eps}{V_{c}}\nu(1-\nu)g_{ij}.
\end{equation}

 \end{document}